\documentclass[11pt,a4paper]{article}
 \pdfoutput=1
\usepackage{jcappub}
\usepackage{amsmath}
\usepackage{amsfonts,color}
\usepackage{amssymb,float}
\usepackage{mathtools}
\usepackage[utf8]{inputenc}
\usepackage{url}
\usepackage{graphicx}
\usepackage{refstyle}
\usepackage{wasysym}
\usepackage{tabularx}
\usepackage{accents}
\usepackage{graphicx}
\usepackage{color}
\usepackage[dvipsnames]{xcolor}
\usepackage{hyperref}

\hypersetup{
    colorlinks=true,
    linkcolor=blue,
    filecolor=magenta,      
    citecolor=red
}

\usepackage{subfigure}		
\usepackage{hyperref} 
\usepackage[normalem]{ulem}

\makeatletter
\long\def\dddddot#1{%
  {\mathop {#1}\limits ^{\vbox to-1.4\ex@ {\kern -\tw@ \ex@ \hbox {\normalfont .....}\vss }}}%
}
\long\def\multidots#1#2{%
  \count@=0
  {{\mathop {#2}\limits ^{\vbox to-1.4\ex@ {\kern -\tw@ \ex@ \hbox {\normalfont %
  \loop%
  \ifnum#1>\count@%
  .%
  \advance\count@ by1%
  \repeat%
  }\vss }}}}%
}
\makeatother

\newcommand{\dd}{{\rm d}}
\newcommand{\udt}[3]{#1^{#2}_{\phantom{#2}#3}}
\newcommand{\udut}[4]{#1^{#2\phantom{#3}#4}_{\phantom{#2}#3\phantom{#4}}}

\newcommand{\dut}[3]{#1_{#2}^{\phantom{#2}#3}}
\newcommand{\dudt}[4]{#1_{#2\phantom{#3}#4}^{\phantom{#2}#3}}

\newcommand{\lc}[1]{\accentset{\circ}{#1}}

\title{\boldmath Well-tempered teleparallel Horndeski cosmology: a teleparallel variation to the cosmological constant problem}

\author[a]{Reginald Christian Bernardo,}
\author[b,c]{Jackson Levi Said,}
\author[b,c]{Maria Caruana,}
\author[d,e]{Stephen Appleby}

\affiliation[a]{National Institute of Physics, University of the Philippines Diliman, Quezon City 1101, Philippines}
\affiliation[b]{Institute of Space Sciences and Astronomy, University of Malta, Malta, MSD 2080}
\affiliation[c]{Department of Physics, University of Malta, Malta, MSD 2080}
\affiliation[d]{Asia Pacific Center for Theoretical Physics, Pohang, 37673, Korea}
\affiliation[e]{Department of Physics, POSTECH, Pohang 37673, Korea}

\emailAdd{rbernardo@nip.upd.edu.ph}
\emailAdd{jackson.said@um.edu.mt}
\emailAdd{maria.caruana.16@um.edu.mt}
\emailAdd{stephen.appleby@apctp.org}

\abstract{
\textbf{Well-tempering is a promising classical method of dynamically screening an arbitrarily large vacuum energy and generating a late-time, low energy de Sitter vacuum state. In this paper, we study for the first time self-tuning in teleparallel gravity and obtain well-tempered cosmological models in the teleparallel gravity analogue of Horndeski theory. This broadens the scope of well-tempered cosmology and teases the potentially far richer cosmological dynamics that could be anchored on teleperallel gravity. We expand the well-tempered recipe to its most general form so far and use it to search for the first well-tempered cosmologies in teleparallel gravity. We also study the cosmological dynamics in a well-tempered model and demonstrate the dynamical stability of the vacuum, the compatibility with a matter era, and the stability of the vacuum through a phase transition.}
}

\begin{document}

\maketitle
\flushbottom

\section{Introduction}
\label{sec:intro}

The standard model of cosmology posits that the current accelerated expansion of the Universe is driven by the vacuum energy of space-time. This scenario raises questions regarding the magnitude of the vacuum energy, which should receive contributions from all propagating degrees of freedom up to the UV cut-off of the effective field theory describing the matter sector. The extreme discrepancy between observed energy scale of the vacuum and the naive expectation from particle physics remains an open question, dubbed the `old Cosmological Constant (CC) problem' \cite{RevModPhys.61.1}. Numerous innovative solutions to this problem have been proposed \cite{Dolgov:1982gh,Kaloper:2013zca,Kaloper:2014fca,Brax:2019fgj,SobralBlanco:2020too,Lombriser:2019jia,Kaloper:2015jra,Arkani-Hamed:2002ukf,Amariti:2019vfv,Evnin:2018zeo}, but so far no consensus has been reached. This has led to a proliferation of alternative dark energy models, which seek to explain the accelerated expansion of the Universe using dynamical fields rather than the vacuum energy.  

In the seminal work \cite{Horndeski:1974wa}, the most general scalar-tensor gravity model possessing second order field equations was constructed within the curvature-based geometric formulation of gravity. The Horndeski model has four free functions of the scalar field and its kinetic term, which opens up a rich phenomenology that can potentially explain both dark energy and dark matter. It was also realised that the scalar field can provide some insight into the old CC problem, as models can be constructed to completely screen the space-time curvature from the net cosmological constant. These so-called self-tuning models, pioneered in Ref. \cite{Charmousis:2011bf}, partially break the Poincare invariance of the vacuum which allows for a dynamical cancellation of the vacuum energy via a property known as {\it degeneracy}. Building on these ideas, a different class of self-tuning models was constructed in Ref. \cite{Appleby:2018yci}, which introduced a different degeneracy condition to generate de Sitter and Minkowski vacuum states \cite{Appleby:2020njl,Appleby:2020dko,Linder:2020xey,Bernardo:2021bsg} screened from an arbitrarily large cosmological constant. Various phenomenological properties of this class of models -- dubbed well-tempered cosmology -- have been studied in the literature \cite{Emond:2018fvv, Bernardo:2021hrz}. The progenitor model (dubbed `Fab Four' due to the particular form of the action) has been extensively studied \cite{Copeland:2012qf,Starobinsky:2016kua,Torres:2018lni,Appleby:2015ysa,Appleby:2012rx}. Gravitational wave experiments \cite{LIGOScientific:2017vwq} with electromagnetic follow-up \cite{Coulter:2017wya,LIGOScientific:2017ync} have revealed that the graviton practically propagates at the speed of light \cite{LIGOScientific:2017zic} which poses a serious challenge to certain scalar-tensor theories including the Fab Four \cite{Lombriser:2015sxa}. However, we note that this narrative has been challenged in Ref. \cite{Babichev:2017lmw}.

Horndeski scalar-tensor gravity can be generalised in a number of interesting ways. One such possibility is that of using teleparallel geometry where the curvature associated with the Levi-Civita connection is replaced with torsion associated with the teleparallel connection \cite{Bahamonde:2021gfp,Aldrovandi:2013wha,Cai:2015emx,Krssak:2018ywd}. Teleparallel gravity (TG) embodies all theories in which the teleparallel connection is utilised \cite{Weitzenbock1923,Bahamonde:2021gfp}, which is curvature-less but satisfies metricity. Thus, the teleparallel Ricci scalar turns out to be identically zero, i.e. $R=0$. Naturally, the regular Ricci scalar $\lc{R}$ (over-circles represent quantities calculated with the Levi-Civita connection) remains nonzero, in general. Now, TG naturally produces a torsion scalar $T$ which is equal to the Ricci scalar $\lc{R}$ up to a boundary term. Thus, the action that is linear in the torsion scalar forms the \textit{Teleparallel equivalent of General Relativity} (TEGR). This division between the torsion and boundary scalars produces a much richer landscape of modified teleparallel theories of gravity as compared with modifications of the Einstein-Hilbert action. This facet of the theory is a result of a much weaker generalised Lovelock theorem in TG \cite{Lovelock:1971yv,Gonzalez:2015sha,Bahamonde:2019shr}.

Following the same reasoning as in $f(\lc{R})$ gravity \cite{Sotiriou:2008rp,Faraoni:2008mf,Capozziello:2011et}, TEGR can readily be generalised to $f(T)$ gravity \cite{Ferraro:2006jd,Ferraro:2008ey,Bengochea:2008gz,Linder:2010py,Chen:2010va,Bahamonde:2019zea, RezaeiAkbarieh:2018ijw}, which have shown promise in meeting the observational demands from recent cosmological surveys \cite{Benetti:2020hxp,Nesseris:2013jea,Anagnostopoulos:2019miu,Nunes:2018evm,Cai:2015emx,Farrugia:2016qqe,Deng:2018ncg}. This can also be generalised to various forms of scalar--tensor gravity \cite{Hohmann:2018vle,Hohmann:2018dqh,Hohmann:2018ijr}. However, during this development, recent observations on the speed of propagation of gravitational waves have severely restricted the rich structure of regular (or curvature-based) Horndeski gravity \cite{Ezquiaga:2017ekz} to a much smaller subset from its original formulation \cite{Horndeski:1974wa}. In this regime, TG is interesting due to its naturally lower-order nature, which has been shown to produce direct generalisations of regular Horndeski gravity \cite{Bahamonde:2019shr}. For this reason, a much richer structure has been developed in this regime of gravity when compared with the regular curvature-based form of Horndeski gravity. Moreover, the teleparallel analogue of Horndeski gravity allows for drastically more models that satisfy the speed of light gravitational wave constraint \cite{Bahamonde:2019ipm,Bahamonde:2021dqn}. Vitally, this means that certain disqualified regular Horndeski gravity models may possibly be revived in this setup.

In this work, we extend the model space of the well-tempered scalar-tensor theories, using the enhanced freedom afforded by Teleparallel gravity. We systematically explore the resulting self-tuning \textit{Tele}parallel Horn\textit{deski} cosmology -- well-tempered \textit{Teledeski} cosmology -- and find, among many other results, the well-tempering in the revived Horndeski and TG sectors, implicit closed-form solution to \textit{all} shift symmetric well-tempered models, and an extension of the no-tempering theorem in the tadpole-free, shift symmetric subclass of the theory (Sec. \ref{sec:teledeski_variations}). A summary of the well-tempered cosmological models singled out in this work is presented in Table \ref{tab:summary}. To the best of our knowledge, this is also the first work on self-tuning in the context of TG. Well-tempered models are an attempt to resolve the scenario of a possibly arbitrarily large vacuum energy using scalar-tensor theories. There are many other considerations to take into account such as observational constraints, weak field behaviour and, the existence of degenerate states, among others; however, well-tempering provides an interesting approach by which arbitrary energies can be tackled in late-time cosmology. In GR, this is not possible, and so we are motivated to study the prospect of teleparallel Horndeski theories since they are second order in nature and continue to satisfy the constraints given by the speed of gravitational wave observations, while still providing a rich framework in which to construct gravitational models.

The outline of this work is as follows. We first introduce Teledeski gravity (Sec. \ref{sec:teleparallel_horndeski}). We then review the ideas of self-tuning and degeneracy, and set up the necessary ingredients in order to achieve well-tempering (Sec. \ref{sec:well_tempered_recipe}). We use the recipe to systematically look for various well-tempered models in the expanded space of Teledeski gravity (Sec. \ref{sec:teledeski_variations}). Finally, we elucidate the well-tempered cosmological dynamics in a particular model (Sec. \ref{sec:dynamics}). We show the dynamical stability of the vacuum (Sec. \ref{subsec:dynamical_stability}), the model's compatibility with a matter universe (Sec. \ref{subsec:compatibility_with_matter}), and the stability of the vacuum through a phase transition of vacuum energy (Sec. \ref{subsec:phase_transitions}).

\textit{Conventions.} The geometrized units $c = 8 \pi G = 1$, where $c$ is the speed of light in vacuum and $G$ is Newton's gravitational constant, will be used throughout. A dot over a variable means differentiation with respect to the cosmic time $t$, e.g., $\dot{\phi} = d \phi/ d t$. Primes on a univariable function denote differentiation with respect to its argument, e.g., $f' = df(\tau)/d\tau$, $\mathcal{I}' = d\mathcal{I}(x)/dx$. The reader interested in recreating the results is highly encouraged to download the Mathematica notebooks from the \href{https://github.com/reggiebernardo/notebooks}{author's github page} \cite{reggie_bernardo_4810864}.

\section{Teledeski Gravity: A Teleparallel Analogue to Horndeski Theory} \label{sec:teleparallel_horndeski}

We provide a concise introduction to teleparallel gravity and the emergence of Teledeski theory (Sec. \ref{subsec:teledeski_theory}). Then, we write down the field equations of Teledeski cosmology (Sec. \ref{subsec:teledeski_cosmology}).

\subsection{Foundations}
\label{subsec:teledeski_theory}

GR is built on the curvature-based geometry in which the Levi-Civita connection $\udt{\lc{\Gamma}}{\sigma}{\mu\nu}$ (over-circles refer to any quantities based on the Levi-Civita connection) is used to build scalar invariants such as the Ricci scalar $\lc{R}$ which is associated with the Einstein-Hilbert action. TG offers an alternative approach in which this is replaced with the torsional teleparallel connection $\udt{\Gamma}{\sigma}{\mu\nu}$ \cite{Aldrovandi:2013wha,Bahamonde:2021gfp,Cai:2015emx,Krssak:2018ywd}.

Another important difference between curvature-based and teleparallel theories is that the metric is replaced as the fundamental dynamical object by a gravitational tetrad $\udt{e}{A}{\mu}$ and an inertial spin connection $\udt{\omega}{B}{C\nu}$, where Greek indices refer to coordinates on the general manifold while Latin ones refer to the local Minkowski spacetime. Thus, the tetrad acts as a soldering agent between these two frames. In practice, they can be used to raise and lower both kinds of indices through 
\begin{align}
    g_{\mu\nu} = \udt{e}{A}{\mu}\udt{e}{B}{\nu} \eta_{AB}\,,& &\text{and}& &\eta_{AB} = \dut{E}{A}{\mu}\dut{E}{B}{\nu} g_{\mu\nu}\,,\label{eq:metr_trans}
\end{align}
while also observing orthogonality conditions
\begin{align}
    \udt{e}{A}{\mu}\dut{E}{B}{\mu} = \delta_B^A\,,& &\text{and}& &\udt{e}{A}{\mu}\dut{E}{A}{\nu} = \delta_{\mu}^{\nu}\,,
\end{align}
where $\dut{E}{A}{\mu}$ is the inverse tetrad. Due to the freedom in choosing inertial frames with the spin connection, there exist an infinite number of tetrads that satisfy these relations. In this background, the tetrad-spin connection pair represent the freedom of the theory and act as the variables of the theory.

On the other hand, the teleparallel connection can be expressed explicitly in terms of these fundamental variables through \cite{Weitzenbock1923,Cai:2015emx,Krssak:2018ywd}
\begin{equation}
    \Gamma^{\lambda}{}_{\nu\mu}=\dut{E}{A}{\lambda}\partial_{\mu}\udt{e}{A}{\nu}+\dut{E}{A}{\lambda}\udt{\omega}{A}{B\mu}\udt{e}{B}{\nu}\,,
\end{equation}
where the spin connection is forced to be flat through the condition \cite{Bahamonde:2021gfp}
\begin{equation}
    \partial_{[\mu}\udt{\omega}{A}{|B|\nu]} + \udt{\omega}{A}{C[\mu}\udt{\omega}{C}{|B|\nu]} \equiv 0\,.
\end{equation}
There also exists frames in which all the spin connection components are compatible with zero, which is called the Weitzenb\"{o}ck gauge \cite{Weitzenbock1923} and is where most calculations are done.

The geometric foundation of TG is based on the replacement of the Levi-Civita connection with the teleparallel connection in the gravitational sector. This means that measures of curvature identically vanish, such as in the case of the Riemann tensor where $\udt{R}{\alpha}{\beta\gamma\epsilon}(\udt{\Gamma}{\sigma}{\mu\nu}) \equiv 0$)\footnote{Obviously, this does not mean that the regular Riemann tensor vanishes in general, i.e. $\udt{\lc{R}}{\alpha}{\beta\gamma\epsilon}(\udt{\lc{\Gamma}}{\sigma}{\mu\nu}) \neq 0$.}. In light of this, we can define a torsion tensor as \cite{Aldrovandi:2013wha,ortin2004gravity}
\begin{equation}
    \udt{T}{A}{\mu\nu} := 2\udt{\Gamma}{A}{[\nu\mu]}\,,
\end{equation}
where square brackets denote the antisymmetric operator, and $\udt{T}{A}{\mu\nu}$ represents the field strength of the theory \cite{Bahamonde:2021gfp}. This tensor transforms covariantly under local Lorentz transformations and diffeomorphisms \cite{Krssak:2015oua}, and can also be decomposed into irreducible parts \cite{PhysRevD.19.3524,Bahamonde:2017wwk}
\begin{align}
    a_{\mu} & :=\frac{1}{6}\epsilon_{\mu\nu\lambda\rho}T^{\nu\lambda\rho}\,,\\[4pt]
    v_{\mu} & :=\udt{T}{\lambda}{\lambda\mu}\,,\\[4pt]
    t_{\lambda\mu\nu} & :=\frac{1}{2}\left(T_{\lambda\mu\nu}+T_{\mu\lambda\nu}\right)+\frac{1}{6}\left(g_{\nu\lambda}v_{\mu}+g_{\nu\mu}v_{\lambda}\right)-\frac{1}{3}g_{\lambda\mu}v_{\nu}\,,
\end{align}
which are the axial, vector, and purely tensorial parts, respectively, and where $\epsilon_{\mu\nu\lambda\rho}$ is the totally antisymmetric Levi-Civita tensor in four dimensions. The decomposition naturally leads to the gravitational scalar invariants \cite{Bahamonde:2015zma}
\begin{align}
    T_{\text{ax}} & := a_{\mu}a^{\mu} = -\frac{1}{18}\left(T_{\lambda\mu\nu}T^{\lambda\mu\nu}-2T_{\lambda\mu\nu}T^{\mu\lambda\nu}\right)\,,\\[4pt]
    T_{\text{vec}} & :=v_{\mu}v^{\mu}=\udt{T}{\lambda}{\lambda\mu}\dut{T}{\rho}{\rho\mu}\,,\\[4pt]
    T{_{\text{ten}}} & :=t_{\lambda\mu\nu}t^{\lambda\mu\nu}=\frac{1}{2}\left(T_{\lambda\mu\nu}T^{\lambda\mu\nu}+T_{\lambda\mu\nu}T^{\mu\lambda\nu}\right)-\frac{1}{2}\udt{T}{\lambda}{\lambda\mu}\dut{T}{\rho}{\rho\mu}\,,
\end{align}
which form the family of the most general scalar invariants that are both parity preserving and involve only quadratic contractions of the torsion tensor. These scalars can be combined to produced the torsion scalar \cite{Bahamonde:2021gfp}
\begin{equation}
    T:=\frac{3}{2}T_{\text{ax}}+\frac{2}{3}T_{\text{ten}}-\frac{2}{3}T{_{\text{vec}}}=\frac{1}{2}\left(E_{A}{}^{\lambda}g^{\rho\mu}E_{B}{}^{\nu}+2E_{B}{}^{\rho}g^{\lambda\mu}E_{A}{}^{\nu}+\frac{1}{2}\eta_{AB}g^{\mu\rho}g^{\nu\lambda}\right)T^{A}{}_{\mu\nu}T^{B}{}_{\rho\lambda}\,,
\end{equation}
which is an important quantity since it can be shown to be equal to the Ricci scalar up to a boundary term
\cite{Bahamonde:2015zma}
\begin{equation}
    R=\lc{R}+T-\frac{2}{e}\partial_{\mu}\left(e\udut{T}{\lambda}{\lambda}{\mu}\right)=0\,,
\end{equation}
where the Ricci scalar calculated with the teleparallel connection, $R$, identically vanishes since this connection is curvature-less, while $e=\text{det}\left(\udt{e}{A}{\mu}\right)=\sqrt{-g}$ is the tetrad determinant. This equivalence can be written explicitly through
\begin{equation}
    \lc{R}=-T+\frac{2}{e}\partial_{\mu}\left(e\udut{T}{\lambda}{\lambda}{\mu}\right):=-T+B\,,
\end{equation}
where $B$ is the total divergence term. Thus, a linear torsion scalar guarantees that the ensuing field equations will be dynamically equivalent to GR, called the teleparallel equivalent to general relativity (TEGR) \cite{Hehl:1994ue,Aldrovandi:2013wha}. Recently, the question of degrees of freedom in TG has come into question. It is known that in addition to being dynamically equivalent to GR, TEGR also has the same number of degrees of freedom \cite{Ferraro:2016wht,Blagojevic:2020dyq}, which is consistent with the known tetrad-spin connection solutions that are compatible with TEGR. However, if we consider a straightforward generalization of TEGR, namely raising the Lagrangian to a generalized $f(T)$ gravity \cite{Ferraro:2006jd,Ferraro:2008ey,Bengochea:2008gz,Linder:2010py,Chen:2010va,Bahamonde:2019zea}, the situation is pointed different. Here, the tetrad-spin connection pair solution for Minkowski space that is in the Weitzenb\"{o}ck gauge turns out to be strongly coupled \cite{Jimenez:2020ofm}. It may be that one of the infinite number of other tetrad-spin connection pairs that produces the Minkowski metric does not express this problem. Another important point to raise is that flat Friedmann–Lema\^{i}tre–Robertson–Walker (FLRW) cosmology does not express extra degrees of freedom for the most popular tetrad-spin connection pair which points to strong coupling. Saying that, these two metric cases are found in different branches of the Hamiltonian analysis \cite{Blagojevic:2020dyq} which has led to a disagreement in the literature on whether this is problematic \cite{Golovnev:2020zpv}. In this context, we are even further motivated to consider the teleparallel analogue of Horndeski gravity which is well known to produce extra degrees of freedom \cite{Bahamonde:2021dqn}.

In the scalar field sector, TG continues to observe the minimal coupling prescription of GR in which the partial derivative is raised to the Levi-Civita covariant derivative for extra fields of this kind, namely \cite{Aldrovandi:2013wha,BeltranJimenez:2020sih}
\begin{equation}
    \partial_{\mu} \rightarrow \mathring{\nabla}_{\mu}\,,
\end{equation}
which applies in the matter sector. At this point, both the gravitational and scalar field are developed enough to consider the recently proposed teleparallel analog of Horndeski gravity \cite{Bahamonde:2019shr,Bahamonde:2019ipm,Bahamonde:2020cfv}, also called Bahamonde-Dialektopoulos-Levi Said (BDLS) theory. The framework of gravity depends on three limiting conditions, namely (i) the field equations must be at most second order in their derivatives of the tetrads; (ii) the scalar invariants will not be parity violating; and (iii) the number of contractions with the torsion tensor is limited to being at most quadratic. It is these conditions that force the ensuing theory to be finite while also appearing as an extension to regular Horndeski gravity that has been recognized in the curvature-based formulation of the theory \cite{Horndeski:1974wa}. This also points to a weaker generalized Lovelock theory in TG \cite{Lovelock:1971yv,Gonzalez:2015sha,Gonzalez:2019tky} which admits more possible terms into the gravitational action.

The combinations of conditions on the theory and the teleparallel geometry construction means that we can construct the following contributing scalar invariant \cite{Bahamonde:2019shr}
\begin{equation}
    I_2 = v^{\mu} \phi_{;\mu}\,,
\end{equation}
where $\phi$ is the scalar field, and which involves a linear contraction with the torsion tensor, while quadratic contractions allow for
\begin{align}
J_{1} & =a^{\mu}a^{\nu}\phi_{;\mu}\phi_{;\nu}\,,\\[4pt]
J_{3} & =v_{\sigma}t^{\sigma\mu\nu}\phi_{;\mu}\phi_{;\nu}\,,\\[4pt]
J_{5} & =t^{\sigma\mu\nu}\dudt{t}{\sigma}{\alpha}{\nu}\phi_{;\mu}\phi_{;\alpha}\,,\\[4pt]
J_{6} & =t^{\sigma\mu\nu}\dut{t}{\sigma}{\alpha\beta}\phi_{;\mu}\phi_{;\nu}\phi_{;\alpha}\phi_{;\beta}\,,\\[4pt]
J_{8} & =t^{\sigma\mu\nu}\dut{t}{\sigma\mu}{\alpha}\phi_{;\nu}\phi_{;\alpha}\,,\\[4pt]
J_{10} & =\udt{\epsilon}{\mu}{\nu\sigma\rho}a^{\nu}t^{\alpha\rho\sigma}\phi_{;\mu}\phi_{;\alpha}\,,
\end{align}
where semicolons represent covariant derivatives with respect to the Levi-Civita connection. Thus, we can write the teleparallel analog of Horndeski gravity as
\begin{equation}\label{action}
    \mathcal{S}_{\text{BDLS}} = \int d^4 x\, e\mathcal{L}_{\text{Tele}} + \sum_{i=2}^{5} \int d^4 x\, e\mathcal{L}_i+ \int d^4x \, e\mathcal{L}_{\rm m}\,,
\end{equation}
where the contributions from regular Horndeski gravity continue to appear as \cite{Horndeski:1974wa}
\begin{align}
\mathcal{L}_{2} & :=G_{2}(\phi,X)\,,\label{eq:LagrHorn1}\\[4pt]
\mathcal{L}_{3} & :=-G_{3}(\phi,X)\mathring{\Box}\phi\,,\\[4pt]
\mathcal{L}_{4} & :=G_{4}(\phi,X)\left(-T+B\right)+G_{4,X}(\phi,X)\left[\left(\mathring{\Box}\phi\right)^{2}-\phi_{;\mu\nu}\phi^{;\mu\nu}\right]\,,\\[4pt]
\mathcal{L}_{5} & :=G_{5}(\phi,X)\mathring{G}_{\mu\nu}\phi^{;\mu\nu}-\frac{1}{6}G_{5,X}(\phi,X)\left[\left(\mathring{\Box}\phi\right)^{3}+2\dut{\phi}{;\mu}{\nu}\dut{\phi}{;\nu}{\alpha}\dut{\phi}{;\alpha}{\mu}-3\phi_{;\mu\nu}\phi^{;\mu\nu}\,\mathring{\Box}\phi\right]\,,\label{eq:LagrHorn5}
\end{align}
which turn out to be identical due to the minimum coupling prescription except that they are calculated using the tetrad, and where
\begin{equation}
\label{eq:LTele}
    \mathcal{L}_{\text{Tele}}:= G_{\text{Tele}}\left(\phi,X,T,T_{\text{ax}},T_{\text{vec}},I_2,J_1,J_3,J_5,J_6,J_8,J_{10}\right)\,,
\end{equation}
where the kinetic term is defined as $X:=-\frac{1}{2}\partial^{\mu}\phi\partial_{\mu}\phi$, $\mathcal{L}_{\rm m}$ is the matter Lagrangian in the Jordan conformal frame, $\lc{G}_{\mu\nu}$ is the standard Einstein tensor, and where commas represent regular partial derivatives. In the limit that $G_{\text{Tele}} = 0$, we recover the regular form of Horndeski gravity, as we would expect. Also, this theory is invariant under local Lorentz transformations and diffeomorphisms due to the way the torsion tensor is constructed.

\subsection{Late Time Cosmology}
\label{subsec:teledeski_cosmology}

The field equations follow by varying the action with respect to the tetrad and the scalar field, and have been presented in Ref.~\cite{Bahamonde:2020cfv} (for brevity's sake we do not reproduce them here). In cosmology, the tetrad can be given in terms of the lapse $N(t)$ and the scale factor $a(t)$, and the scalar field can be written as $\phi(t)$. The modified Friedmann equations and the scalar field equation can be obtained by functional differentiation with respect to $N(t)$, $a(t)$, and $\phi(t)$. The lapse, being non-dynamical, can be set to unity after the variations, which is a choice we take. A flat homogeneous and isotropic background described by the FLRW metric,
\begin{equation}
    \dd s^2 = -N(t)^2 \dd t^2 + a(t)^2(\dd x^2 + \dd y^2 + \dd z^2)\,,
\end{equation}
allows us to choose tetrad components $\udt{e}{a}{\mu} = \textrm{diag}(N(t),a(t),a(t),a(t))$ which is compatible with the Weitzenb\"{o}ck gauge \cite{Krssak:2018ywd,Bahamonde:2021gfp}. This produces a Friedmann equation \cite{Bahamonde:2019shr}
\begin{equation}
    \mathcal{E}_{\rm Tele} + \sum_{i=2}^5 \mathcal{E}_i = 0\,,
\end{equation}
where
\begin{align}
    \mathcal{E}_{\rm Tele} &= 6 H\dot{\phi}\tilde{G}_{6,I_2}+12 H^2 \tilde{G}_{6,T}+2X \tilde{G}_{6,X}-\tilde{G}_{6}\,,\\
    \mathcal{E}_2 &= 2XG_{2,X}-G_2\,,\\
    \mathcal{E}_3 &= 6X\dot\phi HG_{3,X}-2XG_{3,\phi}\,,\\
    \mathcal{E}_4 &= -6H^2G_4+24H^2X(G_{4,X}+XG_{4,XX}) - 12HX\dot\phi G_{4,\phi X}-6H\dot\phi G_{4,\phi }\,,\\
    \mathcal{E}_5 &= 2H^3X\dot\phi\left(5G_{5,X}+2XG_{5,XX}\right) - 6H^2X\left(3G_{5,\phi}+2XG_{5,\phi X}\right)\,,
\end{align}
and
\begin{equation}
    \mathcal{L}_{\rm Tele}=\tilde{G}_6(\phi,X,T,I_{2})\,,
\end{equation}
which represents all the nonvanishing scalars for $G_{\rm{Tele}}$, the Hubble parameter is defined as $H = \dot{a}/a$, and dots denote derivatives with respect to cosmic time. The torsion scalar takes on the form $T = 6H^2$, while $I_2 = 3H\dot{\phi}$ and $X = \frac{1}{2} \dot{\phi}^2$, and commas denote partial derivatives. The second Friedmann equation is then given by
\begin{equation}
    \mathcal{P}_{\rm Tele}+\sum_{i=2}^5 \mathcal{P}_i=0\,,
\end{equation}
where
\begin{align}
    \mathcal{P}_{\rm Tele}&=-3 H\dot{\phi}\tilde{G}_{6,I_2}-12 H^2\tilde{G}_{6,T}-\frac{d}{dt}\Big(4H \tilde{G}_{6,T}+\dot{\phi}\,\tilde{G}_{6,I_2}\Big)+\tilde{G}_6\,,\\
    \mathcal{P}_2&=G_2\,,\\
    \mathcal{P}_3&=-2X\left(G_{3,\phi}+\ddot\phi G_{3,X} \right) \,,\\
    \mathcal{P}_4&=2\left(3H^2+2\dot H\right) G_4 - 12 H^2 XG_{4,X}-4H\dot X G_{4,X} - 8\dot HXG_{4,X}\nonumber\\
    & \phantom{gggg}-8HX\dot X G_{4,XX} +2\left(\ddot\phi+2H\dot\phi\right) G_{4,\phi} + 4XG_{4,\phi\phi} + 4X\left(\ddot\phi-2H\dot\phi\right) G_{4,\phi X}\,,\\
    \mathcal{P}_5&=-2X\left(2H^3\dot\phi+2H\dot H\dot\phi+3H^2\ddot\phi\right)G_{5,X} - 4H^2X^2\ddot\phi G_{5,XX}\nonumber\\
    & \phantom{gggg} +4HX\left(\dot X-HX\right)G_{5,\phi X} + 2\left[2\frac{d}{dt}\left(HX\right)+3H^2X\right]G_{5,\phi} + 4HX\dot\phi G_{5,\phi\phi}\,.
\end{align}

Finally, we can write the scalar field variation which produces the generalised Klein-Gordon equation
\begin{equation}
    \frac{1}{a^3}\frac{d}{dt}\Big[a^3 (J+J_{\rm Tele})\Big]=P_{\phi}+P_{\rm Tele}\,,
\end{equation}
where the standard Horndeski terms appear as $J$ and $P_{\phi}$ which come from the Lagrangian terms $\mathcal{L}_i$, where $i=2,..,5$ and \cite{Kobayashi:2011nu}
\begin{align}
    J &= \dot\phi G_{2,X} +6HXG_{3,X}-2\dot\phi G_{3,\phi} + 6H^2\dot\phi\left(G_{4,X}+2XG_{4,XX}\right)-12HXG_{4,\phi X}\nonumber\\
    & \phantom{gggggggg} +2H^3X\left(3G_{5,X}+2XG_{5,XX}\right) - 6H^2\dot\phi\left(G_{5,\phi}+XG_{5,\phi X}\right)\,,\\
    P_{\phi} &= G_{2,\phi} -2X\left(G_{3,\phi\phi}+\ddot\phi G_{3,\phi X}\right) + 6\left(2H^2+\dot H\right)G_{4,\phi} \nonumber\\
    & \phantom{gggggii} + 6H\left(\dot X+2HX\right)G_{4,\phi X} -6H^2XG_{5,\phi\phi}+2H^3X\dot\phi G_{5,\phi X}\,,
\end{align}
while $J_{\rm Tele}$ and $P_{\rm Tele}$ are new terms related to the teleparallel Horndeski, given by
\begin{align}
    J_{\rm Tele} &= \dot{\phi}\tilde{G}_{6,X}\,,\\
    P_{\rm Tele} &= -9 H^2\tilde{G}_{6,I_2}+\tilde{G}_{6,\phi}-3  \frac{d}{dt}\left(H\tilde{G}_{6,I_2}\right)\,.
\end{align}
For convenience, we rewrite some of these functional forms as 
\begin{align}
    G_2 \left( \phi, X \right) &= V \left( \phi, X \right)\,,\\
    G_3 \left( \phi, X \right) &= G \left( \phi, X \right)\,,\\
\label{eq:G4toA}    G_4 \left( \phi, X \right) &= \left( M_\text{Pl}^2 + A\left( \phi, X \right) \right) / 2\,,\\
\end{align}
and
\begin{equation}
    \tilde{G}_6 \left(\phi,X,T,I_{2}\right)  = \mathcal{G} \left(\phi,X,T,I_{2}\right)\,,
\end{equation}
where $M_\text{Pl}^2 = 1/\left( 8\pi G \right)$. The arguments of $\tilde{G}_6$ do not feature $T_{\rm ax}$ and $T_{\rm ten}$ since they vanish for a flat FLRW background, which also means that we can simply write $T_{\rm vec}$ in terms of the torsion scalar through $T=-(2/3)T_{\rm vec} = 6H^2$. We would like to emphasize that this only occurs at background level and that these quantities do not vanish at perturbative level \cite{Bahamonde:2019ipm}. We also leave out the quintic Horndeski sector for the rest of this work, i.e., $G_5 \left( \phi, X \right) = 0$. We do this firstly for clarity since the resulting models will become quite convoluted with the $G_5$ contribution due to the additional teleparallel contributions, but also to not overly complicate this first work on the topic. It would be interesting to further extend this analysis to include such terms. In what follows, we refer to $V$ and $G$ as Horndeski potentials, $A$ as the conformal Horndeski potential, and $\mathcal{G}$ as the Teledeski potential. Later, we also refer to $A\left(\phi, X \right)$ with the general $X$ dependence as the \textit{revived} Horndeski potential. This Horndeski sector was largely considered ruled out after the gravitational wave observation GW170817 \cite{LIGOScientific:2017vwq} and was only revived later in the broader Teledeski gravity \cite{Bahamonde:2019shr, Bahamonde:2019ipm}.

\section{The Well-Tempered Recipe}
\label{sec:well_tempered_recipe}

In this section, we describe well-tempering and setup the necessary ingredients to introduce sectors of a theory that admit this desirable feature.

Well-tempering is a dynamical screening mechanism designed to protect the spacetime from the influence of an arbitrarily large vacuum energy \cite{Appleby:2018yci, Appleby:2020njl}. This particular kind of self-tuning mechanism has two particular advantages: (1) it does not rely on the tuning of mass scales in the action, and (2) is compatible with the existence of a matter-dominated era. At its core, this dynamical cancellation of vacuum energy works by making the dynamical field equations degenerate on a de Sitter vacuum, or ``on-shell", all while keeping the time dependence in the Hamiltonian constraint (Friedmann equation). We summarize the key ingredients of this in what follows. 

In scalar field theories, the Hubble and scalar field equation can be generically written in the forms
\begin{equation}
\label{eq:Heq_generic}
\dot{H} = \ddot{\phi} \ \mathcal{Z} \left( \phi, \dot{\phi}, H \right) + \mathcal{Y}\left( \phi, \dot{\phi}, H \right)
\end{equation}
and
\begin{equation}
\label{eq:Seq_generic}
0 = \ddot{\phi} \ \mathcal{D} \left( \phi, \dot{\phi}, H \right) + \mathcal{C} \left( \phi, \dot{\phi}, H , \dot{H} \right) 
\end{equation}
where the functions $\mathcal{Y}$, $\mathcal{Z}$, $\mathcal{C}$, and $\mathcal{D}$ are determined by the theory under study. By writing Eqs. (\ref{eq:Heq_generic}) and (\ref{eq:Seq_generic}), we assume that the field equations of the theory is at most second-order in the derivatives of the metric and the scalar field. This is of course expected by design in Horndeski gravity and its teleparallel generalization. These will be revealed for Teledeski gravity in a short while. To get to well-tempered cosmologies, one first evaluates the dynamical equations on a de Sitter vacuum, i.e., $P_\Lambda = -\rho_\Lambda$ and $H\left(t\right) = h$. This ansatz is not generically a solution to the field equations for arbitrary constant $h$, as the choice overconstrains the dynamical system. Well-tempering resolves this by working in the space where the field equations are degenerate on-shell. In particular, Eqs. (\ref{eq:Heq_generic}) and (\ref{eq:Seq_generic}) can be made to be degenerate on-shell provided that
\begin{equation}
\label{eq:ddphi_coef_1}
\mathcal{Z} \sim \mathcal{D} , \ \ \ \ \text{on-shell}
\end{equation}
and 
\begin{equation}
\label{eq:ddphi_coef_0}
\mathcal{Y} \sim \mathcal{C} , \ \ \ \ \text{on-shell} .
\end{equation}
Alternatively, we may write down
\begin{equation}
\label{eq:degeneracy_equation}
\mathcal{Y} \mathcal{D} - \mathcal{C} \mathcal{Z} = 0 , \ \ \ \ \text{on-shell}  .
\end{equation}
We refer to Eq. (\ref{eq:degeneracy_equation}) as the degeneracy equation, and `on-shell' indicates that we have imposed the ansatz $H=h$, $\dot{H} = 0$.  This parametrizes the space of well-tempered cosmologies and its solutions are the ones sought for in practice. In addition to Eq. (\ref{eq:degeneracy_equation}), a few other consistency conditions must be satisfied to make sure that self-tuning occurs via a dynamical cancellation rather than by tuning of mass scales in the action. In particular, the coefficient functions of $\ddot{\phi}$ in Eqs. (\ref{eq:Heq_generic}) and (\ref{eq:Seq_generic}) must be nonzero, namely
\begin{equation}
\label{eq:consistency_1}
\mathcal{D} \neq 0 , \ \ \ \ \text{on-shell}
\end{equation}
and
\begin{equation}
\label{eq:consistency_2}
\mathcal{Z} \neq 0 , \ \ \ \ \text{on-shell} .
\end{equation}
Furthermore, the Hamiltonian constraint must have an \textit{explicit} scalar field time dependence on-shell, or rather, it should be of the form 
\begin{equation}
\label{eq:consistency_3}
3 h^2 = \rho_\Lambda + F \left[ \phi(t), \dot{\phi}\left(t\right) \right] , \ \ \ \ \text{on-shell} ,
\end{equation}
where $F$ is a functional of the self-tuning scalar field. Eqs. (\ref{eq:degeneracy_equation}), (\ref{eq:consistency_1}), (\ref{eq:consistency_2}), and (\ref{eq:consistency_3}) fully flesh out well-tempered cosmologies. We shall now look for its solutions in Teledeski cosmology.

The on-shell coefficient functions $\mathcal{Y}$, $\mathcal{Z}$, $\mathcal{C}$, and $\mathcal{D}$ in the Hubble and scalar field equations can be identified to be
\begin{equation}
\begin{split}
\mathcal{Y} = 
9 h^2 A_{XX} \dot{\phi}^4 & + 9 h^2 A_X \dot{\phi}^2-15 h A_{\phi X} \dot{\phi}^3-3 h A_\phi \dot{\phi}+3 A_{\phi \phi} \dot{ \phi }^2 \\
& +9 h G_X \dot{\phi}^3 -6 G_\phi \dot{ \phi }^2+9 h \mathcal{G}_{I_2} \dot{ \phi } - 12 h \mathcal{G}_{\phi T} \dot{ \phi } + 3 \mathcal{G}_X \dot{\phi}^2-3 \mathcal{G}_{\phi I_2} \dot{ \phi }^2+3 V_X \dot{ \phi }^2          ,
\end{split}
\end{equation}
\begin{equation}
\begin{split}
\mathcal{Z} =
-6 h A_{XX} \dot{\phi}^3 & -6 h A_X \dot{\phi}+3 A_{\phi X} \dot{ \phi }^2+3 A_\phi - 3 G_X \dot{ \phi }^2 \\
&  -3 \mathcal{G}_{I_2} - 36 h^2 \mathcal{G}_{T I_2}-9 h \mathcal{G}_{I_2 I_2} \dot{\phi} - 12 h \mathcal{G}_{X T} \dot{ \phi } - 3 \mathcal{G}_{X I_2} \dot{ \phi }^2             ,
\end{split}
\end{equation}
\begin{equation}
\begin{split}
\mathcal{C} = 
-9 h^3 A_{XX} \dot{ \phi }^3 &-9 h^3 A_X \dot{ \phi } - 3 h^2 A_{\phi X X} \dot{ \phi }^4+9 h^2 A_{\phi X} \dot{ \phi }^2+6 h^2 A_{\phi} \\
&  + 3 h A_{\phi \phi X} \dot{ \phi }^3 - 9 h^2 G_X \dot{ \phi }^2-3 h G_{\phi X} \dot{\phi}^3+6 h G_\phi \dot{ \phi }+G_{\phi \phi} \dot{ \phi }^2+\mathcal{G}_\phi \\
&  -9 h^2 \mathcal{G}_{I_2} -3 h \mathcal{G}_{X} \dot{ \phi } -3 h \mathcal{G}_{\phi I_2} \dot{ \phi } - 3 h V_X \dot{ \phi } - \mathcal{G}_{\phi X} \dot{ \phi }^2 - V_{\phi X} \dot{ \phi }^2+V_\phi     ,
\end{split}
\end{equation}
and
\begin{equation}
\begin{split}
\mathcal{D} = 
 -3 h^2 A_{XXX} \dot{ \phi }^4 & -12 h^2 A_{XX} \dot{ \phi }^2-3 h^2 A_X + 3 h A_{\phi X X} \dot{ \phi }^3 \\
& + 9 h A_{\phi X} \dot{ \phi } - 3 h G_{XX} \dot{ \phi }^3 - 6 h G_{X} \dot{ \phi } +G_{\phi X} \dot{ \phi }^2+2 G_\phi \\
& - \mathcal{G}_X - 9 h^2 \mathcal{G}_{I_2 I_2} - 6 h \mathcal{G}_{X I_2} \dot{ \phi } - \mathcal{G}_{XX} \dot{ \phi }^2-V_{XX} \dot{ \phi }^2 - V_X .
\end{split}
\end{equation}
By substituting these into Eqs. (\ref{eq:degeneracy_equation}), (\ref{eq:consistency_1}), (\ref{eq:consistency_2}), and (\ref{eq:consistency_3}), it can be checked that the degeneracy and consistency conditions correctly reduces to their Horndeski counterparts \cite{Appleby:2020njl}, i.e., in the limit $\mathcal{G} = 0$ and $A = A\left( \phi \right)$ \footnote{In particular, see equations (2.11-14) in Ref. \cite{Appleby:2020njl}}.

The progenitor self-tuning models \cite{Charmousis:2011bf} used a different form of degeneracy to achieve the same goal of tuning the vacuum energy. The alternative mechanism is to impose that the scalar field equation (\ref{eq:Seq_generic}) is identically satisfied on-shell, by demanding ${\mathcal D} = 0$ and ${\mathcal C}=0$. This eliminates one of the equations and the dynamical system is no longer over-constrained. Practically, this approach is simpler as it requires a solution to two linear equations -- ${\mathcal D} = 0$ and ${\mathcal C}=0$ -- rather than the non-linear expression (\ref{eq:degeneracy_equation}). We do not pursue the `trivial scalar field' class of models in this work.

The next section is devoted to a detailed analysis of the solutions of Eqs. (\ref{eq:degeneracy_equation}), (\ref{eq:consistency_1}), (\ref{eq:consistency_2}), and (\ref{eq:consistency_3}) in Teledeski cosmology.

\section{Teledeski Variations}
\label{sec:teledeski_variations}

In this section, we present the main results of this paper. We obtain various well-tempered cosmologies in Teledeski gravity, beginning with the simplest, pure-Teledeski case and going further with increasing complexity, by adding and mixing Horndeski and Teledeski potentials, inspired by regular Horndeski models being revived. A summary of well-tempered models obtained in this paper is presented in Sec. \ref{subsec:summary_wt_teledeski}. The reader uninterested in the detailed derivation of well-tempered cosmological models may proceed directly to Sec. \ref{subsec:summary_wt_teledeski}. On the other hand, we encourage the interested reader to download our Mathematica notebooks where all analytical and numerical results of this paper are derived in full, transparent, detail \cite{reggie_bernardo_4810864}. Each of the following sections shows various possible ways of formulating well-tempered Teledeski models.

\subsection{\texorpdfstring{$\mathcal{G}$}{}: Pure TG}
\label{subsec:GTele}

We consider the pure-Teledeski case where all of the Horndeski potentials vanish, i.e., $V = G = A = 0$. In the absence of the other Horndeski terms that are often associated with curvature-based gravity, this showcases the first self-tuning cosmology solely-anchored on the torsion.

To start, we assume that the Teledeski potential depends only on the linear scalar-torsion coupling $I_2$ as 
\begin{equation}
\label{eq:c1s1}
\mathcal{G} \left( \phi, X , T, I_2 \right) = \mathcal{I}\left( \dfrac{ I_2^2 }{ 2 \left(3 h\right)^2 } \right) . 
\end{equation}
This ansatz does not lead to a well-tempered cosmology. Nonetheless, it is the simplest nontrivial case and it paves the way for the derivation of well-tempered models in this section. By substituting Eq. (\ref{eq:c1s1}) into Eqs. (\ref{eq:degeneracy_equation}), (\ref{eq:consistency_1}), and (\ref{eq:consistency_2}), we obtain
\begin{equation}
\label{eq:degeneracy_eq_c1s1}
x \mathcal{I}'(x) \left(4 x \mathcal{I}''(x)+3 \mathcal{I}'(x)\right)=0 \,,
\end{equation}
\begin{equation}
\label{eq:consistency_1_c1s1}
2 x \mathcal{I}''(x)+\mathcal{I}'(x)\neq 0 \,,
\end{equation}
and
\begin{equation}
\label{eq:consistency_2_c1s1}
\frac{\sqrt{x} \left(x \mathcal{I}''(x)+\mathcal{I}'(x)\right)}{h}\neq 0 \,,
\end{equation}
respectively, where $x = \dot{\phi}^2/2$ and a prime denotes the differentiation of a univariable function with respect to its argument. The general solution to the degeneracy equation (\ref{eq:degeneracy_eq_c1s1}) can be written as
\begin{equation} \label{eq:c1s1sol}
    \mathcal{I}(x)=4 c_1 x^{1/4}+c_2 \,.
\end{equation}
Substituting this back into Eqs.~(\ref{eq:consistency_1_c1s1}) and (\ref{eq:consistency_2_c1s1}) further leads to
\begin{equation}
    \frac{c_1}{2 x^{3/4}}\neq 0\,,
\end{equation}
and
\begin{equation}
    \frac{c_1}{h x^{1/4}}\neq 0\,.
\end{equation}
The consistency conditions are therefore generally satisfied provided that $c_1 \neq 0$ and $| \dot{\phi} | < \infty$. This makes the theory (\ref{eq:c1s1sol}) nearly well-tempered. However, by evaluating the Friedmann constraint (\ref{eq:consistency_3}), we find that
\begin{equation}
    3 h^2  =\rho_\Lambda - c_2 \,.
\end{equation}
Identifying $c_2$ as a constant in the action, this implies that the resulting theory (\ref{eq:c1s1sol}) screens away the vacuum energy $\rho_\Lambda$ via the tuning of mass scales. In other words, the explicit time dependence of the scalar field drops out and so the theory does not well-temper.

To obtain a well-tempered model, we modify the Teledeski potential ansatz to
\begin{equation}
\label{eq:c1s2}
\mathcal{G} \left( \phi, X , T, I_2 \right) = l \phi + \mathcal{I}\left( \dfrac{ I_2^2 }{ 2 \left(3 h\right)^2 } \right)\,,
\end{equation}
where $l$ is a constant. The tadpole term $l \phi$ can also be taken into account through the Horndeski sector of the theory. Alternatively, it can also be sourced through the Teledeski potential. Moving forward, substituting Eq. (\ref{eq:c1s2}) into the degeneracy equation (Eq. (\ref{eq:degeneracy_equation})) leads to
\begin{equation}
\label{eq:degeneracy_eq_c1s2}
\mathcal{I}'(x) \left(12 h x^2 \mathcal{I}''(x)-\sqrt{2} l \sqrt{x}\right)+9 h x \mathcal{I}'(x)^2-\sqrt{2} l x^{3/2} \mathcal{I}''(x) = 0 \,.
\end{equation}
Interestingly, the consistency conditions in Eqs.~(\ref{eq:consistency_1}) and (\ref{eq:consistency_2}) continue to be given by Eqs.~(\ref{eq:consistency_1_c1s1}) and (\ref{eq:consistency_2_c1s1}). Now, Eq.~(\ref{eq:degeneracy_eq_c1s2}) can be recognized as a first-order differential equation for $\mathcal{I}'(x)$. The general solution to this equation has two branches:
\begin{equation}
\label{eq:c1s2sol}
\begin{split}
\mathcal{I}_{\mp}(x) = 
& \frac{1}{3 \sqrt{2} h l \sqrt{x}} \bigg( 
\mp l \sqrt{72 c_1 h^2 x^{3/2}+l^2 x^2} \\
& \pm18 c_1 h^2 \sqrt{x} \left(\ln (x)-2 \ln \left(l \sqrt{72 c_1 h^2 x^{3/2}+l^2 x^2}+36 c_1 h^2 \sqrt{x}+l^2 x\right)\right)+l^2 x \bigg) +c_2\,,
\end{split}
\end{equation}
where $c_1$ and $c_2$ are integration constants. In both branches, it can be confirmed that $c_1 \neq 0$ is generally sufficient to keep the $\ddot{\phi}$ terms in the Hubble and scalar field equations. The consistency conditions given by Eqs. (\ref{eq:consistency_1}) and (\ref{eq:consistency_2}) are therefore satisfied. Most importantly, the Hamiltonian constraint (\ref{eq:consistency_3}) for this solution is given by
\begin{equation}
\begin{split}
3h^2 = \rho_\Lambda & -c_2 \mp \frac{\sqrt{72 c_1 h^2 x^{3/2}+l^2 x^2}}{3 \sqrt{2} h \sqrt{x}} + l \left(\frac{\sqrt{x}}{3 \sqrt{2} h}-\phi (t)\right) \\
& \mp \frac{3 \sqrt{2} c_1 h }{l} \left(\ln (x)-2 \ln \left(l \sqrt{72 c_1 h^2 x^{3/2}+l^2 x^2}+36 c_1 h^2 \sqrt{x}+l^2 x\right)\right)\,.
\end{split}
\end{equation}
The time dependent terms on the right hand side specifically show that $\rho_\Lambda$ is being screened dynamically by the scalar field $\phi \left( t \right)$, and its kinetic density $x = \dot{\phi}^2/2$. The solution given by Eq. (\ref{eq:c1s2sol}) therefore well-tempers.

It must be noted that factoring in the nonzero uncoupled torsion terms $T = 6 h^2$ and $T_\text{vec} = -9 h^2$ into the Teledeski potential will not change its shape in $x$. The dependence arising from the nonzero scalar-torsion coupling $I_2$ is what matters in well-tempering. We proceed in the next sections using similar or slightly-varied forms of the Teledeski potential.

We recall that the quartic Horndeski potential was written as Eq. (\ref{eq:G4toA}). The pure-Teledeski case considered in this section therefore still contains a proper GR limit ($\mathcal{I} \rightarrow 0$) and propagating tensor modes. The consistency conditions (Eqs. (\ref{eq:consistency_1}) and (\ref{eq:consistency_2})) further secure the scalar field as a dynamical degree of freedom.

\subsection{\texorpdfstring{$A(\phi) + \mathcal{G}$}{}: Quartic Horndeski \texorpdfstring{$+$}{} TG}
\label{subsec:G4_phi_GTele}

We consider the Horndeski potentials $V = G = 0$ and $A = A\left(\phi\right)$. In regular Horndeski theory ($\mathcal{G} = 0$), it has been shown that this sector cannot well-temper, owing to the consistency conditions not being satisfied on-shell \cite{Appleby:2020njl}. We show that the addition of $\mathcal{G}$ can activate well-tempering in this sector. This makes the case of teleparallel gravity as an important addition to well-tempered cosmology.

\begin{figure}[ht]
\center
	\subfigure[ $\ln |\mathcal{D}| $ ]{
		\includegraphics[width = 0.47 \textwidth]{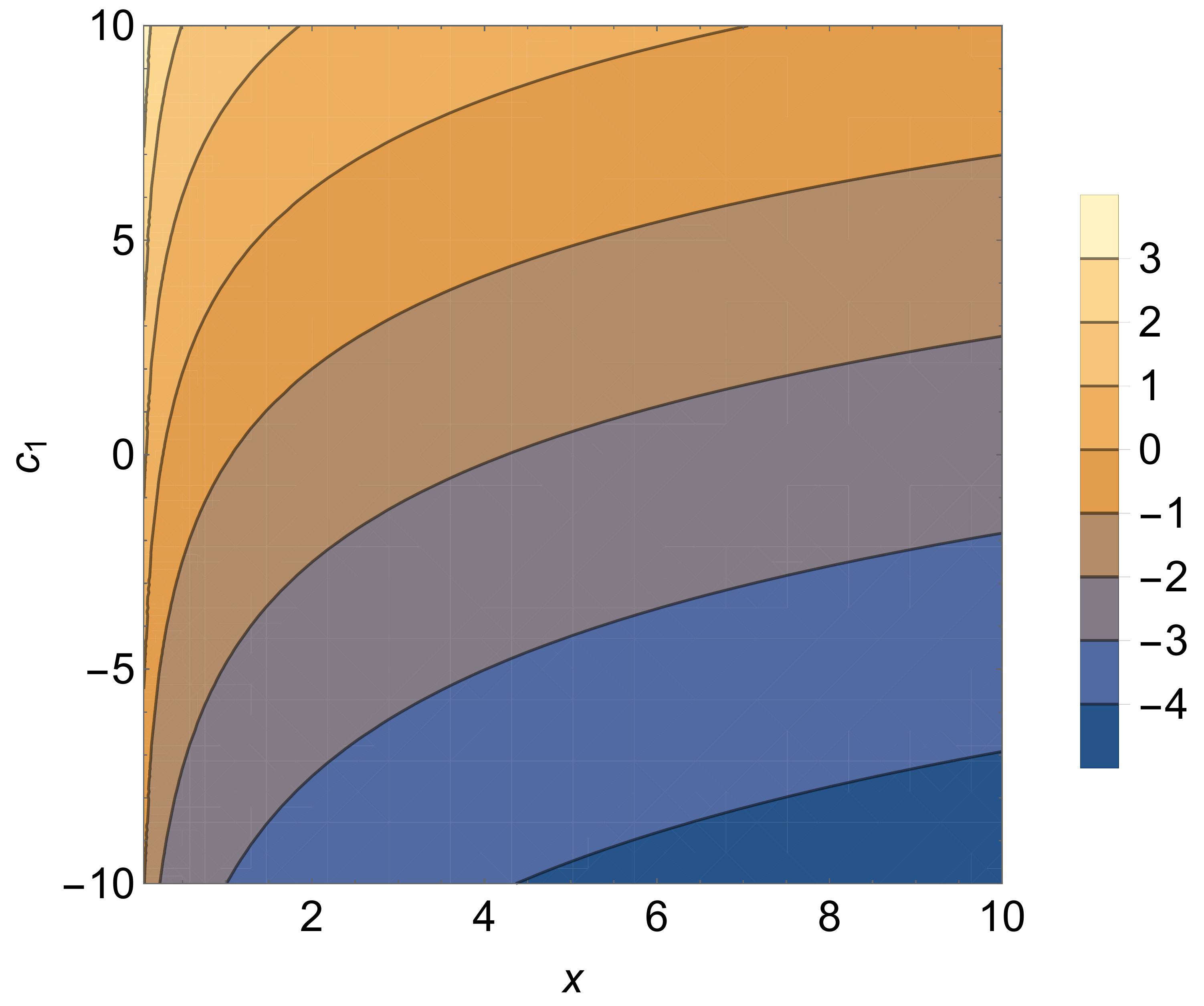}
		}
	\subfigure[ $\ln |\mathcal{Z}| $ ]{
		\includegraphics[width = 0.47 \textwidth]{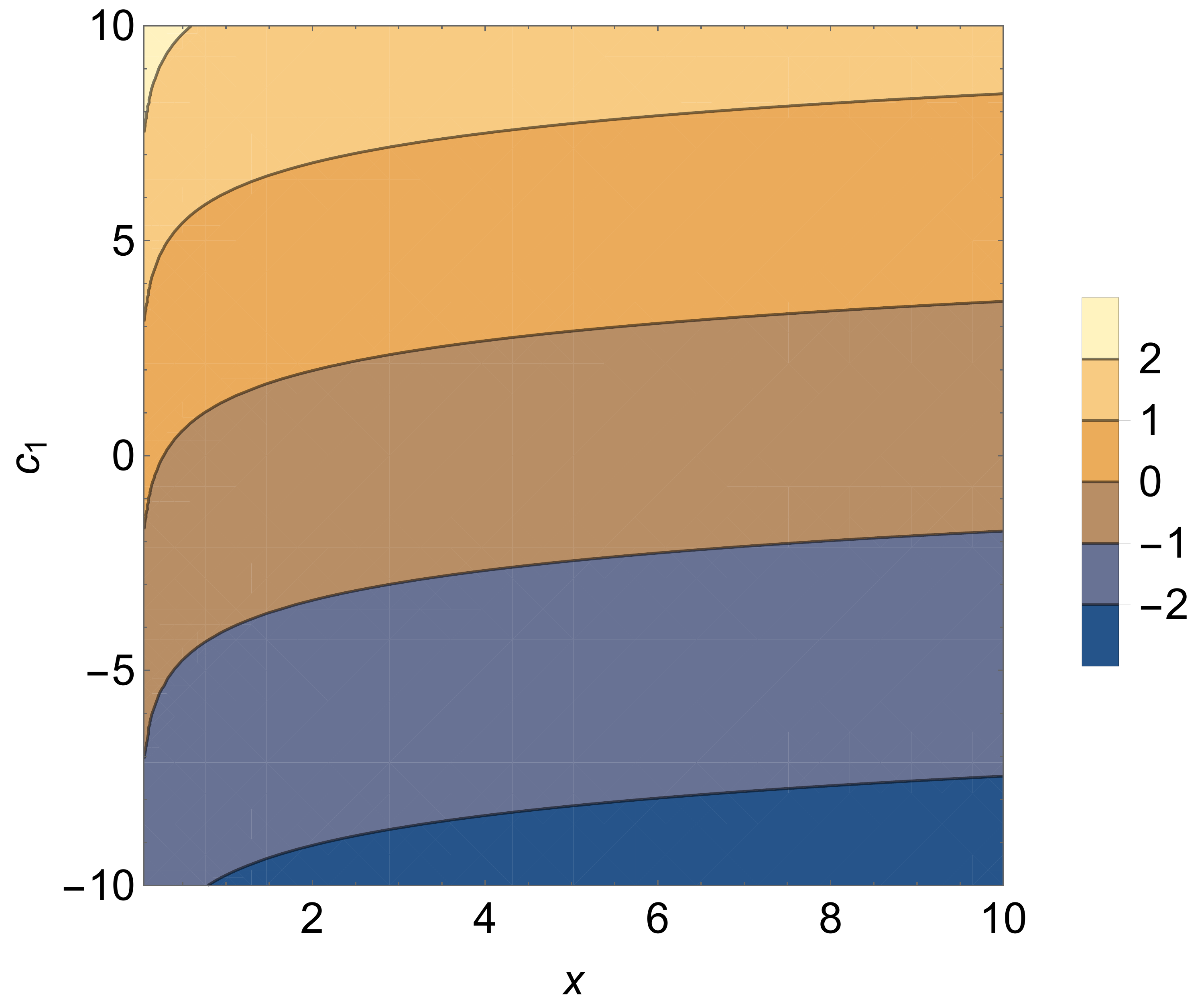}
		}
\caption{Plots of the consistency conditions (Eqs. (\ref{eq:consistency_1}) and (\ref{eq:consistency_2})) for the well-tempered solution given by Eq. (\ref{eq:c2s1sol}) with $h = 1$ and $\alpha = 1$. A remark: $h$ and $\alpha$ may be set to unity without loss of generality as $h$ can set away by defining a dimensionless time coordinate $\tau = h t$ and $\alpha$ by a conformal transformation.}
\label{fig:consistency_c2s1}
\end{figure}

To start, we take the Teledeski ansatz given by Eq. (\ref{eq:c1s1}) and the conformal Horndeski potential to be
\begin{equation}
\label{eq:c2s1}
    A \left( \phi \right) = \alpha \phi \,,
\end{equation}
where $\alpha$ is a constant. In this case, the degeneracy equation and consistency conditions become
\begin{equation}
3 \alpha  h \left(\alpha  h-\sqrt{2} x^{3/2} \mathcal{I}''(x)\right)+\mathcal{I}'(x) \left(4 x^2 \mathcal{I}''(x)-4 \sqrt{2} \alpha  h \sqrt{x}\right)+3 x \mathcal{I}'(x)^2=0\,,
\end{equation}
\begin{equation}
\label{eq:consistency_1_c2s1}
    2 x \mathcal{I}''(x)+\mathcal{I}'(x)\neq 0\,,
\end{equation}
and
\begin{equation}
\label{eq:consistency_2_c2s1}
    3 \alpha  h - 2 \sqrt{2} \sqrt{x} \left(x \mathcal{I}''(x)+\mathcal{I}'(x)\right) \neq 0\,.
\end{equation}
The degeneracy equation can be solved by identifying it as a first-order differential equation in $\mathcal{I}'(x)$. This leads to three distinct branches; but, only one of which corresponds to a real solution. This is given by
\begin{equation}
\label{eq:c2s1sol}
\begin{split}
\mathcal{I}'(x) = & 2^{-5/6} 3^{-2/3} \left(\sqrt{\frac{81 \alpha ^2 e^{c_1} h^2}{x^4}-\frac{12 e^{3 c_1/2}}{x^{9/2}}}+\frac{9 \alpha  e^{c_1/2} h}{x^2}\right)^{1/3} \\
& \phantom{ggg} + \frac{e^{c_1/2}}{2^{1/6} 3^{1/3} x^{3/2}} \left(\sqrt{\frac{81 \alpha ^2 e^{c_1} h^2}{x^4}-\frac{12 e^{3 c_1/2}}{x^{9/2}}}+\frac{9 \alpha  e^{c_1/2} h}{x^2}\right)^{-1/3}+\frac{3 \alpha  h}{\sqrt{2x}}\,,
\end{split}
\end{equation}
where $c_1$ is an integration constant. However, the consistency conditions (Eqs.~(\ref{eq:consistency_1_c2s1}) and (\ref{eq:consistency_2_c2s1})) in this case are too unwieldy, unfit for presentation. We instead show that they are satisfied in a respectable region of the parameter space by numerically plotting them. The results are shown in Fig. \ref{fig:consistency_c2s1}. 

These indeed show that the consistency conditions are satisfied, i.e., $\mathcal{D} \neq 0$ and $\mathcal{Z} \neq 0$, or $\ln|\mathcal{D}| > - \infty$ and $\ln|\mathcal{Z}| > -\infty$.
It is worth pointing out that $c_1$ appears in the exponential in Eq. (\ref{eq:c2s1sol}). In other words, unit changes in $c_1$ correspond to exponential changes in the parameter $e^{c_1}$ which directly affects Eq. (\ref{eq:c2s1sol}). In this way, Fig. \ref{fig:consistency_c2s1} shows an exponential region of the parameter space. The range $x \sim 10$ is also sufficient to reveal the shape dependence of the consistency conditions in $x$. Fig. \ref{fig:consistency_c2s1} shows that a smaller $x$ is preferred to keep the consistency conditions satisfied on-shell.

On the other hand, the Hamiltonian constraint on-shell in this solution becomes
\begin{equation}
\begin{split}
3 h^2 = \rho_\Lambda & -\mathcal{I}(x) + \frac{2^{7/6} x }{3^{2/3}} \left( \sqrt{\frac{81 \alpha ^2 e^{c_1} h^2}{x^4}-\frac{12 e^{3 c_1/2}}{x^{9/2}}}+\frac{9 \alpha  e^{c_1/2} h}{x^2}\right)^{1/3} \\
& +\frac{ 2^{11/6} e^{c_1/2}}{3^{1/3} \sqrt{x} } \left(\sqrt{\frac{81 \alpha ^2 e^{c_1} h^2}{x^4}-\frac{12 e^{3c_1/2}}{x^{9/2}}}+\frac{9 \alpha  e^{c_1/2} h}{x^2}\right)^{-1/3} -3 \alpha  h^2 \phi (t)+3 \sqrt{2} \alpha  h \sqrt{x}\,.
\end{split}
\end{equation}
This shows the dynamical cancellation of $\rho_\Lambda$ through well-tempering. The solution (\ref{eq:c2s1sol}) therefore corresponds to a well-tempered cosmology.

We note that the GR, or rather TEGR, case can be achieved in the limit of vanishing potentials ($A, \mathcal{I} \rightarrow 0$). Propagating tensor degrees of freedom can thus be expected as well as a dynamical, and self-tuning, scalar field owing to the consistency conditions (Eqs. (\ref{eq:consistency_1}) and (\ref{eq:consistency_2})).

\subsection{\texorpdfstring{$A(X) + \mathcal{G}$}{}: Revived Horndeski \texorpdfstring{$+$}{} TG}
\label{subsec:G4_phi_X_GTele}

We consider the Horndeski potentials $V = G = 0$ and $A = A\left(X\right)$. It is worth noting that this sector is largely considered as ruled out in regular Horndeski gravity because of its nonluminally-propagating tensor modes \cite{LIGOScientific:2017vwq, Ezquiaga:2017ekz}. The revival of this sector is one of the main impacts of Teledeski gravity \cite{Bahamonde:2019shr, Bahamonde:2019ipm}. We further show that this revived sector can be endowed with well-tempering.

We proceed with the Teledeski potential
\begin{equation}
\label{eq:c3s2}
    \mathcal{G} \left( \phi, X , T, I_2 \right) = l \phi + g\left( X \right) + \mathcal{I}\left( \dfrac{ I_2^2 }{ 2 \left(3 h\right)^2 } \right)\,,
\end{equation}
where $l$ is a constant and $g$, $\mathcal{I}$ are arbitrary functions of their arguments.
In this case, the degeneracy condition (\ref{eq:degeneracy_equation}) and consistency equations (Eqs. (\ref{eq:consistency_1}) and (\ref{eq:consistency_1})) can be written as
\begin{equation}
\label{eq:degeneracy_eq_c3s2}
\begin{split}
0 = 
& 2 x \left(6 h^2 x A''(x)+3 h^2 A'(x)+g'(x)+\mathcal{I}'(x)\right) \\
& \phantom{ggg} \times \bigg(12 h^2 x^2 A^{(3)}(x)+24 h^2 x A''(x)+3 h^2 A'(x) \\
& \phantom{ggg} \phantom{gggggggggggggggg} +2 x g''(x)+g'(x)+2 x \mathcal{I}''(x)+\mathcal{I}'(x)\bigg) \\
& \phantom{ggg} +\frac{2 \sqrt{2x}}{3 h}  \left(x \left(6 h^2 A''(x)+\mathcal{I}''(x)\right)+3 h^2 A'(x)+\mathcal{I}'(x)\right) \\
& \phantom{ggg} \times \left(18 \sqrt{2} h^3 x^{3/2} A''(x)+9 \sqrt{2} h^3 \sqrt{x} A'(x)+3 \sqrt{2} h \sqrt{x} g'(x)+3 \sqrt{2} h \sqrt{x} \mathcal{I}'(x)-l\right)\,,
\end{split}
\end{equation}
\begin{equation}
\label{eq:consistency_1_c3s2}
-12 h^2 x^2 A^{(3)}(x)-24 h^2 x A''(x)-3 h^2 A'(x)-2 x g''(x)-g'(x)-2 x \mathcal{I}''(x)-\mathcal{I}'(x)\neq 0\,,
\end{equation}
and
\begin{equation}
\label{eq:consistency_2_c3s2}
    -\frac{2 \sqrt{2} \sqrt{x} \left(x \left(6 h^2 A''(x)+\mathcal{I}''(x)\right)+3 h^2 A'(x)+\mathcal{I}'(x)\right)}{3 h}\neq 0\,.
\end{equation}
We solve the degeneracy equation by taking the form of the Teledeski potential $\mathcal{I}$ to be
\begin{equation}
\label{eq:c3s2ansatz}
    \mathcal{I}(x) = c_1 + 3h^2 A(x)-6 h^2 x A'(x) - g(x) + \xi(x)\,,
\end{equation}
where $c_1$ is a constant and $\xi$ is an arbitrary function. The particular form of Eq. (\ref{eq:c3s2ansatz}) with $\xi = 0$ can be determined as the solution to the degeneracy equation for $l = 0$. Pressing forward, substituting Eq. (\ref{eq:c3s2ansatz}) into Eq. (\ref{eq:degeneracy_eq_c3s2}), we obtain
\begin{equation}
\begin{split}
0 = 
& 2 x \xi '(x) \left(2 x \xi ''(x)+\xi '(x)\right) - \frac{2 \sqrt{2x} }{3 h} \left(3 \sqrt{2} h \sqrt{x} \xi '(x)-l\right) \\
& \phantom{ggg} \times \left(x \left(6 h^2 x A^{(3)}(x)+9 h^2 A''(x)+g''(x)-\xi ''(x)\right)+g'(x)-\xi '(x)\right)\,.
\end{split}
\end{equation}
An exact integral solution to this can be expressed in terms of $A''(x)$ as
\begin{equation}
\label{eq:c3s2sol}
\begin{split}
A''(x) = \frac{c_1}{x^{3/2}} + \frac{1}{x^{3/2}} \int^x du \bigg( & \frac{3 h \sqrt{u} \xi '(u) \left(2 u \left(g''(u)-2 \xi ''(u)\right)+2 g'(u)-3 \xi '(u)\right)}{6 \left(\sqrt{2} h^2 l \sqrt{u}-6 h^3 u \xi '(u)\right)} \\
& + \frac{\sqrt{2} l \left(u \left(\xi ''(u)-g''(u)\right)-g'(u)+\xi '(u)\right)}{6 \left(\sqrt{2} h^2 l \sqrt{u}-6 h^3 u \xi '(u)\right)} \bigg) \,.
\end{split}
\end{equation}
By substituting this into the consistency conditions (Eqs. (\ref{eq:consistency_1_c3s2}) and (\ref{eq:consistency_2_c3s2})), one further obtains
\begin{equation}
    -2 x \xi ''(x)-\xi '(x)\neq 0\,,
\end{equation}
and
\begin{equation}
    -\frac{2 \sqrt{2} x \xi '(x) \left(2 x \xi ''(x)+\xi '(x)\right)}{\sqrt{2} l-6 h \sqrt{x} \xi '(x)}\neq 0 \,.
\end{equation}
This can be satisfied for nearly any $\xi$ except for the solution of $2x \xi''(x) + \xi'(x) = 0$, or rather $\xi (x) = 2 \omega \sqrt{x} + \sigma$ where $\omega$ and $\sigma$ are arbitrary constants. The Hamiltonian constraint on-shell becomes
\begin{equation}
\begin{split}
3 h^2 = \rho_\Lambda & - c_1 - 6 h^2 A(x) + x \left(6 h^2 A'(x)-2 g'(x)+4 \xi '(x)\right)-l \phi (t)-\xi (x) \\
&  -12 c_1 h^2 \sqrt{x} -12 h^2 \sqrt{x} \int^x du \bigg( \frac{3 h \sqrt{u} \xi '(u) \left(2 u g''(u)+2 g'(u)-4 u \xi ''(u)-3 \xi '(u)\right)}{6 \left(\sqrt{2} h^2 l \sqrt{u}-6 h^3 u \xi '(u)\right)} \\
& \phantom{ggggggggggggggggggggggggggg} + \frac{\sqrt{2} l \left(-u g''(u)-g'(u)+u \xi ''(u)+\xi '(u)\right)}{6 \left(\sqrt{2} h^2 l \sqrt{u}-6 h^3 u \xi '(u)\right)} \, \bigg) \,.
\end{split}
\end{equation}
Vacuum energy is being cancelled by the time dependence in this equation. Therefore, the solution well-tempers.

We recognize that the above solution may not instantly be reflective of the dynamical cancellation of $\Lambda$ in the Hamiltonian constraint because of the integral. We can make this point explicit by substituting $g (x) = \beta x$ and $\xi(x) = \gamma x$, reminding that the solution is valid for arbitrary functions $g$ and $\xi$. In this case, it can be shown that $\gamma \neq 0$ is sufficient to satisfy the consistency conditions. On the other hand, the Hamiltonian constraint becomes
\begin{equation}
3 h^2 = \rho_\Lambda-c_1 -6 c_2 h^2-\dfrac{l^2 \ln \left(6 \gamma  h \sqrt{x}-\sqrt{2} l\right)}{9 \gamma  h^2}-\dfrac{\sqrt{2} l \sqrt{x}}{3 h}-l \phi (t) .
\end{equation}
This confirms that a cancellation of $\Lambda$ is in action on-shell and undisputably reveals that well-tempering can be accommodated in this revived sector of Horndeski theory.

\subsection{\texorpdfstring{$A(X) + V(X) + G(X) + \mathcal{G}$}{}: The Shift Symmetric Sector}
\label{subsec:shift_symmetric}

In this section, we obtain novel closed-form solutions to shift symmetric scalar-tensor theories extending to the revived Horndeski sector $G_4(X)$ and its broader teleparallel gravity reaches. To do this, consider
\begin{equation}
\label{eq:G2_ss} V\left( \phi, X \right) = l \phi + V \left( X \right)\,,
\end{equation}
\begin{equation}
\label{eq:G3_ss} G\left( \phi, X \right) = G\left( X \right)\,,
\end{equation}
\begin{equation}
\label{eq:G4_ss} A\left( \phi, X \right) = A\left(X\right)\,,
\end{equation}
and
\begin{equation}
\label{eq:GT_ss} \mathcal{G} \left( \phi, X , T, I_2 \right) = \mathcal{Q}\left(X\right) \mathcal{I}\left( \dfrac{ I_2^2 }{ 2 \left(3 h\right)^2 } \right)\,,
\end{equation}
where $l$ is the tadpole, $V$ and $G$ are the Horndeski potentials, $A$ is the revived Horndeski potential, and $\mathcal{Q}$ and $\mathcal{G}$ are the Teledeski potentials. 

Here, we obtain the general solution to the degeneracy equation (\ref{eq:degeneracy_equation}) by following in the footsteps of Ref. \cite{Appleby:2018yci}. We find this approach tailored to conveniently solve \textit{all} well-tempered shift symmetric cosmologies in \textit{closed form}.

The degeneracy equations in this case are regarded as Eqs.~(\ref{eq:ddphi_coef_1}) and (\ref{eq:ddphi_coef_0}). These can be solved by writing down
\begin{eqnarray}
\label{eq:fZD_ss} f\left(\dot{\phi}\right) \mathcal{Z} &=& \mathcal{D}\,, \\
\label{eq:fYC_ss} f\left(\dot{\phi}\right) \mathcal{Y} &=& \mathcal{C}\,,
\end{eqnarray}
where $f \left( x \right)$ is an arbitrary function that parametrizes the space of well-tempered theories. We emphasize that this is equivalent to solving the degeneracy equation (\ref{eq:degeneracy_equation}). This can be seen by simply substituting Eqs. (\ref{eq:fZD_ss}) and (\ref{eq:fYC_ss}) into Eq. (\ref{eq:degeneracy_equation}). The function $f$ can be considered to be a device that allows us to determine the space of well-tempered cosmology in shift symmetric theory.

Moving on, by defining a function $q$ via $f(x) = q\left( x^2/2 \right)/x$ and expressing the Horndeski potentials $V$ and $G$ through the combination
\begin{equation}
\label{eq:VG_wt_ss}
\mathcal{K}(x) = V'(x) + 3h \sqrt{2x} G'(x)\,,
\end{equation}
we can then obtain the degeneracy equations as
\begin{equation}
\label{eq:degeneracy_ss_1}
\begin{split}
0 = & 12 h^2 x^2 A^{(3)}(x)+24 h^2 x A''(x)+3 h^2 A'(x)+2 x \mathcal{K}'(x)+\mathcal{K}(x)+2 x \mathcal{Q}(x) \mathcal{I}''(x) \\
& \phantom{ggg} +4 x \mathcal{Q}'(x) \mathcal{I}'(x) +\mathcal{Q}(x) \mathcal{I}'(x)+2 x \mathcal{I}(x) \mathcal{Q}''(x)+\mathcal{I}(x) \mathcal{Q}'(x) \\
& \phantom{ggg} - \frac{q(x)}{3 h} \bigg( 6 h^2 A'(x)+3 \sqrt{2} h \sqrt{x} G'(x) \\
& \phantom{ggg} \phantom{gggggggg} + 2 \left(6 h^2 x A''(x)+x \mathcal{Q}(x) \mathcal{I}''(x)+x \mathcal{Q}'(x) \mathcal{I}'(x)+\mathcal{Q}(x) \mathcal{I}'(x)\right) \bigg)\,,
\end{split}
\end{equation}
and
\begin{equation}
\label{eq:degeneracy_ss_2}
l = \sqrt{2x} (3 h+q(x)) \left(6 h^2 x A''(x)+3 h^2 A'(x)+\mathcal{K}(x)+\mathcal{Q}(x) \mathcal{I}'(x)+\mathcal{I}(x) \mathcal{Q}'(x)\right)\,.
\end{equation}
These can be solved for the Horndeski potentials $V$ and $G$ as follows. First, the algebraic constraint (\ref{eq:degeneracy_ss_2}) is solved for $\mathcal{K}$, giving
\begin{equation}
    \mathcal{K}(x)=\frac{l}{\sqrt{2x} (3 h+q(x))} -6 h^2 x A''(x)-3 h^2 A'(x)-\mathcal{Q}(x) \mathcal{I}'(x)-\mathcal{I}(x) \mathcal{Q}'(x)\,.
\end{equation} 
This is then substituted into Eq. (\ref{eq:degeneracy_ss_1}) in order to determine the potential $G$. This also consequently determines $V$ via the constraint Eq. (\ref{eq:VG_wt_ss}). The exact solution is given by
\begin{equation}
\label{eq:V_wt}
\begin{split}
V'(x) = & \frac{l \left(6 h x q'(x)+3 h q(x)+q(x)^2\right)}{\sqrt{2x} q(x) (3 h+q(x))^2} + 6 h^2 x A''(x)+3 h^2 A'(x) \\
& \phantom{ggg} +2 x \mathcal{Q}(x) \mathcal{I}''(x)+\left(2 x \mathcal{Q}'(x)+\mathcal{Q}(x)\right) \mathcal{I}'(x)-\mathcal{I}(x) \mathcal{Q}'(x)\,,
\end{split}
\end{equation}
and
\begin{equation}
\label{eq:G_wt}
\begin{split}
G'(x)= & -\frac{l q'(x)}{q(x) (3 h+q(x))^2} -2 \sqrt{2} h \sqrt{x} A''(x)-\frac{\sqrt{2} h A'(x)}{\sqrt{x}} \\
& \phantom{ggg} -\frac{\sqrt{2} \sqrt{x} \mathcal{Q}(x) \mathcal{I}''(x)}{3 h}-\frac{\sqrt{2} \left(x \mathcal{Q}'(x)+\mathcal{Q}(x)\right) \mathcal{I}'(x)}{3 h \sqrt{x}}\,.
\end{split}
\end{equation}
Eqs. (\ref{eq:V_wt}) and (\ref{eq:G_wt}) should be taken as functionals $V \left[ q, A, \mathcal{Q}, \mathcal{I} \right]$ and $G \left[ q, A, \mathcal{Q}, \mathcal{I} \right]$ of arbitrary functions of $q$, the revived Horndeski potential $A$, and the Teledeski potentials $\mathcal{Q}$ and $\mathcal{I}$. We refer to $q$ as the well-tempered potential. This is the most general well-tempered cosmology in shift symmetric theory.

The result (Eqs. (\ref{eq:V_wt}) and (\ref{eq:G_wt})) can also be understood as follows. We started with \textit{five} free functions ($V$, $G$, $A$, $\mathcal{Q}$, and $\mathcal{I}$). The recipe takes one away as a price of well-tempering. So, in Eqs. (\ref{eq:V_wt}) and (\ref{eq:G_wt}), the Horndeski potentials $V$ and $G$ can no longer be specified freely, but rather, they turn into functionals of the \textit{four} arbitrary functions $q$, $A$, $\mathcal{Q}$, and $\mathcal{I}$. In other words, the well-tempered model defined by the theory potentials $V[q, A, \mathcal{Q}, \mathcal{I}]$, $G[q, A, \mathcal{Q}, \mathcal{I}]$, $A$, $\mathcal{Q}$, and $\mathcal{I}$ is determined by four arbitrary functions.

We now proceed further to the on-shell field equations. By substituting the above solution into the on-shell Hamiltonian constraint, we obtain
\begin{equation}
\begin{split}
3 h^2 =  \rho_\Lambda & -l \phi (t) +2 x V'(x)-V(x) +6 \sqrt{2} h x^{3/2} G'(x) + 12 h^2 x^2 A''(x) \\
& +12 h^2 x A'(x)-3 h^2 A(x)+4 x \mathcal{Q}(x) \mathcal{I}'(x)+2 x \mathcal{I}(x) \mathcal{Q}'(x)-\mathcal{Q}(x) \mathcal{I}(x)\,.
\end{split}
\end{equation}
This clearly shows the dynamical cancellation of $\Lambda$ on-shell. Substituting the well-tempered potentials (Eqs. (\ref{eq:V_wt}) and (\ref{eq:G_wt})) into the on-shell Hubble and scalar field equations, we obtain
\begin{equation}
\dot{x} q'\left(x\right) + 3h q\left(x\right) + q\left(x\right)^2 = 0\,.
\end{equation}
Both dynamical equations reduce to the same equation because of well-tempering. By noting that $\dot{x} = \dot{\phi}\ddot{\phi}$ and $\dot{q} = \dot{x} q'\left(x\right)$, this result can in fact be recognized as a Riccati equation
\begin{equation}
\label{eq:riccati}
\dot{y} \left( t \right) + y \left( t \right) \left( 3 h + y \left( t \right) \right) = 0 \,.
\end{equation}
The solution to this is given by
\begin{equation}
\label{eq:riccati_sol}
q \left( \dfrac{ \dot{ \phi }^2 }{2} \right) = \dfrac{ 3 h }{ \exp\left( 3h \left( t - \mathcal{T} \right) \right) -1 }\,,
\end{equation}
where $y\left( t \right) = q\left( \dot{\phi}^2/2 \right)$ and $\mathcal{T}$ is an integration constant. This provides an \textit{implicit} closed-form solution to well-tempered dynamics in shift symmetric theory, i.e., by choosing $q(x)$, the scalar field velocity $\dot{\phi}(t)$ can be immediately solved algebraically through Eq. (\ref{eq:riccati_sol}). This important result was first recognized in Ref. \cite{Bernardo:2021hrz} and now its extension to the revived Horndeski and teleparallel gravity sectors are also provided.

\subsection{The No-Tempering Theorem}
\label{subsec:no_tempering_theorem}

We continue our focus on the shift symmetric sector, but this time, on its tadpole-free ($l = 0$) subclass. It was shown in Ref. \cite{Bernardo:2021hrz} that the Horndeski sector known as kinetic gravity braiding ($A = 0$ and $\mathcal{G} = 0$) does not well-temper. We now check whether this no-tempering theorem extends to the broader shift symmetric sector admitting the revived Horndeski sector and teleparallel gravity terms.

We start by substituting Eqs. (\ref{eq:G2_ss}), (\ref{eq:G3_ss}), (\ref{eq:G4_ss}), and (\ref{eq:GT_ss}) into the conditions in Eqs. (\ref{eq:degeneracy_equation}), (\ref{eq:consistency_1}), and (\ref{eq:consistency_2}). The degeneracy equation then becomes
\begin{equation}
\begin{split}
0 =
& \left[ \sqrt{2} \left(3 h^2 \left(2 x A''(x)+A'(x)\right)+V'(x)+\mathcal{Q}(x) \mathcal{I}'(x)+\mathcal{I}(x) \mathcal{Q}'(x)\right)+6 h \sqrt{x} G'(x) \right] \\
& \times \bigg[ 3 \sqrt{2} h^2 \left(3 A'(x)+4 x \left(x A^{(3)}(x)+3 A''(x)\right)\right) \\
& \phantom{gggg} + 6 h \sqrt{x} \left(2 x G''(x)+3 G'(x)\right) \\
& \phantom{gggg} +\sqrt{2} \bigg( 2 x \left(V''(x)+3 \mathcal{Q}'(x) \mathcal{I}'(x)+\mathcal{I}(x) \mathcal{Q}''(x)\right)  \\
& \phantom{gggggggggg} +V'(x)+\mathcal{Q}(x) \left(4 x \mathcal{I}''(x)+3 \mathcal{I}'(x)\right)+\mathcal{I}(x) \mathcal{Q}'(x) \bigg) \bigg]\,.
\end{split}
\end{equation}
The solution to this admits two branches, namely
\begin{equation}
\label{eq:no_temp_b1}
\mathcal{I}(x)=\frac{c_1}{\mathcal{Q}(x)} + \dfrac{1}{\mathcal{Q}(x)} \int^x du \left( -3 h \left(2 h u A''(u)+h A'(u)+\sqrt{2} \sqrt{u} G'(u)\right)-V'(u) \right)\,,
\end{equation}
and
\begin{equation}
\label{eq:no_temp_b2}
\begin{split}
\mathcal{I}(x) = \frac{x^{1/4}}{\sqrt{\mathcal{Q}(x)}} \bigg[ c_2 + \int^x \dfrac{du}{ 4 u^{5/4} \sqrt{ 2 \mathcal{Q}(u)}}
\bigg( & 3 \sqrt{2} h^2 \left(A(u)-4 u \left(u A''(u)+A'(u)\right)\right) \\
& +c_1-12 h u^{3/2} G'(u)-2 \sqrt{2} u V'(u)+\sqrt{2} V(u) \bigg) \bigg] \,,
\end{split}
\end{equation}
where $c_1$ and $c_2$ are constants, or mass scales, that enter in the action. However, the first branch (\ref{eq:no_temp_b1}) leads to $\mathcal{D} = 0$, which is a clear violation of one of the consistency conditions. Hence, the first branch does not well-temper. On the other hand, the second branch generally satisfies the consistency conditions $\mathcal{D} \neq 0$ and $\mathcal{Z} \neq 0$. Pressing forward, it can be shown that the Hamiltonian constraint on-shell in the second branch becomes
\begin{equation}
\label{eq:consistency_3_no_temp_b2}
    3 h^2 = \rho_\Lambda + \left( \sqrt{2} c_1 \right) / 2\,.
\end{equation}
This reveals that on-shell, the self-tuning in the second branch-model  (\ref{eq:no_temp_b2}) occurs through the tuning of mass scales in the action, rather than well-tempering. Therefore, the theory does well-temper. 

This calculation extends the no-tempering theorem to the revived $G_4(X)$-Horndeski and Teledeski sectors: In the shift symmetric sector, without the tadpole, it is not possible to well-temper.

To end, we note that three assumptions entered the above proof. First, the degeneracy equation (\ref{eq:degeneracy_equation}) must be satisfied. Second, the coefficient functions of $\ddot{\phi}$, on-shell, in the Hubble and scalar field equations must not vanish, or rather Eqs. (\ref{eq:consistency_1}) and (\ref{eq:consistency_2}) must hold. Third, the scalar field time dependence must appear explicitly in the Hamiltonian constraint on-shell (Eq. (\ref{eq:consistency_3})). If any one of the above assumptions is violated, then we dismiss the solution, on the grounds that it does not well-temper.

\subsection{A Summary of Well-Tempered Teledeski Models}
\label{subsec:summary_wt_teledeski}

We present a summary of the well-tempered models obtained in this section in Table \ref{tab:summary}.

\begin{table}[H]
\center
\caption{A summary of well-tempered Teledeski potentials singled out in this paper.}
\begin{tabular}{| c | c |}
\hline
Sector & Teledeski Potential \\ \hline \hline
$\mathcal{G}$ & Eq. (\ref{eq:c1s2sol}) \\  \hline
$A(\phi) + \mathcal{G}$ & Eq. (\ref{eq:c2s1sol}) \\ \hline
$A(X) + \mathcal{G}$ &  Eqs. (\ref{eq:c3s2ansatz}) and (\ref{eq:c3s2sol}) \\ \hline
$A(X) + V(X) + G(X) + \mathcal{G}$ & Eqs. (\ref{eq:V_wt}) and (\ref{eq:G_wt}) \\ \hline
\end{tabular}
\label{tab:summary}
\end{table}

We stress that in all of these models, the degeneracy and consistency conditions are \textit{strictly} met. In other words, on-shell, the above well-tempered models guarantee a dynamical cancellation of $\Lambda$ through the scalar field $\phi(t)$.

The consistency conditions may be relaxed to discover a broader class of degenerate theories. See, for example, Table 1 in Ref. \cite{Appleby:2020njl} for nearly well-tempered Horndeski theories that completely and incompletely satisfies the degeneracy conditions. We have also encountered some nearly well-tempered Teledeski cosmologies, e.g., Eq. (\ref{eq:c1s1sol}) in Sec. \ref{subsec:GTele}, Eqs. (\ref{eq:c3s2ansatz}) and (\ref{eq:c3s2sol}) with $\xi = 0$ in Sec. \ref{subsec:G4_phi_X_GTele}, and Eq. (\ref{eq:no_temp_b1}) in Sec. \ref{subsec:no_tempering_theorem}. In this work we have not considered `trivial scalar field' self-tuning as first described in Ref. \cite{Charmousis:2011bf}, although we expect degenerate vacuum states of this type to be present for a wide range of Teledeski actions. 

We also emphasize that all of the theories considered in this section contain a proper (TE)GR limit and hence admit propagating tensor degrees of freedom. The well-tempering conditions (Eqs. (\ref{eq:degeneracy_equation}), (\ref{eq:consistency_1}), (\ref{eq:consistency_2}), and (\ref{eq:consistency_3})) further guarantee that the scalar field is not just a dynamical degree of freedom, but also one which eats up an arbitrarily large cosmological constant in order to deliver a late-time, low energy de Sitter vacuum state. All of the theories listed in Table \ref{tab:summary} guarantee this desirable feature. We shall see this in action in the following section.

\section{Dynamics in a Well-Tempered Teledeski Cosmology}
\label{sec:dynamics}

In this section we flesh out the dynamics in well-tempered cosmology. We do so in a $A(X) + K(X) + G(X) + \mathcal{G}$ well-tempered model with the revived Horndeski and Teledeski potentials. We discuss the dynamical stability of the self-tuning vacuum, its compatibility with a matter universe, and stability through phase transitions.

\subsection{The \texorpdfstring{$A(X) + V(X) + G(X) + \mathcal{G}$}{} Model}
\label{subsec:the_model}

We consider a model in Sec. \ref{subsec:shift_symmetric}. In this case, recall that the well-tempered theory is given by the functional forms of $V\left[ q(x), A(x), \mathcal{Q}(x), \mathcal{I}(x) \right]$ (\ref{eq:V_wt}) and $G\left[ q(x), A(x), \mathcal{Q}(x), \mathcal{I}(x) \right]$ (\ref{eq:G_wt}). We fully specify the model by considering the potentials
\begin{equation}
    q(x) = - \dfrac{ 3 \gamma h }{ \gamma - \sqrt{x} }\,,
\end{equation}
\begin{equation}
    A(x) = \alpha x\,,
\end{equation}
\begin{equation}
    \mathcal{Q}(x) = 1\,,
\end{equation}
and
\begin{equation}
    \mathcal{I}(x) = \beta x\,,
\end{equation}
where $\alpha$, $\beta$, and $\gamma$ are constants. The quantity $h$ is the Hubble parameter on the well-tempered vacuum. Thus, we are considering a theory with the usual Horndeski sector (defined by $V$ and $G$), a revived Horndeski sector (defined by $A$), and a teleparallel sector (defined by $\mathcal{I}$). In what follows, we refer to $\alpha$, $\beta$, and $\gamma$ as the revived Horndeski, Teledeski, and well-tempered parameters, respectively. The limit $\alpha \rightarrow 0$ and $\beta \rightarrow 0$ corresponds to the Horndeski case discussed in Ref. \cite{Appleby:2018yci}. We focus on the broader sector ($\alpha \neq 0$ and $\beta \neq 0$).

The \textit{gravitational} action for the Teledeski theory is given by
\begin{equation}
\label{eq:action_model}
\begin{split}
    \mathcal{S} \left[ e, \phi \right] = \int d^4 x \, e
    \bigg[ & l \phi + \left( 3h^2 \alpha + \beta \right) X \\
    & - \left( \frac{l}{18 h^2} \left(\frac{2 \gamma }{\sqrt{X}}+\ln (X)\right) - \frac{2 \sqrt{2} \left( 3 h^2 \alpha   + \beta \right) }{3 h} \sqrt{X} \right) \Box \phi \\ 
    & + \dfrac{M_{\text{Pl}}^2 + \alpha X }{2} \left( -T + B \right) + \dfrac{ \beta }{18 h^2} I_2^2 \bigg]\,,
\end{split}
\end{equation}
where the term $l \phi(t)$ is the tadpole, $X = - \left( \partial^\mu \phi \right) \left( \partial_\mu \phi \right) / 2$, $T$ is the torsion scalar, $T^{\lambda \ \mu}_{\ \lambda }$ is the torsion vector, $I_2 = T^{\lambda \ \mu}_{\ \lambda } \phi_{; \mu}$, and $B$ is a boundary term (defined in Sec.~\ref{subsec:teledeski_theory}).

The field equations of the theory are given by the following. The Friedmann constraint is given by
\begin{equation}
    3 H^2 = \rho_\Lambda + \rho + \frac{l }{3 h^2} \left(\sqrt{2} H \left(\sqrt{X}-\gamma \right)-3 h^2 \phi \right) +\frac{X}{h^2} (h-H) (h-3 H) \left(3 h^2 \alpha  + \beta \right)\,,
\end{equation}
where $\rho_\Lambda$ is the vacuum energy and $\rho$ are matter fields, e.g., dark matter, visible matter. On the other hand, the Hubble and scalar field equations are given by
\begin{equation}
\begin{split}
0 =
& \dot{H} \left(36 h^2  X- 12 X^2 \left(3 h^2 \alpha  + \beta\right) \right) \\
& \phantom{ggg} +\dot{X} \left( 12 X (h-H) \left(3 h^2 \alpha  + \beta\right) + \sqrt{2} l \left(\gamma -\sqrt{X}\right) \right) \\
& \phantom{ggg} + 36 X^2 (h-H)^2 \left(3 h^2 \alpha  + \beta\right) + 6 \sqrt{2} H l X \left(\sqrt{X}-\gamma \right) + 18 h^2 X (\rho + P) \,,
\end{split}
\end{equation}
and
\begin{equation}
\begin{split}
0 =
& \dot{X} \left( 3 \sqrt{2} X (h-H)^2 \left(3 h^2 \alpha  + \beta\right) +\gamma  H l \right) \\
& \phantom{ggg} + \dot{H} \left( l \left(2 X^{3/2}-2 \gamma  X\right) - 12 \sqrt{2} X^2 (h-H) \left(3 h^2 \alpha  + \beta\right) \right) \\
& \phantom{ggg} + l \left(6 X^{3/2} \left(H^2-h^2\right)-6 \gamma  H^2 X\right)+18 \sqrt{2} H X^2 (h-H)^2 \left(3 h^2 \alpha  + \beta\right)\,,
\end{split}
\end{equation}
respectively. We also consider the matter fields which evolve via the conservation law
\begin{equation}
\label{eq:Meq_conservation}
    \dot{\rho} + 3 H \left( \rho + P \right) = 0 \,,
\end{equation}
where $\rho$ and $P$ are the energy density and the pressure, respectively.

Now, we prepare the dimensionless version of the field equations for the numerical analysis. For the vacuum energy and the matter fields, we simply write down
\begin{eqnarray}
\rho_\Lambda &=& 3 h^2 \lambda\,, \\
\rho (t) &=& 3 H(t)^2 \Omega(t)\,, \\
P(t) &=& w \rho(t)\,, 
\end{eqnarray}
where $w$ is the matter equation of state, e.g., $w = 0$ for visible and dark matter and $w = 1/3$ for radiation. Using the Hubble parameter $h$ of the self-tuning vacuum as a length scale, we define the dimensionless gravitational fields using the following transformation
\begin{eqnarray}
t &=& \tau / h\,, \\
H (t) &=& h y (\tau)\,, \\
X(t) &=& h^2 \chi(\tau)\,, \\
\label{eq:alpha_nd} \alpha &=& \bar{ \alpha } / h^2\,, \\
\label{eq:gamma_nd} \gamma &=& h \bar{\gamma}\,, \\
\label{eq:l_nd} l &=& h^2 \bar{l}\,.
\end{eqnarray}
After this transformation, the fields $\left( H, \phi, \rho \right)$ are then represented by their dimensionless counterparts $\left( y, \phi, \Omega \right)$. On the parameters, in what follows, we omit the bars on top of the dimensionless parameters for notational simplicity, e.g., write $\bar{\alpha}$ as $\alpha$, and instead put up a warning whenever a notational distinction between the dimensionful and dimensionless versions of the parameters is necessary. The dimensionless Friedmann constraint is given by
\begin{equation}
\label{eq:FeqM1nd}
3 \left(\chi  (3 \alpha +\beta )+ 3 \lambda +3 y^2 (\chi  (3 \alpha +\beta )+ \Omega -1)\right)-3 l \phi =\sqrt{2} l y \left(\gamma -\sqrt{\chi }\right)+12 \chi  y (3 \alpha +\beta )\,.
\end{equation}
The dimensionless Hubble, scalar field, and matter equations are
\begin{equation}
\label{eq:HeqM1nd}
\begin{split}
0 =
& 12 \chi  y' (3 -\chi  (3 \alpha +\beta ))+ \chi ' \left(\sqrt{2} l \left(\gamma -\sqrt{\chi }\right)-12 \chi  (y-1) (3 \alpha +\beta )\right) \\
& \phantom{ggg} +6 \sqrt{2} l \chi  y \left(\sqrt{\chi }-\gamma \right)+36 \chi  \left( \frac{3}{2} (w+1) y^2 \Omega +\chi  (y-1)^2 (3 \alpha +\beta )\right)
\,,
\end{split}
\end{equation}
\begin{equation}
\label{eq:SeqM1nd}
\begin{split}
0 =
& l y' \left(2 \chi ^{3/2}-2 \gamma  \chi \right) + \chi ' \left(\gamma  l y+3 \sqrt{2} \chi  (y-1)^2 (3 \alpha +\beta )\right) \\
& \phantom{ggg} + 6 l \chi  \left(\sqrt{\chi } \left(y^2-1\right)-\gamma  y^2\right)+18 \sqrt{2} \chi ^2 (y-1)^2 y (3 \alpha +\beta )+ (12 \sqrt{2} \chi ^2 (y-1) (3 \alpha +\beta ) \,,
\end{split}
\end{equation}
and
\begin{equation}
\label{eq:MeqM1nd}
3 (w+1) y^3 \Omega + y^2 \Omega '+ 2 y \Omega  y'=0\,.
\end{equation}
In the dimensionless equations, a prime denotes a derivative with respect to the dimensionless time $\tau = h t$, e.g., $\chi' = d \chi / d \tau$ and $y' = d y/d \tau$.

In the remaining sections, we shall use these dimensionless equations (Eqs. (\ref{eq:FeqM1nd}), (\ref{eq:HeqM1nd}), (\ref{eq:SeqM1nd}), and (\ref{eq:MeqM1nd})) to explore the dynamical stability of the self-tuning vacuum, assess the compatibility of the well-tempered model with the existence of a matter era, and establish the stability of the vacuum through phase transitions.

\subsection{Dynamical Stability of the Well-Tempered Vacuum}
\label{subsec:dynamical_stability}

We prove the dynamical stability of the self-tuning vacuum here. We establish this analytically by examining the on-shell solution and then solving for the linearized perturbations on-shell. Then, we look at the global dynamical stability of the well-tempered vacuum by drawing phase portraits of the dynamical system.

We emphasize for clarity that the perturbations in this section do not come from a scalar-vector-tensor decomposition for each of the components of the torsion and the scalar, but instead considers a dynamical system analysis (Eqs. (\ref{eq:y_pert_def}) and (\ref{eq:phi_pert_def})). This simple analysis admits an analytical solution (Eqs. (\ref{eq:deltaH}) and (\ref{eq:deltaphi_linear})) and hints at the dynamical stability of the well-tempered vacuum. Furthermore, the linearized solution provides an objective way to assess when a numerical solution finally ends up on the well-tempered vacuum state.

Taking the on-shell limit ($y\left(\tau\right) \rightarrow 1$ and $\Omega \left(\tau\right) \rightarrow 0$) of the Hamiltonian constraint and the scalar field equation leads to
\begin{equation}
    9  (\lambda -1)+l \left(\phi'-\sqrt{2} \gamma \right)=3 l \phi\,,
\end{equation}
and
\begin{equation}
    \phi'' - 3 \phi' = 0\,.
\end{equation}
This shows that the scalar field on-shell behaves as
\begin{equation}
\label{eq:phi_M1_on_shell}
    \phi \left( \tau \right) = \dfrac{a_1}{3} e^{3\tau} + a_2 \,,
\end{equation}
where $a_1$ and $a_2$ are integration constants characterizing the dynamics of the scalar field. The Hamiltonian constraint on-shell is given by
\begin{equation}
    l \left(3 a_2+\sqrt{2} \gamma \right)= 9 (\lambda -1)\,.
\end{equation}
The dynamical self-tuning of the vacuum energy is reflected here as the entry of the constant $a_2$ characterizing the dynamics of the scalar field.

Now, we perturb the on-shell Hubble and scalar fields as
\begin{equation}
\label{eq:y_pert_def}    y(\tau) = 1 + \delta y(\tau)\,,
\end{equation}
and
\begin{equation}
\label{eq:phi_pert_def}    \phi(\tau) = \left( \dfrac{a_1}{3} e^{3\tau} + a_2 \right) + \delta \phi(\tau)\,.
\end{equation}
Substituting these perturbations into the Friedmann constraint and the scalar field equation leads to the dynamical equations for $\left( \delta y (\tau), \delta \phi (\tau) \right)$
\begin{equation}
\label{eq:FeqM1ndO1}
\delta y \left(3 a_1^2 e^{6 \tau } (3 \alpha +\beta )+l \left(a_1 e^{3 \tau }-\sqrt{2} \gamma \right)-18 \right)+l \left(\delta \phi '-3 \delta \phi \right)=0\,.
\end{equation}
and
\begin{equation}
\label{eq:SeqM1ndO1}
a_1 e^{3 \tau } \left(\sqrt{2} a_1 e^{3 \tau }-2 \gamma \right) \delta y'+6 a_1 e^{3 \tau } \delta y \left(\sqrt{2} a_1 e^{3 \tau }-\gamma \right)+2 \gamma  \left(\delta \phi '' -3 \delta \phi ' \right) = 0\,.
\end{equation}
This can be uncoupled by differentiating the constraint (\ref{eq:FeqM1ndO1}) and then eliminating $\phi$ in Eq. (\ref{eq:SeqM1ndO1}). Hence, it leads to a differential equation for $\delta y$ which can be solved for
\begin{equation}
\label{eq:deltaH}
    \delta y(\tau) = -\dfrac{b_1}{a_1^2 e^{6 \tau } \left(\sqrt{2} l-6 \gamma  (3 \alpha +\beta )\right)-4 a_1 \gamma  l e^{3 \tau }+2 \gamma  \left(18 +\sqrt{2} \gamma  l\right)}\,,
\end{equation}
where $b_1$ is an integration constant. Clearly, at late times, $\delta y \sim e^{-6 \tau}$. This exponential descent to the well-tempered vacuum will present itself later as a characteristic late-time drop $\ln |(H/h) - 1| \sim - 6 \tau$ in plots of the Hubble function (Figs. \ref{fig:fields_mat_m1}(a) and \ref{fig:logHphi_pt}(a)). The Hubble perturbation thus vanishes as the state approaches the well-tempered vacuum. On the other hand, the scalar perturbation equation becomes
\begin{equation}
\delta \phi '(\tau )-3 \delta \phi (\tau )=\dfrac{b_1 \left(3 a_1^2 e^{6 \tau } (3 \alpha +\beta )+a_1 l e^{3 \tau }-18 -\sqrt{2} \gamma  l\right)}{l \left(a_1^2 e^{6 \tau } \left(\sqrt{2} l-6 \gamma  (3 \alpha +\beta )\right)-4 a_1 \gamma  l e^{3 \tau }+2 \gamma  \left(18 +\sqrt{2} \gamma  l\right)\right)}\,.
\end{equation}
This is a first-order differential equation for $\delta \phi$ and can be solved analytically. However, for the purposes of the stability analysis, we shall concern ourselves only with the late-time asymptotic limit. Expanding the nonhomogeneous term at late times, we find that
\begin{equation}
\delta \phi '(\tau )-3 \delta \phi (\tau ) = \frac{3 b_1 (3 \alpha +\beta )}{l \left(-18 \alpha  \gamma -6 \beta  \gamma +\sqrt{2} l\right)} + \mathcal{O} \left( e^{-3\tau} \right)\,.
\end{equation}
The asymptotic solution can therefore be written as
\begin{equation}
\label{eq:deltaphi_linear}
\delta \phi (\tau )=c_1 e^{3 \tau }-\frac{b_1 (3 \alpha +\beta )}{l \left(\sqrt{2} l-6 \gamma  (3 \alpha +\beta )\right)} + \mathcal{O} \left( e^{-3\tau} \right)\,,
\end{equation}
where $c_1$ is an integration constant. This shows that the leading-order asymptotic solution is of the exact same form as the on-shell scalar field (\ref{eq:phi_M1_on_shell}). Moreover, the subdominant terms vanish at late times. Therefore, we find that the Hubble perturbation vanishes and that the scalar perturbation can merely be reabsorbed into the generally-divergent scalar field on-shell. This shows that at late times the perturbed state inevitably returns on-shell, and thus proves the dynamical stability of the well-tempered vacuum.

The divergent scalar field on-shell is often a characteristic of the well-tempered Horndeski cosmologies considered so far \cite{Appleby:2018yci, Appleby:2020njl, Emond:2018fvv}. Our results show that it continues to hold even in the much larger class of Teledeski cosmology.

The conclusions of the linear perturbation analysis (Eqs. (\ref{eq:deltaH}) and (\ref{eq:deltaphi_linear})) should also hold even with matter fields because of the equivalence principle. In this case, any matter field will have the late-time behavior $\Omega \sim e^{-3(1 + w) \tau}$, where $w$ is the matter field's equation of state. For $w \geq 0$, or rather, any visible or dark matter, the near-exponential expansion simply drains the matter's energy, or in symbols $\Omega \rightarrow 0$, leaving behind only the metric and the scalar on-shell. This will be supported in Sec. \ref{subsec:compatibility_with_matter}.

We can go beyond a perturbative analysis of the dynamical stability by writing down the system as a two-dimensional dynamical system. To do so, we solve for $\phi$ in the Hamiltonian constraint and then use it with the constraint $\chi = \phi'^2/2$. The resulting equation and the scalar field equation then forms an explicit two-dimensional dynamical system for the vector field $\left( \chi, y \right)$:
\begin{equation}
\label{eq:chip_m1}
\begin{split}
\chi' = - \dfrac{2}{\mathcal{D}\left[ \chi, y \right]} \bigg(& 6 \sqrt{2} l \chi ^{3/2} \left( - l \gamma +l \sqrt{\chi }+6 \sqrt{2} \chi  (y-1) (3 \alpha +\beta )\right) \\
& -6 \chi  \left(l \left(\sqrt{\chi } \left(y^2-1\right)-\gamma  y^2\right) +3 \sqrt{2} \chi  y (y-1)^2 (3 \alpha +\beta )\right) \\
& \times \left(\sqrt{2} \gamma  l-\sqrt{2} l \sqrt{\chi }+6 \chi  (2-3 y) (3 \alpha +\beta )+18  y\right) \bigg)\,,
\end{split}
\end{equation}
and
\begin{equation}
\label{eq:yp_m1}
\begin{split}
y' = \dfrac{3 \sqrt{\chi } (y-1)}{ \mathcal{H}\left[ \chi, y \right] } \bigg( & 18 \sqrt{2} \chi ^{3/2} (3 \alpha +\beta )^2+\sqrt{2} \gamma  l^2-\sqrt{2} l^2 \sqrt{\chi } \\
& +y \bigg(-90 \sqrt{2} \chi ^{3/2} (3 \alpha +\beta )^2+\sqrt{2} \gamma  l^2 \\
& \phantom{ggggg} -l \sqrt{\chi } \left(6 \gamma  (3 \alpha +\beta )+\sqrt{2} l\right)+12 l \chi  (3 \alpha +\beta )\bigg) \\
& +6 \sqrt{\chi } y^2 (3 \alpha +\beta ) \left(21 \sqrt{2} \chi  (3 \alpha +\beta )+3 \gamma  l-4 l \sqrt{\chi }\right)\\
& +12 l \chi  (3 \alpha +\beta ) -54 \sqrt{2} \chi ^{3/2} y^3 (3 \alpha +\beta )^2 \bigg) \,.
\end{split}
\end{equation}
In these equations, the functionals $\mathcal{D}$ and $\mathcal{H}$ appearing in the denominators are given by
\begin{equation}
\label{eq:Hc_m1}
\begin{split}
\mathcal{D}\left[ \chi, y \right] = 2 l y \bigg( & -\sqrt{2} l \left(\gamma -\sqrt{\chi }\right)^2-\frac{54 \sqrt{2} }{l} \chi  (y-1)^2 (3 \alpha +\beta ) (\chi  (3 \alpha +\beta )+ 1 ) \\
& +6 \left(\gamma  \chi  (5 y-4) (3 \alpha +\beta )-4 \chi ^{3/2} (y-1) (3 \alpha +\beta )- 3 \gamma  y\right)\bigg)\, ,
\end{split}
\end{equation}
and
\begin{equation}
\mathcal{H}\left[ \chi, y \right] = - \mathcal{D}\left[ \chi, y \right] / \left( 2 y \right)\,.
\end{equation}
Two notable features of this dynamical system stands out in relation to self-tuning. First, it is independent of $\lambda$. The is characteristic of a well-tempered cosmology. It implies that the attractor(/s) of the system will be approached regardless of the size of vacuum energy. Second, the system is degenerate in the revived Horndeski and teleparallel sectors, i.e., the dynamics \textit{does not} depend on $\alpha$ and $\beta$ separately, but rather only through their combination $3 \alpha + \beta$. The keen reader may have already noticed this earlier through the field equations (Eqs. (\ref{eq:FeqM1nd}), (\ref{eq:HeqM1nd}), and (\ref{eq:SeqM1nd})). This particularly points to the potentially-infinite set of well-tempered Teledeski models ($3\alpha + \beta = 0$) that are tangent to the same Horndeski model ($\alpha = 0$ and $\beta = 0$). Most importantly, it means that away from this tangent point, or $3 \alpha + \beta \neq 0$, new cosmological dynamics can be achieved through the interplay between the revived Horndeski and Teledeski sectors.

This is also a good place to remind that the parameters $\alpha$, $\gamma$, and $l$ in Eqs. (\ref{eq:chip_m1}), (\ref{eq:yp_m1}), and (\ref{eq:Hc_m1}) are their dimensionless counterparts using the Hubble length scale of the well-tempered vacuum state as a reference (recall Eqs. (\ref{eq:alpha_nd}), (\ref{eq:gamma_nd}), and (\ref{eq:l_nd})). This implies for example that $\alpha \sim 1$, $\gamma \sim 1$, and $l \sim 1$ corresponds the dimensionfull quantities $\alpha \sim 1/h^2$, $\gamma \sim h$, and $l \sim h^2$. The dimensionless fields and parameters are utilized for numerical convenience.

\begin{figure}[ht]
\center
	\subfigure[ $3 \alpha + \beta = -3$ ]{
		\includegraphics[width = 0.47 \textwidth]{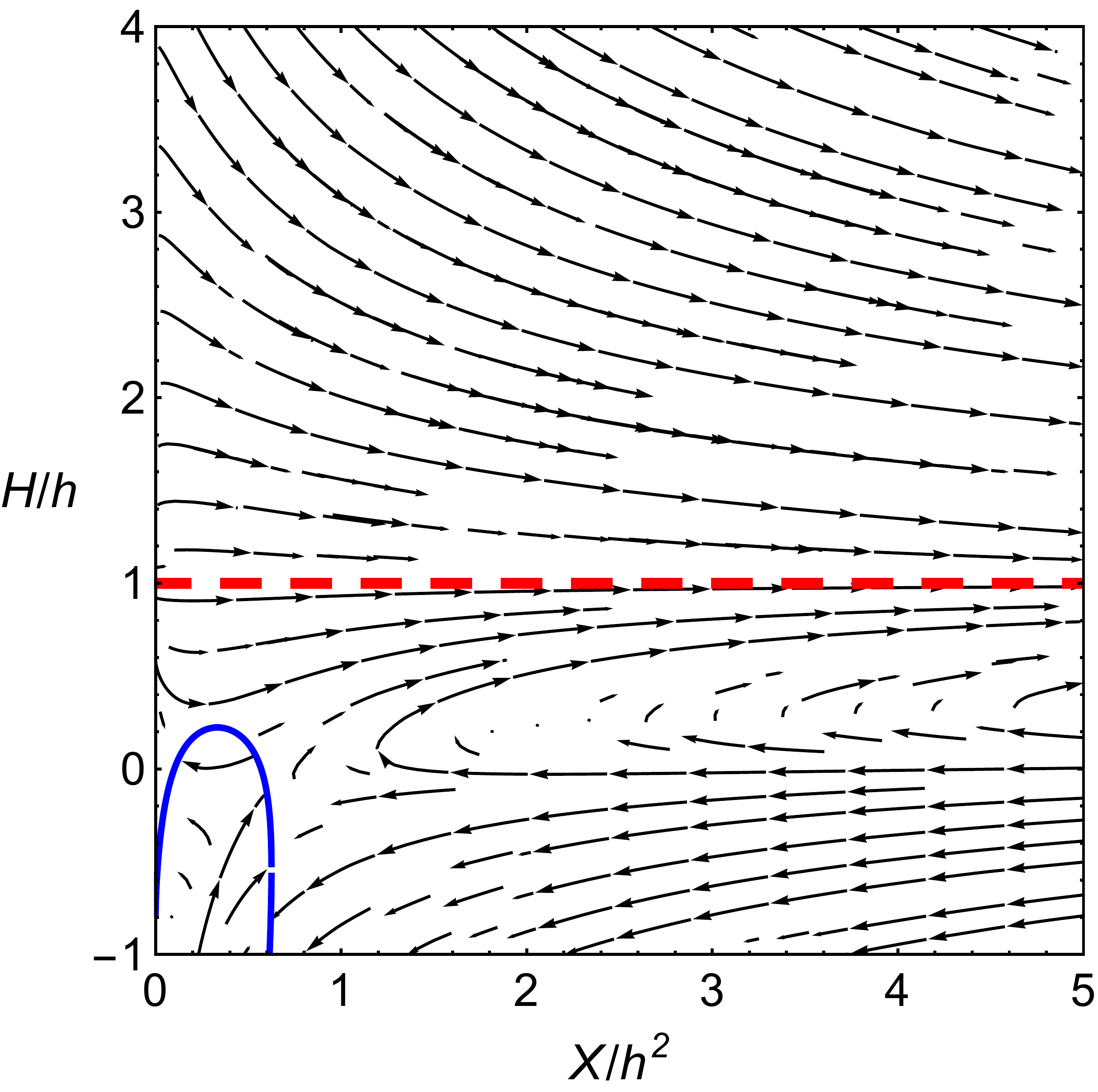}
		}
	\subfigure[ $3 \alpha + \beta = -2$ ]{
		\includegraphics[width = 0.47 \textwidth]{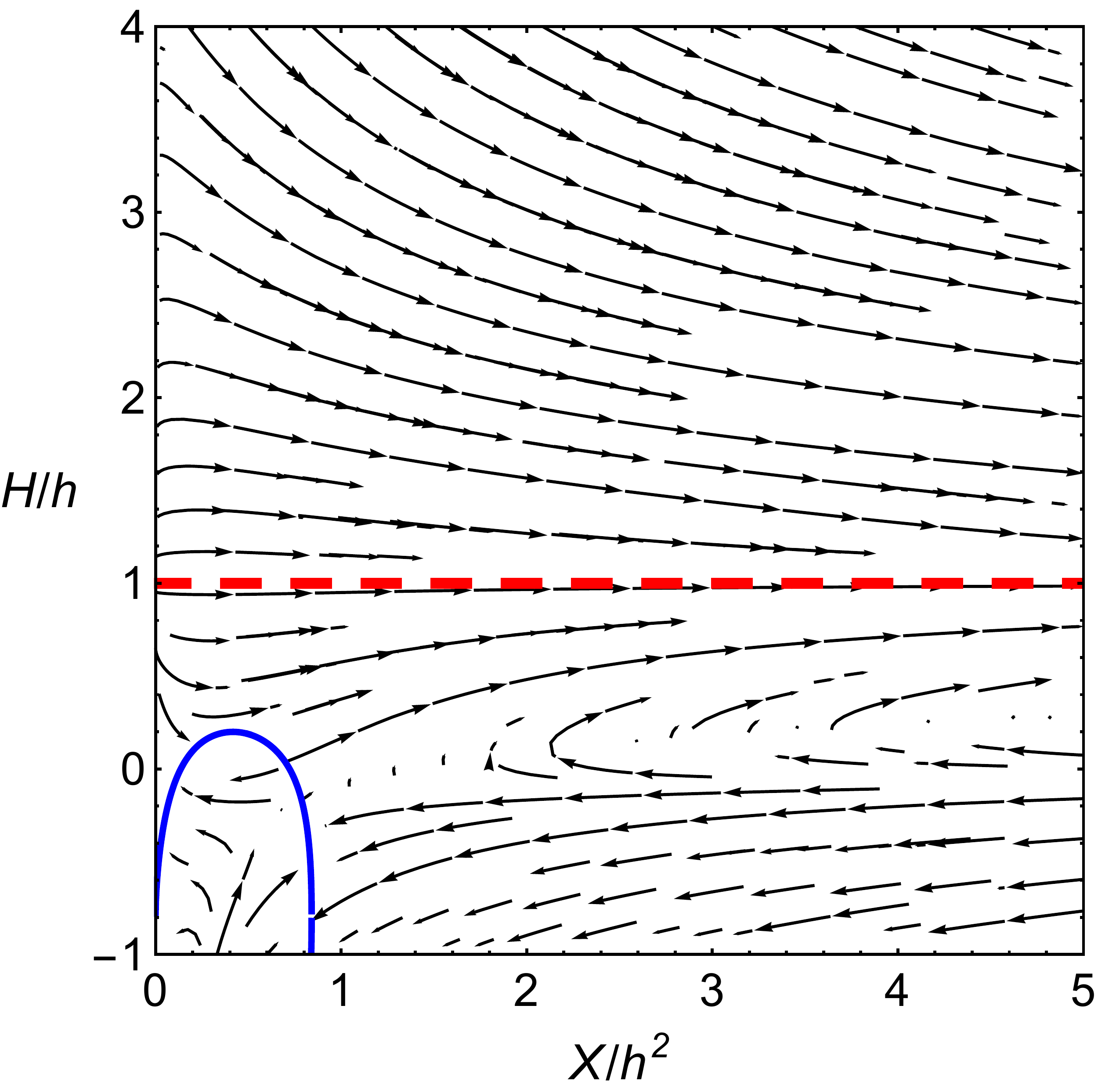}
		}
	\subfigure[ $3 \alpha + \beta = 1$ ]{
		\includegraphics[width = 0.47 \textwidth]{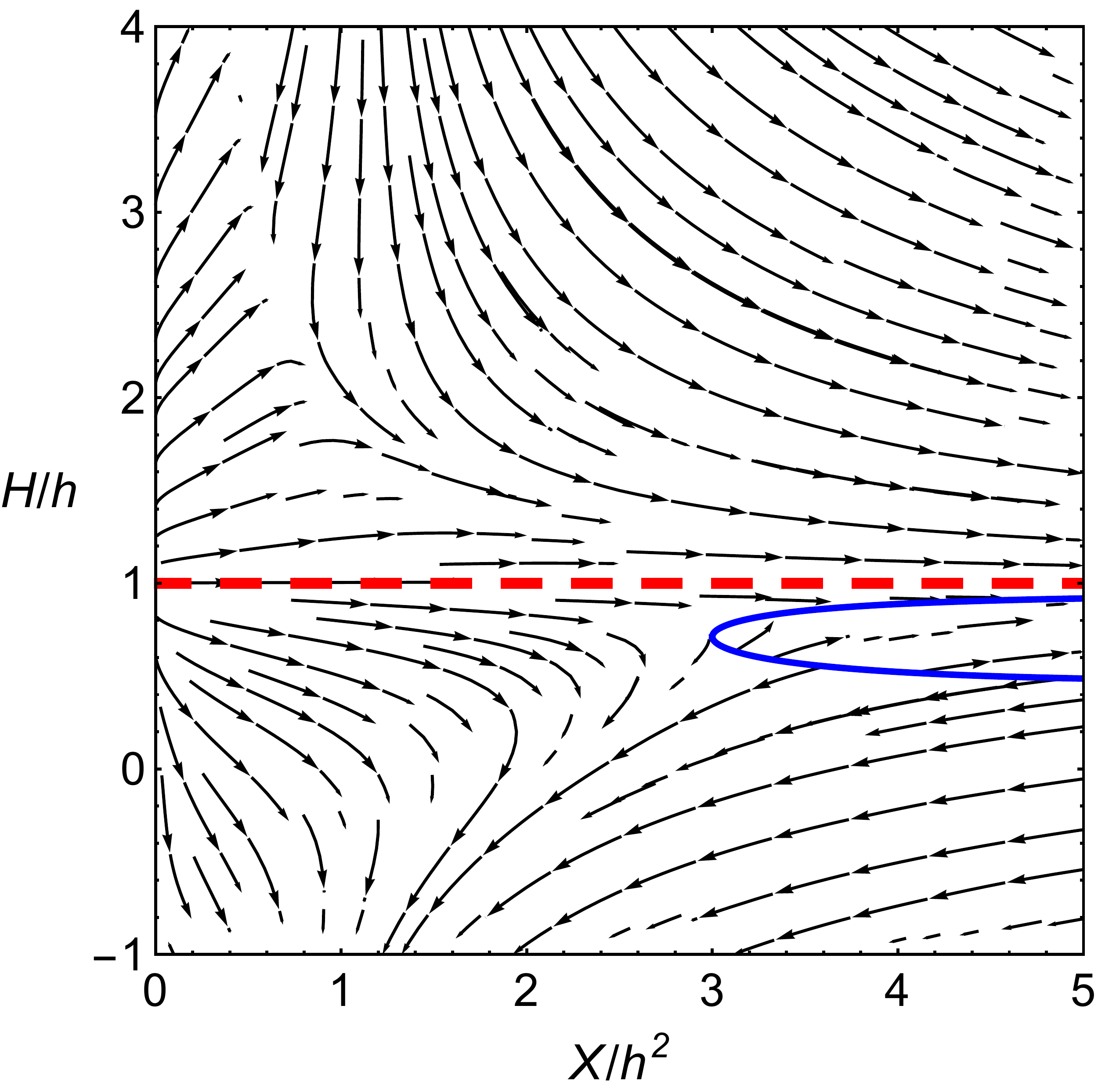}
		}
	\subfigure[ $3 \alpha + \beta = 7$ ]{
		\includegraphics[width = 0.47 \textwidth]{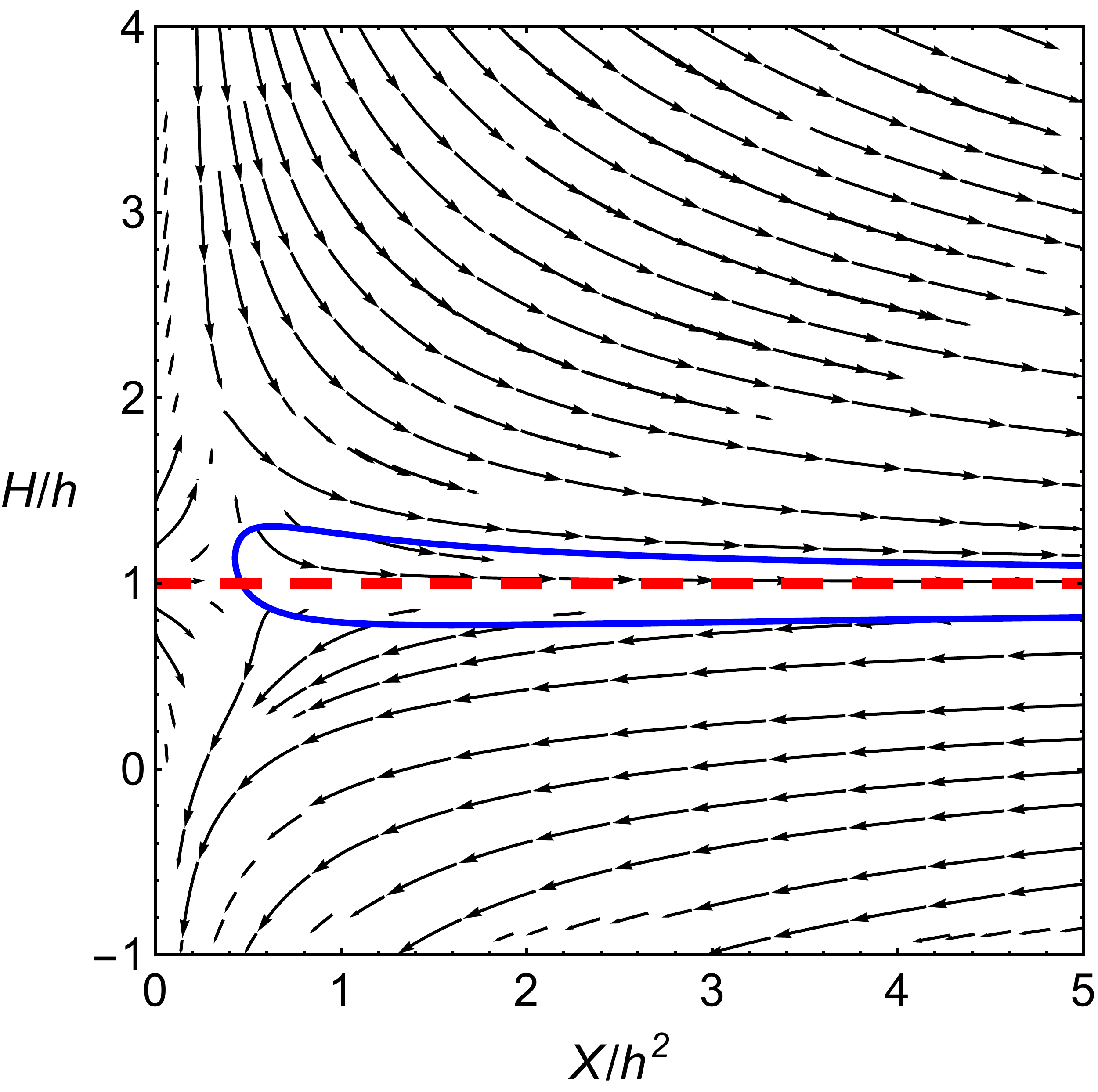}
		}
\caption{Phase portraits for the dynamical system given by Eqs. (\ref{eq:chip_m1}) and (\ref{eq:yp_m1}) for a fixed well-tempering parameter $\gamma = 1$ and tadpole $l = 10$. The red dashed line denotes to well-tempered vacuum and the blue solid line corresponds to the curve $\mathcal{D}\left[\chi, y \right] = 0$ (Eq. (\ref{eq:Hc_m1})).}
\label{fig:portrait_m1}
\end{figure}

In studying the dynamical system (Eqs. (\ref{eq:chip_m1}) and (\ref{eq:yp_m1})), we focus on the interplay between the revived Horndeski ($\alpha$) and Teledeski ($\beta$) parameters. For this purpose, we fix the well-tempered parameter ($\gamma = 1$) and the tadpole ($l = 10$) in units of the self-tuning Hubble constant $h$, i.e., in their dimensionfull forms, $\gamma \sim h$ and $l \sim 10 h^2$. We then draw phase portraits for various $\alpha$ and $\beta$. Results are shown in Fig. \ref{fig:portrait_m1}.

These phase portraits echo the earlier conclusion on the stability of the self-tuning vacuum (red-dashed line) and show that the dynamics in the physically relevant region of the phase space ($H > h$) mostly ends up on-shell. This supports the case of well-tempered cosmology as a viable phenomenological theory of dark energy in the sense that there will be smooth transition from a matter to a de Sitter state due to the low energy de Sitter state being an attractor from above. Some more notable features of the dynamical system also show up in Fig. \ref{fig:portrait_m1}. We first point out the existence of a critical curve $H_C$ defined by the functional $\mathcal{D}\left[ \chi, y \right] = 0$ where the vector field $\left( \chi', y' \right)$ becomes undefined. It must be noted that this curve does not correspond to a physical divergence. Often, it points out that there is a better choice of the coordinates in phase space. Second, there is an apparent attractor in the phase region $H \ll h$. This region of phase space is not physically interesting. Nonetheless, this shows that well-tempering does not imply that the self-tuning vacuum will be the only cosmological attractor of the system. Third, there is a notable change in the overall behavior of the vector field $\left( \chi', y' \right)$ depending on the sign of $3 \alpha + \beta$. This is strongly reflected in the location of the critical curve $H_C$. In Figs. \ref{fig:portrait_m1}(a-b), characterized by $3 \alpha + \beta < 0$, the curve $H_C$ (blue-solid curve) is drawn below the self-tuning vacuum (red-dashed line). On the other hand, in Figs. \ref{fig:portrait_m1}(c-d), characterized by $3 \alpha + \beta > 0$, the curve $H_C$ becomes more horizontally-oriented and in Fig. \ref{fig:portrait_m1}(d) even surrounds a portion of the well-tempered vacuum.

The theory space $3 \alpha + \beta = 0$ coincides with the well-tempered Horndeski theory ($\alpha = 0$ and $\beta = 0$). This can be recognized by looking back at the field equations (Eqs. (\ref{eq:FeqM1nd}), (\ref{eq:HeqM1nd}), and (\ref{eq:SeqM1nd})). This shows that there is an infinite number of well-tempered Teledeski models that are indistinguishable from the same Horndeski theory. Moreover, the distinct phase portraits shown for the two cases -- $3 \alpha + \beta < 0$ and $3 \alpha + \beta > 0$ -- is a testament to the far richer cosmological dynamics that can be anchored in a well-tempered Teledeski cosmology.

\subsection{Compatibility with a Matter Era}
\label{subsec:compatibility_with_matter}

We will show the compatibility of well-tempered cosmology with a matter era, i.e., the dynamics admits a matter-dominated era followed by period of late-time cosmic acceleration. This is an important test for any potential dark energy theory. In this context, the dark energy itself is the self-tuning field, protecting the cosmic expansion from the influence of a possibly infinite vacuum energy.

It is also worth noting that the traditional Fab Four theories, including their related self-tuning cosmologies based on a trivial scalar approach, do not overcome this test as they accommodate a vacuum that screens all kinds of energy, including that of visible and dark matter. 

The numerical integration is described as follows. We integrate Eqs. (\ref{eq:HeqM1nd}), (\ref{eq:SeqM1nd}), and (\ref{eq:MeqM1nd}) for $a(\tau)$, $\phi(\tau)$, and $\Omega(\tau)$ where $a$ is defined by $y(\tau) = a'(\tau)/a(\tau)$. By solving Eq. (\ref{eq:MeqM1nd}) for $\Omega\left( \tau \right) = \omega a (\tau)^{- (1 + 3 w)}/a'(\tau)^2$, the dependent variables can be reduced to $a(\tau)$ and $\phi(\tau)$. Moving forward, we fix the initial conditions on the Hamiltonian constraint (\ref{eq:FeqM1nd}) and choose the initial conditions $a\left( \tau \right) \sim a_0 \tau^{2/3}$ and $\Omega \sim 1$ (\textit{matter era}) provided a set of theory parameters $\left( \alpha, \beta, \gamma, l \right)$. The matter equation of state $w = 0$ is fixed. We consider the independent phase space variables to be $\left( a_0 = a(\tau_0), \phi_0' = \phi'(\tau_0) \right)$ and determine $\phi_0 = \phi(\tau_0)$ using the Hamiltonian constraint at the initial time $\tau_0$. The integration is then carried forward in time up to some $\tau = \tau_f > \tau_0$, all while making sure that the solution stays on the Hamiltonian constraint throughout the evolution.

\begin{figure}[ht]
\center
	\subfigure[ log-Hubble function ]{
		\includegraphics[width = 0.47 \textwidth]{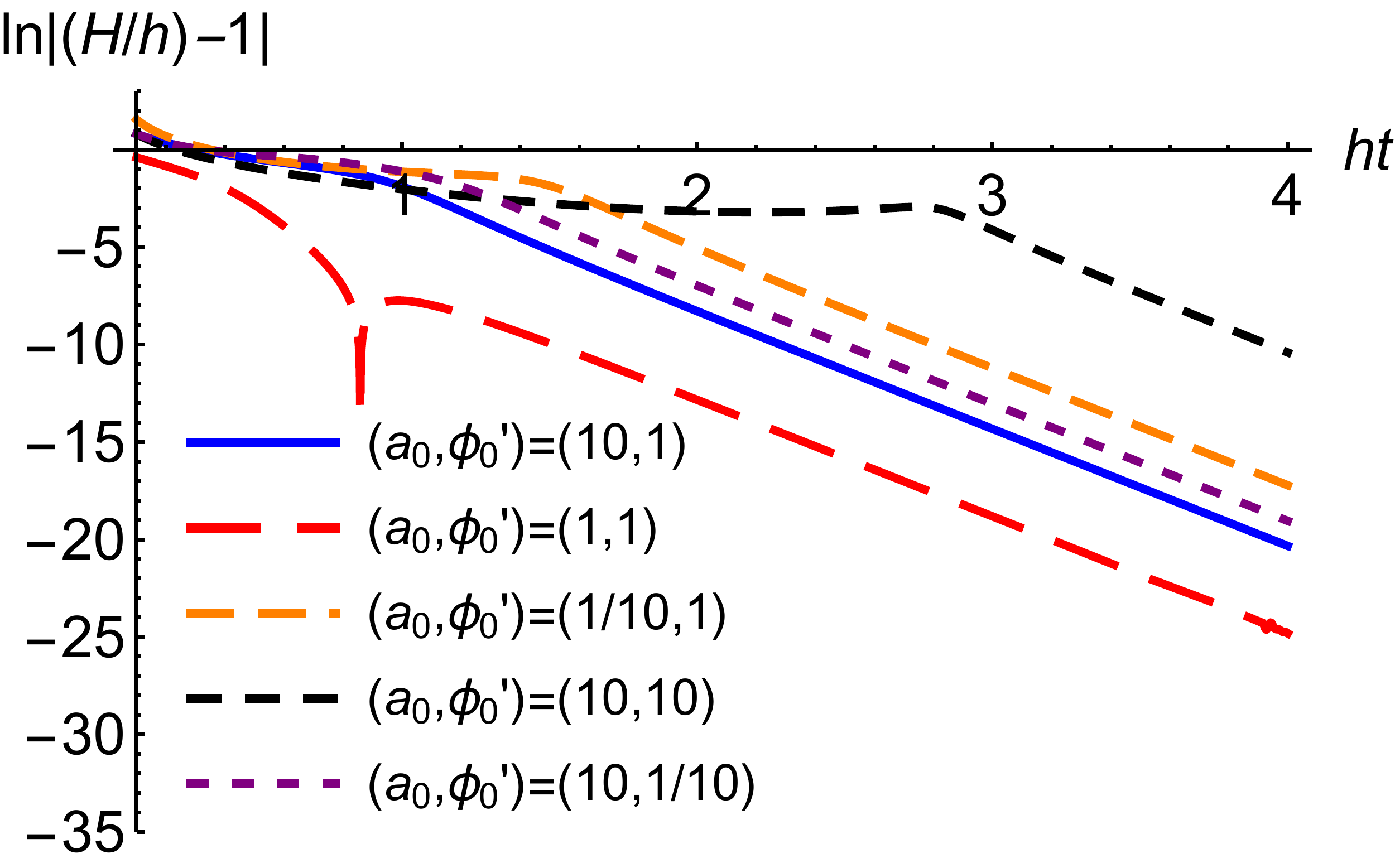}
		}
	\subfigure[ log-scalar field ]{
		\includegraphics[width = 0.47 \textwidth]{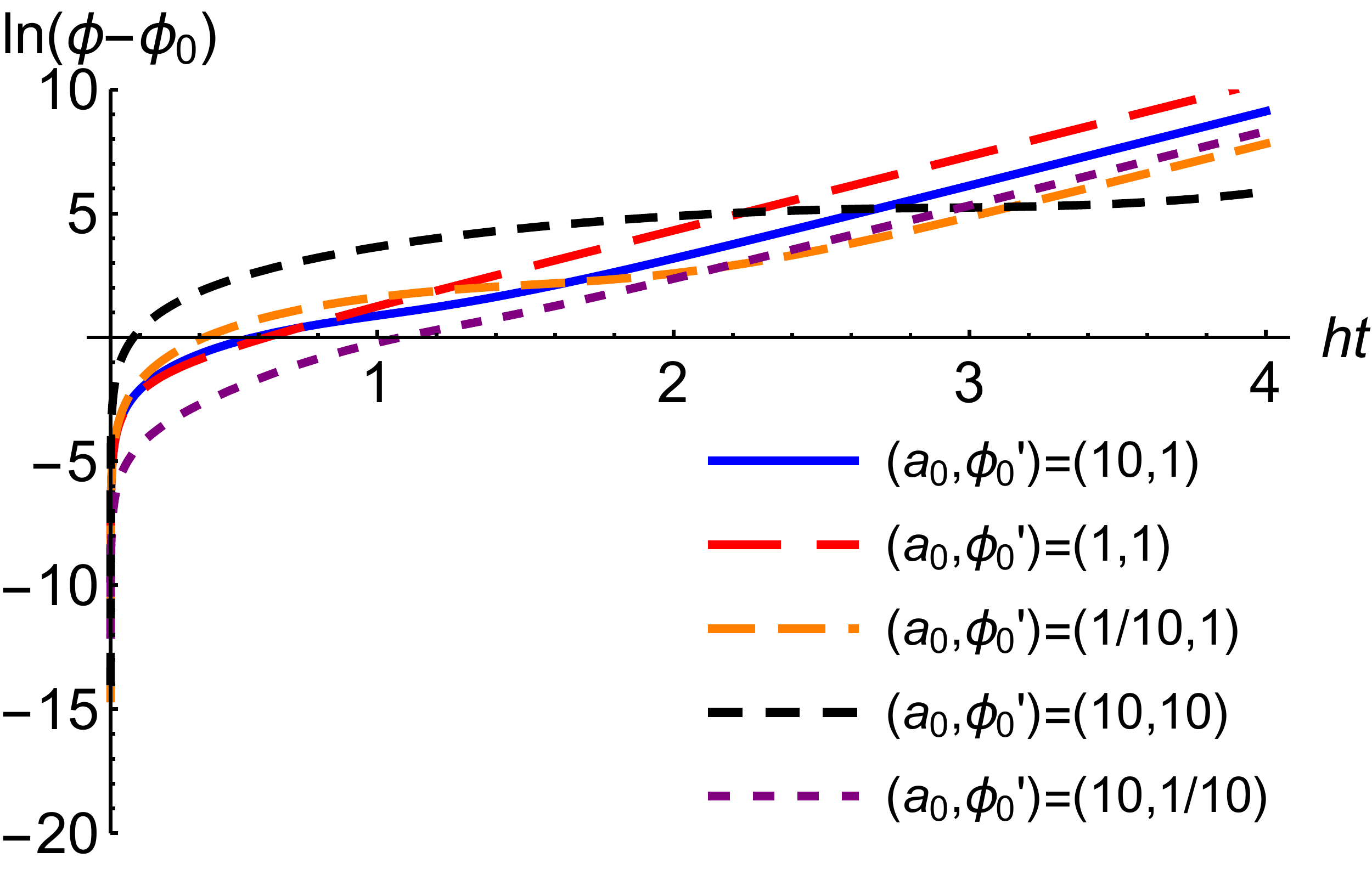}
		}
	\subfigure[ matter density parameter (Eq. (\ref{eq:Om_m1})) ]{
		\includegraphics[width = 0.47 \textwidth]{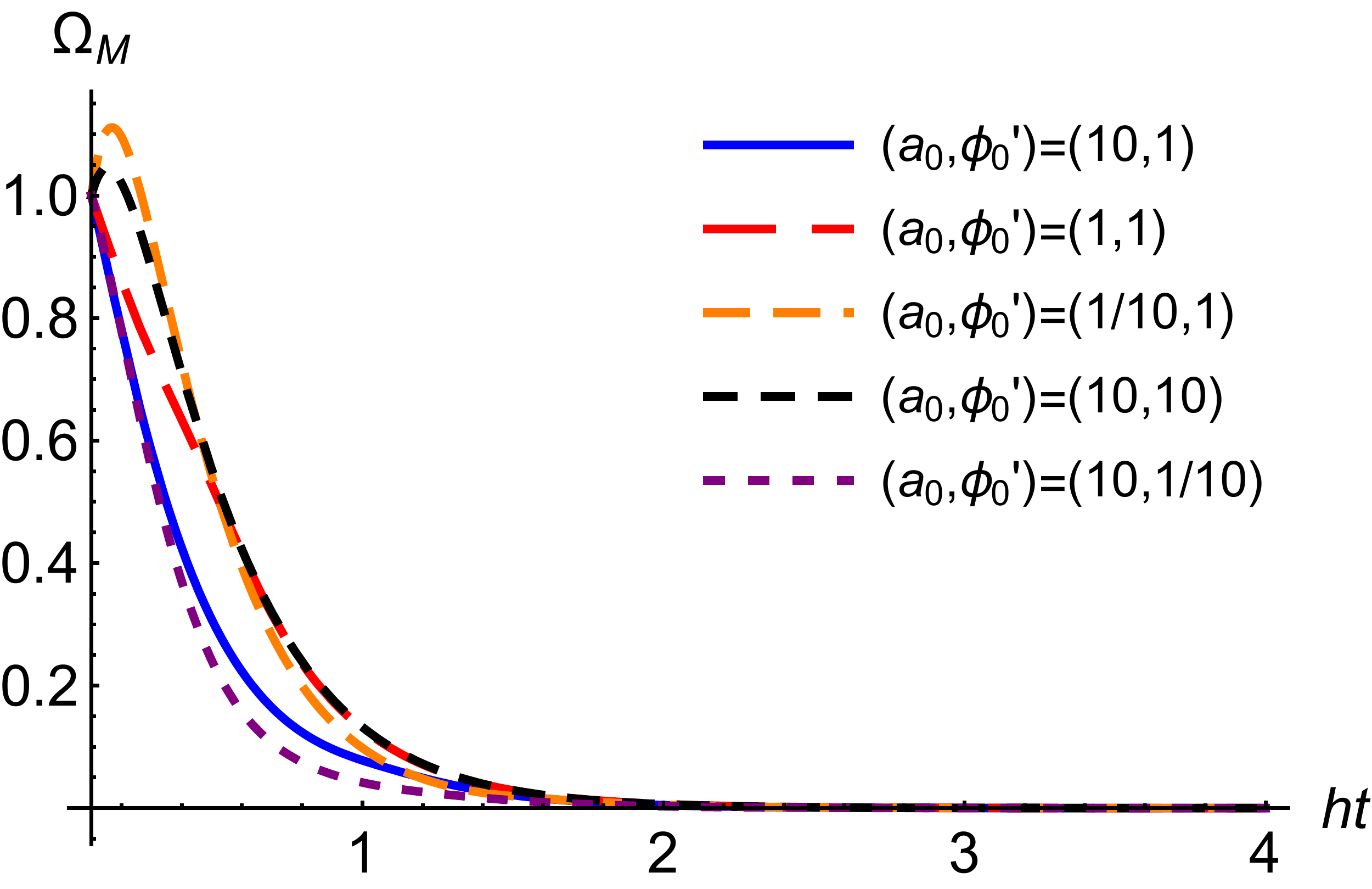}
		}
	\subfigure[ dark energy density parameter (Eq. (\ref{eq:Ode_m1})) ]{
		\includegraphics[width = 0.47 \textwidth]{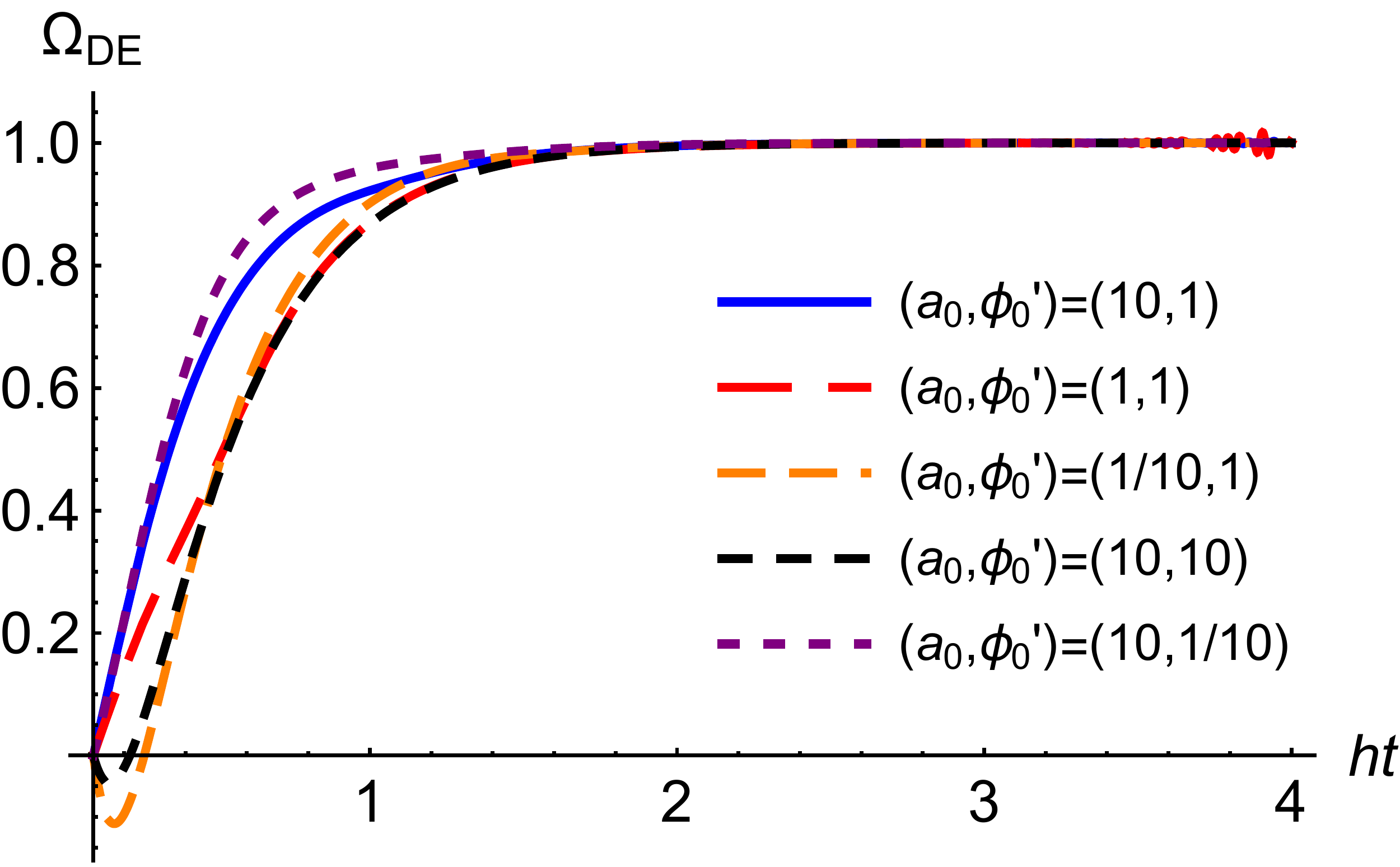}
		}
\caption{Results of the numerical integration of Eqs. (\ref{eq:HeqM1nd}), (\ref{eq:SeqM1nd}), and (\ref{eq:MeqM1nd}) with $\lambda = 10^{10}$ and $\left( \alpha = -1, \beta = 1, \gamma = 1, l = 10\right)$ for several initial conditions $\left( a_0, \phi_0' \right)$. Shown are (a) the logarithm of the Hubble evolution minus the well-tempered Hubble constant, $\ln | (H/h) - 1 |$, (b) the logarithm of the scalar field, $\ln ( \phi - \phi_0 )$, (c) the matter density parameter, and (d) the dark energy density parameter.}
\label{fig:fields_mat_m1}
\end{figure}

The results of the numerical integration for various initial conditions $(a_0, \phi_0')$ are shown in Fig. \ref{fig:fields_mat_m1}. Here, we take a vacuum energy parameter $\lambda = 10^{10}$ and the theory parameters $\left( \alpha = -1, \beta = 1, \gamma = 1, l = 10\right)$, same as in Fig. \ref{fig:portrait_m1}(b). Similar conclusions can be achieved with other theory parameters. We note that this choice of parameters motivate the problem of self-tuning cosmology, i.e., to screen an arbitrarily large vacuum energy to deliver a late-time, low energy de Sitter state. In this case, we demonstrate this in the following numerical results with a vacuum energy that is a staggering ten(!) orders of magnitude larger than the Hubble scale with only theory parameters kept at order unity; unequivocally, there is no fine tuning needed to cancel out the cosmological constant.

The numerical results support the consistency with a matter era and, moreover, a matter-dark energy era. Fig. \ref{fig:fields_mat_m1}(a) shows that the cosmological dynamics at some point exponentially-drops to the well-tempered vacuum. The onset to the vacuum is characterized by the late-time solution $\delta y \sim^{-6 \tau}$ (Eq. (\ref{eq:deltaH})), or rather the line $\ln \left( \delta y \right) = \ln | (H/h) - 1 | \sim - 6 \tau$ in Fig. \ref{fig:fields_mat_m1}(a). The descent to the vacuum is accompanied by the diverging scalar as shown in Fig. \ref{fig:fields_mat_m1}(b).

On the other hand, the density parameters for matter ($\Omega_M$) and dark energy ($\Omega_\text{DE}$) can be obtained by referring to the Hamiltonian constraint. This leads to
\begin{equation}
\label{eq:Om_m1}
\Omega_{\text{M}} = \dfrac{\omega }{a a'^2}\,,
\end{equation}
\begin{equation}
\label{eq:OHor_m1}
\Omega_\text{H}^{(\phi)} = -\dfrac{l a \left(a' \left(\sqrt{2} \gamma -\phi ' \right)+3 a \phi  \right)}{ 9 a'^2 }\,,
\end{equation}
\begin{equation}
\label{eq:OTel_m1}
\Omega_\text{T}^{(\phi)} = \dfrac{\sigma  \left(-4 a a'+3 a'^2+a^2\right) \phi'^2}{6 a'^2}\,,
\end{equation}
and
\begin{equation}
    \Omega_\Lambda = \dfrac{\lambda  a^2}{a'^2}\,.
\end{equation}
In these equations, $\Omega_\text{H}^{(\phi)}$, $\Omega_\text{T}^{(\phi)}$, and $\Omega_\Lambda$ are the regular Horndeski, the revived Horndeski-Teledeki, and the vacuum energy contributions, respectively, to dark energy. The dark energy density parameter is identified as
\begin{equation}
\label{eq:Ode_m1}
    \Omega_\text{DE} = \Omega_\Lambda + \Omega_\text{H}^{(\phi)} + \Omega_\text{T}^{(\phi)}\,.
\end{equation}
The matter and dark energy densities are shown in Figs. \ref{fig:fields_mat_m1}(c-d). These confirm a smooth transition initially from a matter era to a well-tempered dark energy era. An interesting feature also shows up. In some cases, at finite intervals, the dark energy density temporarily evolves to negative values, causing the matter densities to also go above the critical expansion density. This is evident in the cases $(a_0, \phi_0') = (1/10, 1)$ and $(10, 10)$ in Figs. \ref{fig:fields_mat_m1}(c-d) at times $\tau \sim 0.1$. This period is short lived. Nonetheless, it is interesting to investigate in future work whether this would have observational consequences. 

We lastly note that Figs. \ref{fig:fields_mat_m1}(c-d) reveal that $\Omega_\text{M}$ + $\Omega_\text{DE} = 1$. This shows that the Hamiltonian constraint is satisfied throughout the numerical integration.

\subsection{Stability through Phase Transitions}
\label{subsec:phase_transitions}

It is also interesting to check the stability of the vacuum under phase transitions. This was considered in well-tempered Horndeski cosmology \cite{Appleby:2018yci, Emond:2018fvv}. Now, we perform the same test for the broader class of well-tempered Teledeski models.

\begin{figure}[ht]
\center
	\subfigure[ ]{
		\includegraphics[width = 0.47 \textwidth]{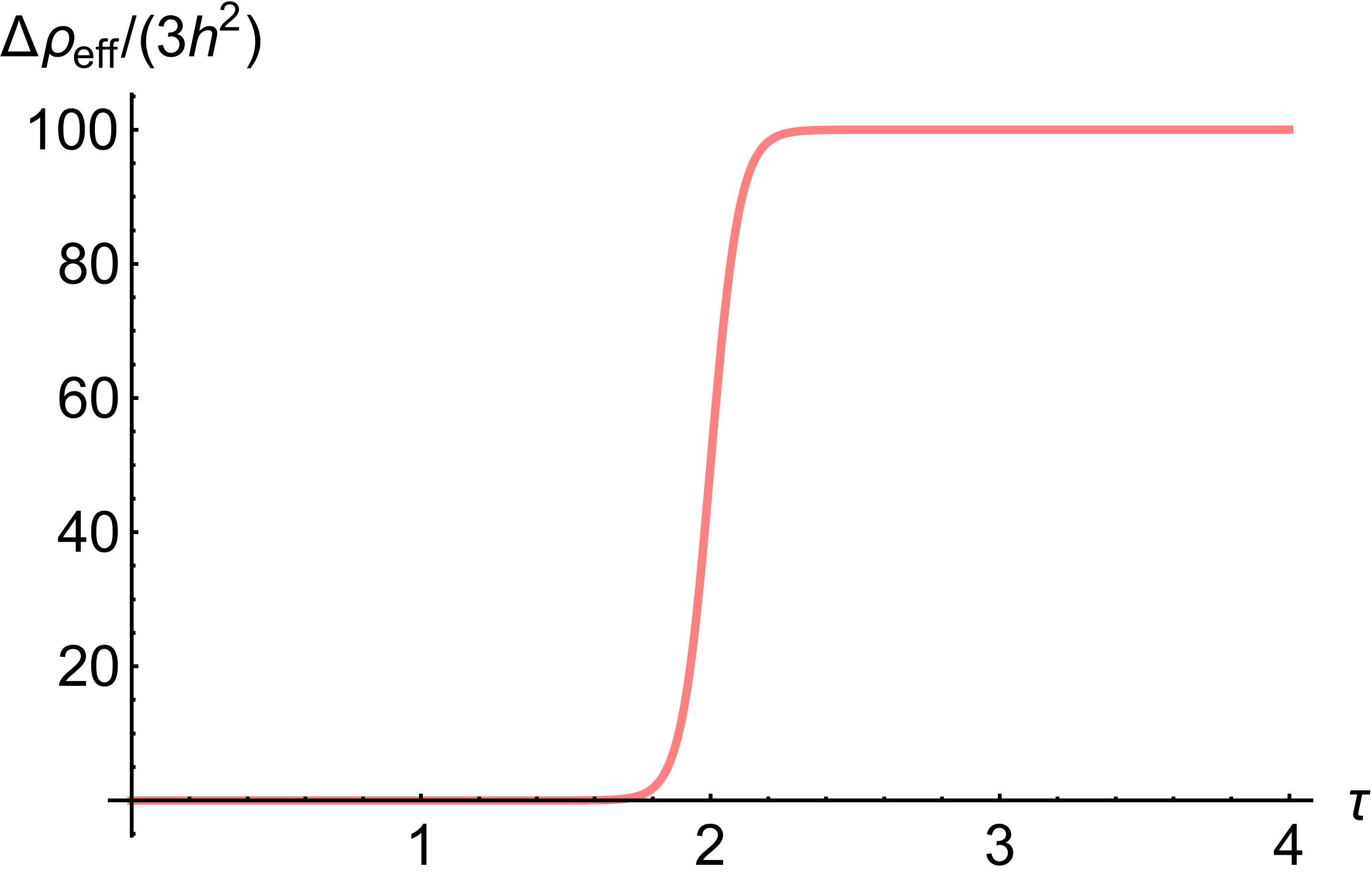}
		}
	\subfigure[ ]{
		\includegraphics[width = 0.47 \textwidth]{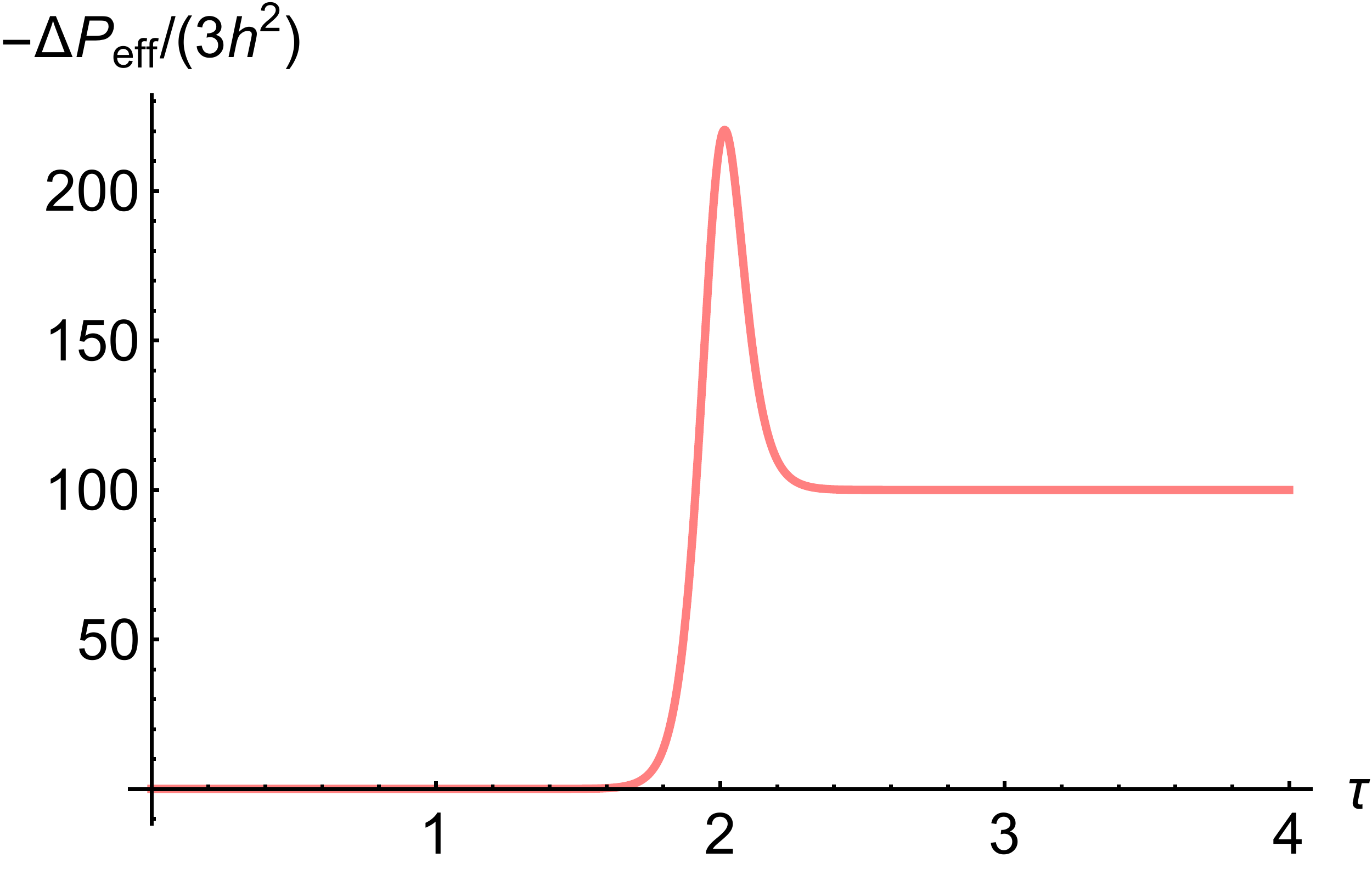}
		}
\caption{The effective (a) energy density $\rho_{\text{eff}}$ and (b) pressure $P_{\text{eff}}$ for a phase transition of two orders of magnitude in the energy scale of the self-tuning vacuum, i.e., $\Delta \rho_\Lambda \sim 10^2 h^2$.}
\label{fig:pt_source}
\end{figure}

We implement the phase transition through the effective energy density and pressure given by \cite{Appleby:2018yci, Emond:2018fvv}
\begin{equation}
\label{eq:rho_eff}
\rho_{\text{eff}} = \rho_\Lambda + \dfrac{ \Delta \rho_\Lambda }{2} \tanh \left( \dfrac{t - T}{\Delta T} \right)\,,
\end{equation}
and
\begin{equation}
\label{eq:P_eff}
P_\text{eff} = - \rho_\text{eff} - \dfrac{ \dot{\rho}_\text{eff} }{ 3H }\,,
\end{equation}
respectively. We stress that $T$ and $\Delta T$ in Eqs. (\ref{eq:rho_eff}) and (\ref{eq:P_eff}) are phase transition time scales and should not be confused with the torsion throughout this section. This describes a vacuum energy in smooth transition from an initial energy density $\rho_\Lambda - \Delta \rho_\Lambda/2$ ($t \ll T$) to $\rho + \Delta \rho_\Lambda/2$ ($t \gg T$) in an interval $\Delta T$ at the time $t \sim T$. This is shown in Fig. \ref{fig:pt_source} where $\tau = h t$, $h T = 2$, $h \Delta T = 1/10$, and $\Delta \rho_\Lambda / 3h^2 = 10^2$.

The negative pressure, $-P_{\text{eff}}$, transitions from $\rho_\Lambda - \Delta \rho_\Lambda/2$ to $\rho_\Lambda + \Delta \rho_\Lambda/2$ in the time interval $h \Delta T$ in the top hat manner as shown in Fig. \ref{fig:pt_source}(b). The fluid clearly acts like vacuum energy, i.e., $P/\rho = -1$, before and after the transition takes place ($t \ll T$ and $t \gg T$).

\begin{figure}[ht]
\center
	\subfigure[ ]{
		\includegraphics[width = 0.47 \textwidth]{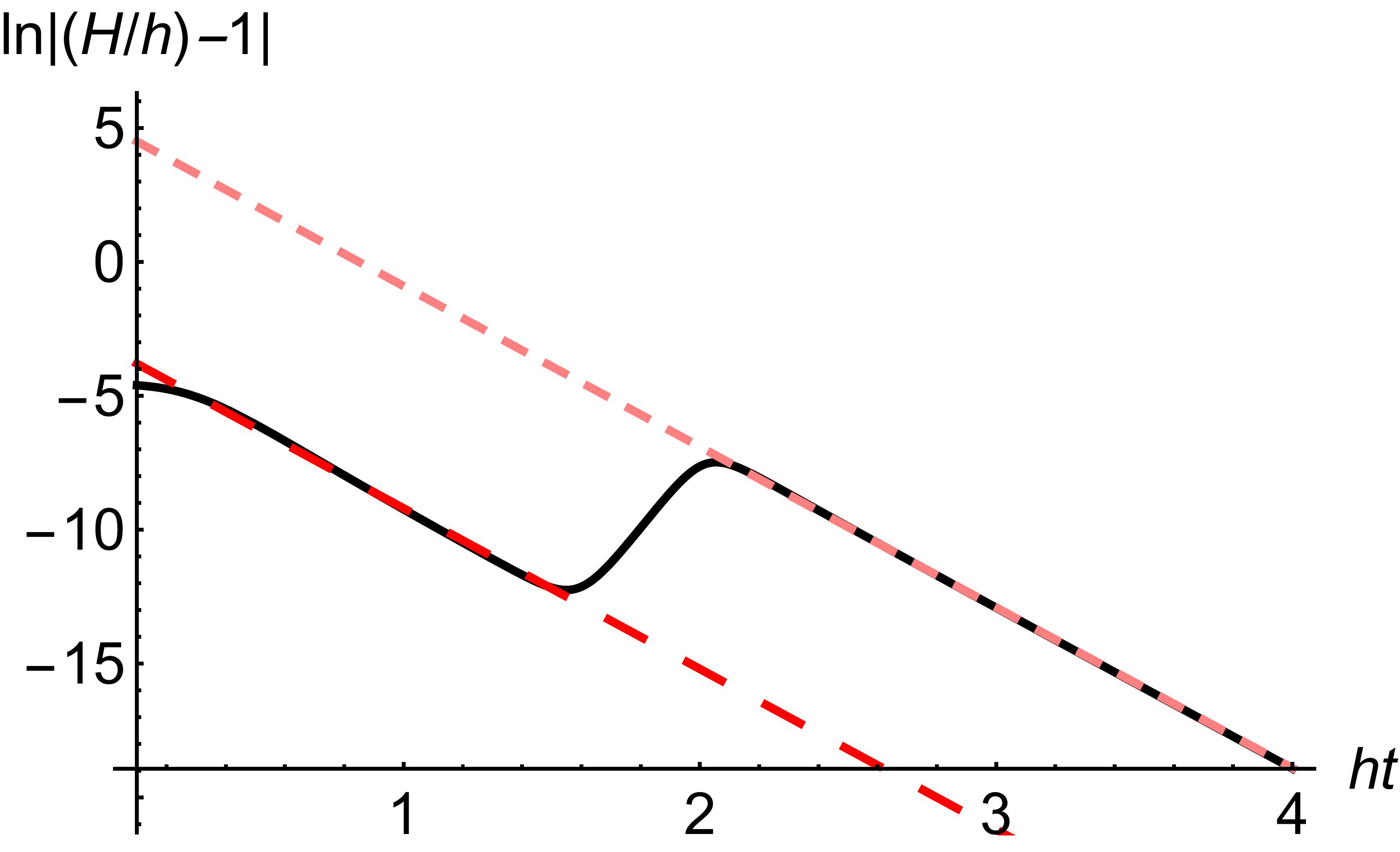}
		}
	\subfigure[ ]{
		\includegraphics[width = 0.47 \textwidth]{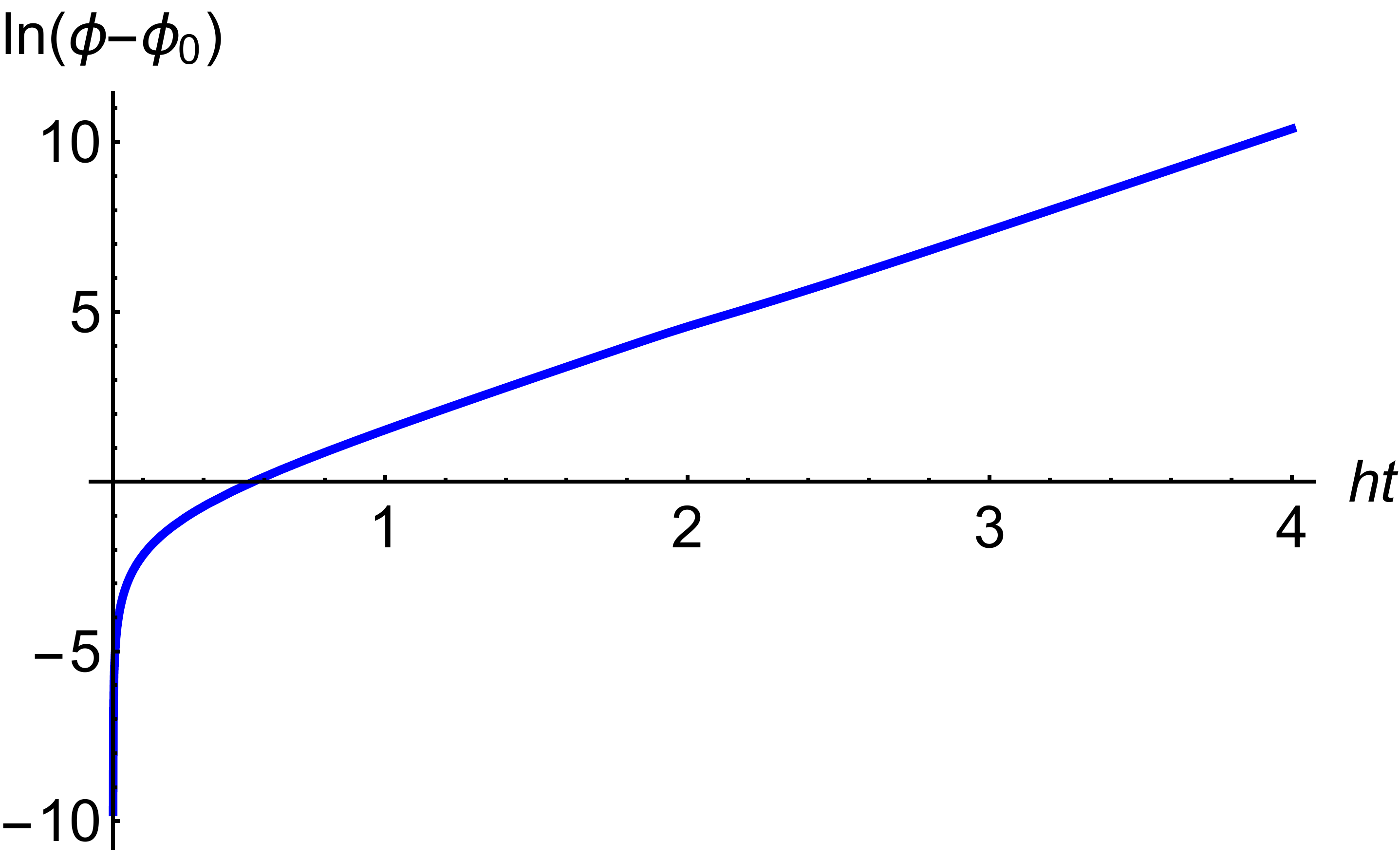}
		}
\caption{The cosmological solution for the phase transition in Fig. \ref{fig:pt_source}: (a) Hubble function and (b) scalar field. The theory parameters are $\left( \alpha = -1, \beta = 1, \gamma = 1, l = 10\right)$. The initial vacuum energy is $\rho_\Lambda/ 3 h^2 \sim 10^{10}$ and the phase transition parameters are $h T = 2$, $h \Delta T = 1/10$, and $\Delta \rho_\Lambda / 3h^2 = 10^2$. The dashed lines in (a) describe the asymptotic solution $\delta H/h \sim e^{- 6 h t}$ (\ref{eq:deltaH}) at the well-tempered vacuum. The red-long-dashed line describes the initial approach to the vacuum (before the phase transition) and the pink-short-dashed line corresponds to the final approach to the vacuum (after the phase transition).}
\label{fig:logHphi_pt}
\end{figure}

Substituting the effective source (Eqs. (\ref{eq:rho_eff}) and (\ref{eq:P_eff})) into the field equations, we can now numerically integrate for the cosmological dynamics and test the stability of the well-tempered vacuum under a phase transition. We consider an initial condition $H \sim h$ and $\phi_0' \sim 1$ fixed on the Hamiltonian constraint and evolve the solution forward. The cosmological solution is shown in Fig. \ref{fig:logHphi_pt} for the theory $\left( \alpha = -1, \beta = 1, \gamma = 1, l = 10\right)$ and the phase transition parameters $h T = 2$, $h \Delta T = 1/10$, and $\Delta \rho_\Lambda / 3h^2 = 10^2$ (Fig. \ref{fig:pt_source}).

The response of the cosmological solution to the phase transition is a clear manifestation of the stability. In Fig. \ref{fig:logHphi_pt}, the transition occurs at the time $h t \sim 2$ for an interval $h t \sim 1/10$. Before the transition effectively takes place, the system is approaching the well-tempered vacuum. This is evident in the asymptotic solution $\delta H/h \sim e^{- 6 h t}$ (\ref{eq:deltaH}) appearing as the red-long-dashed line in Fig. \ref{fig:logHphi_pt}(a). The intervention of the phase transition then disturbs the dynamics. The Hubble function responds in the expected way to the onset of a more negative pressure. The system then relaxes back to the well-tempered vacuum. The asymptotic solution $\delta H/h \sim e^{- 6 h t}$ (\ref{eq:deltaH}) shown as the pink-short-dashed line in Fig. \ref{fig:logHphi_pt}(a) supports this final descent to the well-tempered vacuum. On the other hand, it should be mentioned that the scalar field also responds to the phase transition. However, as the scalar field already well-tempers a titanic vacuum energy $\rho_\Lambda / h^2 \sim 10^{10}$, the effect cannot be seen in Fig. \ref{fig:logHphi_pt}(b).

Our analysis confirms the stability of the well-tempered vacuum under a phase transition of the vacuum energy. This extends the previous work on Horndeski cosmology to Teledeski cosmology.

\section{Conclusions}
\label{sec:conclusions}

We have studied well-tempered cosmology in Teledeski gravity and presented the general approach for obtaining well-tempering in this broader class of scalar-tensor theories. A summary of the various models singled out in this paper is presented in Table \ref{tab:summary}. To the best of our knowledge, this is the first time that a self-tuning mechanism is considered in the context of teleparallel gravity.

We want to particularly highlight the closed-form analytical results in the shift symmetric sector (Sec. \ref{subsec:shift_symmetric}) and the extension of the no-tempering theorem in the tadpole-free, shift symmetric sector (Sec. \ref{subsec:no_tempering_theorem}).

We have also shown the dynamical stability of the vacuum, established the compatibility with a matter era, and demonstrated the vacuum's stability through a phase transition in a well-tempered cosmology (Sec. \ref{sec:dynamics}). This extends the previous work in Horndeski theory and teases the far richer cosmological dynamics that can be achieved in Teledeski cosmology.

Teledeski gravity offers several advantages which are principally sourced by a more general gravitational wave propagation equation giving a direct avenue for reviving previously disqualified models in regular Horndeski gravity. This is presented in Ref.~\cite{Bahamonde:2019ipm} where the speed of propagation is explicitly derived. In Ref.~\cite{Bahamonde:2021dqn} the polarisation modes for various classes of Teledeski gravity are presented together with their propagating degrees of freedom. This is crucial to understanding the dynamics of each class of models of the teleparallel analogue of Horndeski gravity. As future work, it would be interesting to assess precise cosmological models that are both well-tempered and continue to satisfy the gravitational wave propagation speed constraint which comes from observations.

We leave some final remarks that may be addressed in future investigations. First, we have not touched on the soundness of the resulting theory, or rather, whether the kinetic and gradient terms of the perturbations in the quadratic action have the correct signs. This has been addressed in Horndeski theory; however, the cosmological perturbations in Teledeski gravity remains to be calculated. Second, we have left out the quintic Horndeski sector for simplicity. It remains to be seen what this sector adds to well-tempered cosmology. Third, the strong gravity regime, e.g., black holes, of a well-tempered cosmology continues to be untouched territory. This is true in Horndeski theory and more so in its broader teleparallel gravity extension. Fourth, besides well-tempering, self-tuning in scalar-tensor cosmologies can also be achieved via a trivial scalar approach \textit{a la} `Fab Four' \cite{Charmousis:2011bf}. In this way, the over-constrained dynamical system on a de Sitter vacuum is resolved by making the scalar field equation trivially-satisfied on-shell. There may be other kinds of self-tuning yet to be discovered. Lastly, well-tempered cosmology, so far, has not been constrained with cosmological data. In the abundance of data from current and planned cosmological surveys, a step toward this should be considered soon.

\begin{acknowledgments}\label{sec:acknowledgements}
JLS would like to acknowledge networking support by the COST Action CA18108 and funding support from Cosmology@MALTA which is supported by the University of Malta. JLS would also like to acknowledge funding from ``The Malta Council for Science and Technology'' in project IPAS-2020-007. SA is supported by an appointment to the JRG Program at the APCTP through the Science and Technology Promotion Fund and Lottery Fund of the Korean Government, and was also supported by the Korean Local Governments in Gyeongsangbuk-do Province and Pohang City. MC would like to acknowledge funding by the Tertiary Education Scholarship Scheme (TESS, Malta).
\end{acknowledgments}

\appendix


\begin{thebibliography}{10}

\bibitem{RevModPhys.61.1}
S.~Weinberg, \emph{The cosmological constant problem},
  \href{https://doi.org/10.1103/RevModPhys.61.1}{\emph{Rev. Mod. Phys.}
  {\bfseries 61} (1989) 1}.

\bibitem{Dolgov:1982gh}
A.~D. Dolgov, \emph{{An attempt to get rid of the Cosmological Constant}},  in
  \emph{{Nuffield Workshop on the Very Early Universe}}, 1982.

\bibitem{Kaloper:2013zca}
N.~Kaloper and A.~Padilla, \emph{{Sequestering the Standard Model Vacuum
  Energy}}, \href{https://doi.org/10.1103/PhysRevLett.112.091304}{\emph{Phys.
  Rev. Lett.} {\bfseries 112} (2014) 091304}
  [\href{https://arxiv.org/abs/1309.6562}{{\ttfamily 1309.6562}}].

\bibitem{Kaloper:2014fca}
N.~Kaloper and A.~Padilla, \emph{{Sequestration of Vacuum Energy and the End of
  the Universe}},
  \href{https://doi.org/10.1103/PhysRevLett.114.101302}{\emph{Phys. Rev. Lett.}
  {\bfseries 114} (2015) 101302}
  [\href{https://arxiv.org/abs/1409.7073}{{\ttfamily 1409.7073}}].

\bibitem{Brax:2019fgj}
P.~Brax and P.~Valageas, \emph{{Cosmological cancellation of the vacuum energy
  density}}, \href{https://doi.org/10.1103/PhysRevD.99.123506}{\emph{Phys. Rev.
  D} {\bfseries 99} (2019) 123506}
  [\href{https://arxiv.org/abs/1903.04825}{{\ttfamily 1903.04825}}].

\bibitem{SobralBlanco:2020too}
D.~Sobral~Blanco and L.~Lombriser, \emph{{Local self-tuning mechanism for the
  cosmological constant}},
  \href{https://doi.org/10.1103/PhysRevD.102.043506}{\emph{Phys. Rev. D}
  {\bfseries 102} (2020) 043506}
  [\href{https://arxiv.org/abs/2003.04303}{{\ttfamily 2003.04303}}].

\bibitem{Lombriser:2019jia}
L.~Lombriser, \emph{{On the cosmological constant problem}},
  \href{https://doi.org/10.1016/j.physletb.2019.134804}{\emph{Phys. Lett. B}
  {\bfseries 797} (2019) 134804}
  [\href{https://arxiv.org/abs/1901.08588}{{\ttfamily 1901.08588}}].

\bibitem{Kaloper:2015jra}
N.~Kaloper, A.~Padilla, D.~Stefanyszyn and G.~Zahariade, \emph{{Manifestly
  Local Theory of Vacuum Energy Sequestering}},
  \href{https://doi.org/10.1103/PhysRevLett.116.051302}{\emph{Phys. Rev. Lett.}
  {\bfseries 116} (2016) 051302}
  [\href{https://arxiv.org/abs/1505.01492}{{\ttfamily 1505.01492}}].

\bibitem{Arkani-Hamed:2002ukf}
N.~Arkani-Hamed, S.~Dimopoulos, G.~Dvali and G.~Gabadadze, \emph{{Nonlocal
  modification of gravity and the cosmological constant problem}},
  \href{https://arxiv.org/abs/hep-th/0209227}{{\ttfamily hep-th/0209227}}.

\bibitem{Amariti:2019vfv}
A.~Amariti, C.~Charmousis, D.~Forcella, E.~Kiritsis and F.~Nitti, \emph{{Brane
  cosmology and the self-tuning of the cosmological constant}},
  \href{https://doi.org/10.1088/1475-7516/2019/10/007}{\emph{JCAP} {\bfseries
  10} (2019) 007} [\href{https://arxiv.org/abs/1904.02727}{{\ttfamily
  1904.02727}}].

\bibitem{Evnin:2018zeo}
O.~Evnin and K.~Nguyen, \emph{{Graceful exit for the cosmological constant
  damping scenario}},
  \href{https://doi.org/10.1103/PhysRevD.98.124031}{\emph{Phys. Rev. D}
  {\bfseries 98} (2018) 124031}
  [\href{https://arxiv.org/abs/1810.12336}{{\ttfamily 1810.12336}}].

\bibitem{Horndeski:1974wa}
G.~W. Horndeski, \emph{{Second-order scalar-tensor field equations in a
  four-dimensional space}},
  \href{https://doi.org/10.1007/BF01807638}{\emph{Int. J. Theor. Phys.}
  {\bfseries 10} (1974) 363}.

\bibitem{Charmousis:2011bf}
C.~Charmousis, E.~J. Copeland, A.~Padilla and P.~M. Saffin, \emph{{General
  second order scalar-tensor theory, self tuning, and the Fab Four}},
  \href{https://doi.org/10.1103/PhysRevLett.108.051101}{\emph{Phys. Rev. Lett.}
  {\bfseries 108} (2012) 051101}
  [\href{https://arxiv.org/abs/1106.2000}{{\ttfamily 1106.2000}}].

\bibitem{Appleby:2018yci}
S.~Appleby and E.~V. Linder, \emph{{The Well-Tempered Cosmological Constant}},
  \href{https://doi.org/10.1088/1475-7516/2018/07/034}{\emph{JCAP} {\bfseries
  07} (2018) 034} [\href{https://arxiv.org/abs/1805.00470}{{\ttfamily
  1805.00470}}].

\bibitem{Appleby:2020njl}
S.~Appleby and E.~V. Linder, \emph{{The Well-Tempered Cosmological Constant:
  The Horndeski Variations}},
  \href{https://doi.org/10.1088/1475-7516/2020/12/036}{\emph{JCAP} {\bfseries
  12} (2020) 036} [\href{https://arxiv.org/abs/2009.01720}{{\ttfamily
  2009.01720}}].

\bibitem{Appleby:2020dko}
S.~Appleby and E.~V. Linder, \emph{{The well-tempered cosmological constant:
  fugue in B$^\flat$}},
  \href{https://doi.org/10.1088/1475-7516/2020/12/037}{\emph{JCAP} {\bfseries
  12} (2020) 037} [\href{https://arxiv.org/abs/2009.01723}{{\ttfamily
  2009.01723}}].

\bibitem{Linder:2020xey}
E.~V. Linder and S.~Appleby, \emph{{An Expansion of Well Tempered Gravity}},
  \href{https://doi.org/10.1088/1475-7516/2021/03/074}{\emph{JCAP} {\bfseries
  03} (2021) 074} [\href{https://arxiv.org/abs/2012.03965}{{\ttfamily
  2012.03965}}].

\bibitem{Bernardo:2021bsg}
R.~C. Bernardo, J.~L. Said, M.~Caruana and S.~Appleby, \emph{{Well-Tempered
  Minkowski Solutions in Teleparallel Horndeski Theory}},
  \href{https://arxiv.org/abs/2108.02500}{{\ttfamily 2108.02500}}.

\bibitem{Emond:2018fvv}
W.~T. Emond, C.~Li, P.~M. Saffin and S.-Y. Zhou, \emph{{Well-Tempered
  Cosmology}}, \href{https://doi.org/10.1088/1475-7516/2019/05/038}{\emph{JCAP}
  {\bfseries 05} (2019) 038}
  [\href{https://arxiv.org/abs/1812.05480}{{\ttfamily 1812.05480}}].

\bibitem{Bernardo:2021hrz}
R.~C. Bernardo, \emph{{Self-tuning kinetic gravity braiding: Cosmological
  dynamics, shift symmetry, and the tadpole}},
  \href{https://doi.org/10.1088/1475-7516/2021/03/079}{\emph{JCAP} {\bfseries
  03} (2021) 079} [\href{https://arxiv.org/abs/2101.00965}{{\ttfamily
  2101.00965}}].

\bibitem{Copeland:2012qf}
E.~J. Copeland, A.~Padilla and P.~M. Saffin, \emph{{The cosmology of the
  Fab-Four}}, \href{https://doi.org/10.1088/1475-7516/2012/12/026}{\emph{JCAP}
  {\bfseries 12} (2012) 026} [\href{https://arxiv.org/abs/1208.3373}{{\ttfamily
  1208.3373}}].

\bibitem{Starobinsky:2016kua}
A.~A. Starobinsky, S.~V. Sushkov and M.~S. Volkov, \emph{{The screening
  Horndeski cosmologies}},
  \href{https://doi.org/10.1088/1475-7516/2016/06/007}{\emph{JCAP} {\bfseries
  06} (2016) 007} [\href{https://arxiv.org/abs/1604.06085}{{\ttfamily
  1604.06085}}].

\bibitem{Torres:2018lni}
I.~Torres, J.~C. Fabris and O.~F. Piattella, \emph{{Classical and quantum
  cosmology of Fab Four John theories}},
  \href{https://doi.org/10.1016/j.physletb.2019.135003}{\emph{Phys. Lett. B}
  {\bfseries 798} (2019) 135003}
  [\href{https://arxiv.org/abs/1811.08852}{{\ttfamily 1811.08852}}].

\bibitem{Appleby:2015ysa}
S.~Appleby, \emph{{Self Tuning Scalar Fields in Spherically Symmetric
  Spacetimes}},
  \href{https://doi.org/10.1088/1475-7516/2015/05/009}{\emph{JCAP} {\bfseries
  05} (2015) 009} [\href{https://arxiv.org/abs/1503.06768}{{\ttfamily
  1503.06768}}].

\bibitem{Appleby:2012rx}
S.~A. Appleby, A.~De~Felice and E.~V. Linder, \emph{{Fab 5: Noncanonical
  Kinetic Gravity, Self Tuning, and Cosmic Acceleration}},
  \href{https://doi.org/10.1088/1475-7516/2012/10/060}{\emph{JCAP} {\bfseries
  10} (2012) 060} [\href{https://arxiv.org/abs/1208.4163}{{\ttfamily
  1208.4163}}].

\bibitem{LIGOScientific:2017vwq}
{\scshape LIGO Scientific, Virgo} collaboration, \emph{{GW170817: Observation
  of Gravitational Waves from a Binary Neutron Star Inspiral}},
  \href{https://doi.org/10.1103/PhysRevLett.119.161101}{\emph{Phys. Rev. Lett.}
  {\bfseries 119} (2017) 161101}
  [\href{https://arxiv.org/abs/1710.05832}{{\ttfamily 1710.05832}}].

\bibitem{Coulter:2017wya}
D.~A. Coulter et~al., \emph{{Swope Supernova Survey 2017a (SSS17a), the Optical
  Counterpart to a Gravitational Wave Source}},
  \href{https://doi.org/10.1126/science.aap9811}{\emph{Science} {\bfseries 358}
  (2017) 1556} [\href{https://arxiv.org/abs/1710.05452}{{\ttfamily
  1710.05452}}].

\bibitem{LIGOScientific:2017ync}
B.~P. Abbott
  et~al.\href{https://doi.org/10.3847/2041-8213/aa91c9}{\emph{Astrophys. J.
  Lett.} {\bfseries 848} (2017) L12}
  [\href{https://arxiv.org/abs/1710.05833}{{\ttfamily 1710.05833}}].

\bibitem{LIGOScientific:2017zic}
{\scshape LIGO Scientific, Virgo, Fermi-GBM, INTEGRAL} collaboration,
  \emph{{Gravitational Waves and Gamma-rays from a Binary Neutron Star Merger:
  GW170817 and GRB 170817A}},
  \href{https://doi.org/10.3847/2041-8213/aa920c}{\emph{Astrophys. J. Lett.}
  {\bfseries 848} (2017) L13}
  [\href{https://arxiv.org/abs/1710.05834}{{\ttfamily 1710.05834}}].

\bibitem{Lombriser:2015sxa}
L.~Lombriser and A.~Taylor, \emph{{Breaking a Dark Degeneracy with
  Gravitational Waves}},
  \href{https://doi.org/10.1088/1475-7516/2016/03/031}{\emph{JCAP} {\bfseries
  03} (2016) 031} [\href{https://arxiv.org/abs/1509.08458}{{\ttfamily
  1509.08458}}].

\bibitem{Babichev:2017lmw}
E.~Babichev, C.~Charmousis, G.~Esposito-Far\`ese and A.~Leh\'ebel,
  \emph{{Stability of Black Holes and the Speed of Gravitational Waves within
  Self-Tuning Cosmological Models}},
  \href{https://doi.org/10.1103/PhysRevLett.120.241101}{\emph{Phys. Rev. Lett.}
  {\bfseries 120} (2018) 241101}
  [\href{https://arxiv.org/abs/1712.04398}{{\ttfamily 1712.04398}}].

\bibitem{Bahamonde:2021gfp}
S.~Bahamonde, K.~F. Dialektopoulos, C.~Escamilla-Rivera, G.~Farrugia, V.~Gakis,
  M.~Hendry et~al., \emph{{Teleparallel Gravity: From Theory to Cosmology}},
  \href{https://arxiv.org/abs/2106.13793}{{\ttfamily 2106.13793}}.

\bibitem{Aldrovandi:2013wha}
R.~Aldrovandi and J.~G. Pereira, \emph{{Teleparallel Gravity}}, vol.~173.
  Springer, Dordrecht, 2013,
  \href{https://doi.org/10.1007/978-94-007-5143-9}{10.1007/978-94-007-5143-9}.

\bibitem{Cai:2015emx}
Y.-F. Cai, S.~Capozziello, M.~De~Laurentis and E.~N. Saridakis, \emph{{$f(T)$
  teleparallel gravity and cosmology}},
  \href{https://doi.org/10.1088/0034-4885/79/10/106901}{\emph{Rept. Prog.
  Phys.} {\bfseries 79} (2016) 106901}
  [\href{https://arxiv.org/abs/1511.07586}{{\ttfamily 1511.07586}}].

\bibitem{Krssak:2018ywd}
M.~Kr{\v{s}}{\v{s}}{\'{a}}k, R.~van~den Hoogen, J.~Pereira, C.~B{\"{o}}hmer and
  A.~Coley, \emph{{Teleparallel theories of gravity: illuminating a fully
  invariant approach}},
  \href{https://doi.org/10.1088/1361-6382/ab2e1f}{\emph{Class. Quant. Grav.}
  {\bfseries 36} (2019) 183001}
  [\href{https://arxiv.org/abs/1810.12932}{{\ttfamily 1810.12932}}].

\bibitem{Weitzenbock1923}
R.~Weitzenb\"{o}ck, \emph{`Invariantentheorie'}. Noordhoff, Gronningen, 1923.

\bibitem{Lovelock:1971yv}
D.~Lovelock, \emph{{The Einstein tensor and its generalizations}},
  \href{https://doi.org/10.1063/1.1665613}{\emph{J. Math. Phys.} {\bfseries 12}
  (1971) 498}.

\bibitem{Gonzalez:2015sha}
P.~Gonzalez and Y.~Vasquez, \emph{{Teleparallel Equivalent of Lovelock
  Gravity}}, \href{https://doi.org/10.1103/PhysRevD.92.124023}{\emph{Phys. Rev.
  D} {\bfseries 92} (2015) 124023}
  [\href{https://arxiv.org/abs/1508.01174}{{\ttfamily 1508.01174}}].

\bibitem{Bahamonde:2019shr}
S.~Bahamonde, K.~F. Dialektopoulos and J.~Levi~Said, \emph{{Can Horndeski
  Theory be recast using Teleparallel Gravity?}},
  \href{https://doi.org/10.1103/PhysRevD.100.064018}{\emph{Phys. Rev. D}
  {\bfseries 100} (2019) 064018}
  [\href{https://arxiv.org/abs/1904.10791}{{\ttfamily 1904.10791}}].

\bibitem{Sotiriou:2008rp}
T.~P. Sotiriou and V.~Faraoni, \emph{{f(R) Theories Of Gravity}},
  \href{https://doi.org/10.1103/RevModPhys.82.451}{\emph{Rev. Mod. Phys.}
  {\bfseries 82} (2010) 451} [\href{https://arxiv.org/abs/0805.1726}{{\ttfamily
  0805.1726}}].

\bibitem{Faraoni:2008mf}
V.~Faraoni, \emph{{f(R) gravity: Successes and challenges}},  in \emph{{18th
  SIGRAV Conference}}, 10, 2008,
  \href{https://arxiv.org/abs/0810.2602}{{\ttfamily 0810.2602}}.

\bibitem{Capozziello:2011et}
S.~Capozziello and M.~De~Laurentis, \emph{{Extended Theories of Gravity}},
  \href{https://doi.org/10.1016/j.physrep.2011.09.003}{\emph{Phys. Rept.}
  {\bfseries 509} (2011) 167}
  [\href{https://arxiv.org/abs/1108.6266}{{\ttfamily 1108.6266}}].

\bibitem{Ferraro:2006jd}
R.~Ferraro and F.~Fiorini, \emph{{Modified teleparallel gravity: Inflation
  without inflaton}},
  \href{https://doi.org/10.1103/PhysRevD.75.084031}{\emph{Phys. Rev.}
  {\bfseries D75} (2007) 084031}
  [\href{https://arxiv.org/abs/gr-qc/0610067}{{\ttfamily gr-qc/0610067}}].

\bibitem{Ferraro:2008ey}
R.~Ferraro and F.~Fiorini, \emph{{On Born-Infeld Gravity in Weitzenbock
  spacetime}}, \href{https://doi.org/10.1103/PhysRevD.78.124019}{\emph{Phys.
  Rev.} {\bfseries D78} (2008) 124019}
  [\href{https://arxiv.org/abs/0812.1981}{{\ttfamily 0812.1981}}].

\bibitem{Bengochea:2008gz}
G.~R. Bengochea and R.~Ferraro, \emph{{Dark torsion as the cosmic speed-up}},
  \href{https://doi.org/10.1103/PhysRevD.79.124019}{\emph{Phys. Rev.}
  {\bfseries D79} (2009) 124019}
  [\href{https://arxiv.org/abs/0812.1205}{{\ttfamily 0812.1205}}].

\bibitem{Linder:2010py}
E.~V. Linder, \emph{{Einstein's Other Gravity and the Acceleration of the
  Universe}}, \href{https://doi.org/10.1103/PhysRevD.81.127301,
  10.1103/PhysRevD.82.109902}{\emph{Phys. Rev.} {\bfseries D81} (2010) 127301}
  [\href{https://arxiv.org/abs/1005.3039}{{\ttfamily 1005.3039}}].

\bibitem{Chen:2010va}
S.-H. Chen, J.~B. Dent, S.~Dutta and E.~N. Saridakis, \emph{{Cosmological
  perturbations in f(T) gravity}},
  \href{https://doi.org/10.1103/PhysRevD.83.023508}{\emph{Phys. Rev.}
  {\bfseries D83} (2011) 023508}
  [\href{https://arxiv.org/abs/1008.1250}{{\ttfamily 1008.1250}}].

\bibitem{Bahamonde:2019zea}
S.~Bahamonde, K.~Flathmann and C.~Pfeifer, \emph{{Photon sphere and perihelion
  shift in weak $f(T)$ gravity}},
  \href{https://doi.org/10.1103/PhysRevD.100.084064}{\emph{Phys. Rev. D}
  {\bfseries 100} (2019) 084064}
  [\href{https://arxiv.org/abs/1907.10858}{{\ttfamily 1907.10858}}].

\bibitem{RezaeiAkbarieh:2018ijw}
A.~Rezaei~Akbarieh and Y.~Izadi, \emph{{Tachyon Inflation in Teleparallel
  Gravity}}, \href{https://doi.org/10.1140/epjc/s10052-019-6819-z}{\emph{Eur.
  Phys. J. C} {\bfseries 79} (2019) 366}
  [\href{https://arxiv.org/abs/1812.06649}{{\ttfamily 1812.06649}}].

\bibitem{Benetti:2020hxp}
M.~Benetti, S.~Capozziello and G.~Lambiase, \emph{{Updating constraints on f(T)
  teleparallel cosmology and the consistency with Big Bang Nucleosynthesis}},
  \href{https://doi.org/10.1093/mnras/staa3368}{\emph{Mon. Not. Roy. Astron.
  Soc.} {\bfseries 500} (2020) 1795}
  [\href{https://arxiv.org/abs/2006.15335}{{\ttfamily 2006.15335}}].

\bibitem{Nesseris:2013jea}
S.~Nesseris, S.~Basilakos, E.~N. Saridakis and L.~Perivolaropoulos,
  \emph{{Viable $f(T)$ models are practically indistinguishable from
  $\Lambda$CDM}}, \href{https://doi.org/10.1103/PhysRevD.88.103010}{\emph{Phys.
  Rev.} {\bfseries D88} (2013) 103010}
  [\href{https://arxiv.org/abs/1308.6142}{{\ttfamily 1308.6142}}].

\bibitem{Anagnostopoulos:2019miu}
F.~K. Anagnostopoulos, S.~Basilakos and E.~N. Saridakis, \emph{{Bayesian
  analysis of $f(T)$ gravity using $f\sigma_8$ data}},
  \href{https://doi.org/10.1103/PhysRevD.100.083517}{\emph{Phys. Rev.}
  {\bfseries D100} (2019) 083517}
  [\href{https://arxiv.org/abs/1907.07533}{{\ttfamily 1907.07533}}].

\bibitem{Nunes:2018evm}
R.~C. Nunes, S.~Pan and E.~N. Saridakis, \emph{{New observational constraints
  on $f(T)$ gravity through gravitational-wave astronomy}},
  \href{https://doi.org/10.1103/PhysRevD.98.104055}{\emph{Phys. Rev. D}
  {\bfseries 98} (2018) 104055}
  [\href{https://arxiv.org/abs/1810.03942}{{\ttfamily 1810.03942}}].

\bibitem{Farrugia:2016qqe}
G.~Farrugia and J.~L. Said, \emph{{Stability of the flat FLRW metric in $f(T)$
  gravity}}, \href{https://doi.org/10.1103/PhysRevD.94.124054}{\emph{Phys.
  Rev.} {\bfseries D94} (2016) 124054}
  [\href{https://arxiv.org/abs/1701.00134}{{\ttfamily 1701.00134}}].

\bibitem{Deng:2018ncg}
X.-M. Deng, \emph{{Probing f(T) gravity with gravitational time advancement}},
  \href{https://doi.org/10.1088/1361-6382/aad391}{\emph{Class. Quant. Grav.}
  {\bfseries 35} (2018) 175013}.

\bibitem{Hohmann:2018vle}
M.~Hohmann, \emph{{Scalar-torsion theories of gravity I: general formalism and
  conformal transformations}},
  \href{https://doi.org/10.1103/PhysRevD.98.064002}{\emph{Phys. Rev. D}
  {\bfseries 98} (2018) 064002}
  [\href{https://arxiv.org/abs/1801.06528}{{\ttfamily 1801.06528}}].

\bibitem{Hohmann:2018dqh}
M.~Hohmann and C.~Pfeifer, \emph{{Scalar-torsion theories of gravity II: $L(T,
  X, Y, \phi)$ theory}},
  \href{https://doi.org/10.1103/PhysRevD.98.064003}{\emph{Phys. Rev. D}
  {\bfseries 98} (2018) 064003}
  [\href{https://arxiv.org/abs/1801.06536}{{\ttfamily 1801.06536}}].

\bibitem{Hohmann:2018ijr}
M.~Hohmann, \emph{{Scalar-torsion theories of gravity III: analogue of
  scalar-tensor gravity and conformal invariants}},
  \href{https://doi.org/10.1103/PhysRevD.98.064004}{\emph{Phys. Rev. D}
  {\bfseries 98} (2018) 064004}
  [\href{https://arxiv.org/abs/1801.06531}{{\ttfamily 1801.06531}}].

\bibitem{Ezquiaga:2017ekz}
J.~M. Ezquiaga and M.~Zumalac\'arregui, \emph{{Dark Energy After GW170817: Dead
  Ends and the Road Ahead}},
  \href{https://doi.org/10.1103/PhysRevLett.119.251304}{\emph{Phys. Rev. Lett.}
  {\bfseries 119} (2017) 251304}
  [\href{https://arxiv.org/abs/1710.05901}{{\ttfamily 1710.05901}}].

\bibitem{Bahamonde:2019ipm}
S.~Bahamonde, K.~F. Dialektopoulos, V.~Gakis and J.~Levi~Said, \emph{{Reviving
  Horndeski theory using teleparallel gravity after GW170817}},
  \href{https://doi.org/10.1103/PhysRevD.101.084060}{\emph{Phys. Rev. D}
  {\bfseries 101} (2020) 084060}
  [\href{https://arxiv.org/abs/1907.10057}{{\ttfamily 1907.10057}}].

\bibitem{Bahamonde:2021dqn}
S.~Bahamonde, M.~Caruana, K.~F. Dialektopoulos, V.~Gakis, M.~Hohmann,
  J.~Levi~Said et~al., \emph{{Gravitational Wave Propagation and Polarizations
  in the Teleparallel analog of Horndeski Gravity}},
  \href{https://arxiv.org/abs/2105.13243}{{\ttfamily 2105.13243}}.

\bibitem{reggie_bernardo_4810864}
R.~Bernardo, ``{reggiebernardo/notebooks: dark energy research notebooks}.''
  \href{https://doi.org/10.5281/zenodo.4810864}{10.5281/zenodo.4810864}, 2021.

\bibitem{ortin2004gravity}
T.~Ort{\'\i}n, \emph{Gravity and Strings}, Cambridge Monographs on Mathematical
  Physics. Cambridge University Press, 2004.

\bibitem{Krssak:2015oua}
M.~Kr{\v{s}}{\v{s}}{\'{a}}k and E.~N. Saridakis, \emph{{The covariant
  formulation of f(T) gravity}},
  \href{https://doi.org/10.1088/0264-9381/33/11/115009}{\emph{Class. Quant.
  Grav.} {\bfseries 33} (2016) 115009}
  [\href{https://arxiv.org/abs/1510.08432}{{\ttfamily 1510.08432}}].

\bibitem{PhysRevD.19.3524}
K.~Hayashi and T.~Shirafuji, \emph{New general relativity},
  \href{https://doi.org/10.1103/PhysRevD.19.3524}{\emph{Phys. Rev. D}
  {\bfseries 19} (1979) 3524}.

\bibitem{Bahamonde:2017wwk}
S.~Bahamonde, C.~G. B{\"{o}}hmer and M.~Kr{\v{s}}{\v{s}}{\'{a}}k, \emph{{New
  classes of modified teleparallel gravity models}},
  \href{https://doi.org/10.1016/j.physletb.2017.10.026}{\emph{Phys. Lett.}
  {\bfseries B775} (2017) 37}
  [\href{https://arxiv.org/abs/1706.04920}{{\ttfamily 1706.04920}}].

\bibitem{Bahamonde:2015zma}
S.~Bahamonde, C.~G. B{\"{o}}hmer and M.~Wright, \emph{{Modified teleparallel
  theories of gravity}},
  \href{https://doi.org/10.1103/PhysRevD.92.104042}{\emph{Phys. Rev.}
  {\bfseries D92} (2015) 104042}
  [\href{https://arxiv.org/abs/1508.05120}{{\ttfamily 1508.05120}}].

\bibitem{Hehl:1994ue}
F.~W. Hehl, J.~D. McCrea, E.~W. Mielke and Y.~Ne'eman, \emph{{Metric affine
  gauge theory of gravity: Field equations, Noether identities, world spinors,
  and breaking of dilation invariance}},
  \href{https://doi.org/10.1016/0370-1573(94)00111-F}{\emph{Phys. Rept.}
  {\bfseries 258} (1995) 1}
  [\href{https://arxiv.org/abs/gr-qc/9402012}{{\ttfamily gr-qc/9402012}}].

\bibitem{Ferraro:2016wht}
R.~Ferraro and M.~J. Guzm\'an, \emph{{Hamiltonian formulation of teleparallel
  gravity}}, \href{https://doi.org/10.1103/PhysRevD.94.104045}{\emph{Phys. Rev.
  D} {\bfseries 94} (2016) 104045}
  [\href{https://arxiv.org/abs/1609.06766}{{\ttfamily 1609.06766}}].

\bibitem{Blagojevic:2020dyq}
M.~Blagojevi\'c and J.~M. Nester, \emph{{Local symmetries and physical degrees
  of freedom in $f(T)$ gravity: a Dirac Hamiltonian constraint analysis}},
  \href{https://doi.org/10.1103/PhysRevD.102.064025}{\emph{Phys. Rev. D}
  {\bfseries 102} (2020) 064025}
  [\href{https://arxiv.org/abs/2006.15303}{{\ttfamily 2006.15303}}].

\bibitem{Jimenez:2020ofm}
J.~B. Jim\'enez, A.~Golovnev, T.~Koivisto and H.~Veerm\"ae, \emph{{Minkowski
  space in $f(T)$ gravity}},
  \href{https://doi.org/10.1103/PhysRevD.103.024054}{\emph{Phys. Rev. D}
  {\bfseries 103} (2021) 024054}
  [\href{https://arxiv.org/abs/2004.07536}{{\ttfamily 2004.07536}}].

\bibitem{Golovnev:2020zpv}
A.~Golovnev and M.-J. Guzm\'an, \emph{{Foundational issues in f(T) gravity
  theory}}, \href{https://doi.org/10.1142/S0219887821400077}{\emph{Int. J.
  Geom. Meth. Mod. Phys.} {\bfseries 18} (2021) 2140007}
  [\href{https://arxiv.org/abs/2012.14408}{{\ttfamily 2012.14408}}].

\bibitem{BeltranJimenez:2020sih}
J.~Beltrán~Jiménez, L.~Heisenberg and T.~Koivisto, \emph{{The coupling of
  matter and spacetime geometry}},
  \href{https://doi.org/10.1088/1361-6382/aba31b}{\emph{Class. Quant. Grav.}
  {\bfseries 37} (2020) 195013}
  [\href{https://arxiv.org/abs/2004.04606}{{\ttfamily 2004.04606}}].

\bibitem{Bahamonde:2020cfv}
S.~Bahamonde, K.~F. Dialektopoulos, M.~Hohmann and J.~Levi~Said,
  \emph{{Post-Newtonian limit of Teleparallel Horndeski gravity}},
  \href{https://doi.org/10.1088/1361-6382/abc441}{\emph{Class. Quant. Grav.}
  {\bfseries 38} (2020) 025006}
  [\href{https://arxiv.org/abs/2003.11554}{{\ttfamily 2003.11554}}].

\bibitem{Gonzalez:2019tky}
P.~González, S.~Reyes and Y.~Vásquez, \emph{{Teleparallel Equivalent of
  Lovelock Gravity, Generalizations and Cosmological Applications}},
  \href{https://doi.org/10.1088/1475-7516/2019/07/040}{\emph{JCAP} {\bfseries
  07} (2019) 040} [\href{https://arxiv.org/abs/1905.07633}{{\ttfamily
  1905.07633}}].

\bibitem{Kobayashi:2011nu}
T.~Kobayashi, M.~Yamaguchi and J.~Yokoyama, \emph{{Generalized G-inflation:
  Inflation with the most general second-order field equations}},
  \href{https://doi.org/10.1143/PTP.126.511}{\emph{Prog. Theor. Phys.}
  {\bfseries 126} (2011) 511}
  [\href{https://arxiv.org/abs/1105.5723}{{\ttfamily 1105.5723}}].

\end{thebibliography}

\providecommand{\href}[2]{#2}\begingroup\raggedright\endgroup

\end{document}